\documentclass[prd,onecolumn,nofootinbib,aps,tightenlines,preprintnumbers,notitlepage,longbibliography,superscriptaddress]{revtex4-1}
\usepackage{graphicx}
\usepackage{color}
\usepackage{amsmath}
\usepackage{amsfonts}
\usepackage[colorlinks=true,citecolor=blue,urlcolor=blue]{hyperref}

\newcommand{\Planck}{\textit{Planck}}
\newcommand{\CLASS}{\texttt{CLASS}}
\newcommand{\Recfast}{\texttt{Recfast}}
\newcommand{\hyrec}{\texttt{HyRec}}

\newcommand{\cm}{\mathrm{cm}}

\newcommand{\Mpc}{\mathrm{Mpc}}

\newcommand{\MeV}{\mathrm{MeV}}
\newcommand{\GeV}{\mathrm{GeV}}

\newcommand{\Vrms}{V_\textrm{RMS}}
\newcommand{\Vflow}{V_\textrm{flow}}
\newcommand{\vth}{\bar{v}_\textrm{th}}
\newcommand{\Vvar}{\langle V_{\chi b}^2 \rangle}
\newcommand{\Vvarcb}{\langle V_{cb}^2 \rangle}

\DeclareMathOperator{\erf}{Erf}

\graphicspath{ {plots/} }

\begin{document}
\title{A Critical Assessment of CMB Limits on Dark Matter--Baryon Scattering:\\ New Treatment of the Relative Bulk Velocity}
\author{Kimberly K. Boddy}
\affiliation{Department of Physics \& Astronomy, Johns Hopkins University, Baltimore, MD 21218, USA}
\author{Vera Gluscevic}
\affiliation{School of Natural Sciences, Institute for Advanced Study, Einstein Drive, Princeton, NJ 08540, USA}
\affiliation{Department of Physics, University of Florida, Gainesville, Florida 32611, USA}
\author{Vivian Poulin}
\affiliation{Department of Physics \& Astronomy, Johns Hopkins University, Baltimore, MD 21218, USA}
\author{Ely D. Kovetz}
\affiliation{Department of Physics \& Astronomy, Johns Hopkins University, Baltimore, MD 21218, USA}
\author{Marc Kamionkowski}
\affiliation{Department of Physics \& Astronomy, Johns Hopkins University, Baltimore, MD 21218, USA}
\author{Rennan Barkana}
\affiliation{Raymond and Beverly Sackler School of Physics and Astronomy, Tel-Aviv University, Tel-Aviv 69978, Israel}

\begin{abstract}
We perform an improved cosmic microwave background (CMB) analysis to search for dark matter--proton scattering with a momentum-transfer cross section of the form $\sigma_0v^n$ for $n\!=\!-2$ and $n\!=\!-4$.
In particular, we present a new and robust prescription for incorporating the relative bulk velocity between the dark matter and baryon fluids into the standard linear Boltzmann calculation.
Using an iterative procedure, we self-consistently include the effects of the bulk velocities in a cosmology in which dark matter interacts with baryons.
With this prescription, we derive CMB bounds on the cross section, excluding $\sigma_0 \! > \! 2.3 \! \times\! 10^{-33}~\mathrm{cm}^2$ for $n\!=\!-2$ and $\sigma_0 \! > \! 1.7 \! \times \! 10^{-41}~\mathrm{cm}^2$ for $n\!=\!-4$ at 95\% confidence, for dark matter masses below 10~MeV.
Furthermore, we investigate how these constraints change when only a subcomponent of dark matter is interacting.
We show that \textit{Planck} limits vanish if $\lesssim \! 0.4\%$ of dark matter is tightly coupled to baryons.
We discuss the implications of our results for present and future cosmological observations.
\end{abstract}
\maketitle

\section{Introduction}

Cosmological observables provide a unique avenue to search for evidence of non-gravitational interactions between dark matter (DM) and the Standard Model particles, and thereby gain insight into the unknown physical nature of DM.
In particular, elastic scattering between DM and baryons transfers heat and momentum between the two fluids.
The time evolution for the rate of momentum transfer depends on how the interaction cross section scales with the relative particle velocities, and the effects of scattering can be important at different cosmological epochs.
If scattering is efficient before recombination, it affects the temperature, polarization, and lensing anisotropies of the cosmic microwave background (CMB), as well as the linear matter power spectrum on small angular scales~\cite{Chen:2002yh,Sigurdson:2004zp,Dvorkin:2013cea,Gluscevic:2017ywp,Boddy:2018kfv,Xu:2018efh}.
If scattering is significant in the post-recombination Universe, it can result in anomalous late-time heating or cooling of the baryon gas, altering the 21-cm signal from neutral hydrogen at redshifts prior to the Epoch of Reionization~\cite{Tashiro:2014tsa,Munoz:2015bca,Barkana:2018lgd}.

In a $\Lambda$CDM Universe, there is a relative bulk velocity between the cold DM and baryon fluids, which results in supersonic coherent flows of the baryons post recombination~\cite{Tseliakhovich:2010bj}.
If DM and baryons interact, but the rate of momentum transfer is low, the drag force between the two fluids may not efficiently dissipate their relative bulk velocity, allowing it to dominate over the thermal particle motions, once the Universe is sufficiently cooled.
Furthermore, if the relative bulk velocity is significant prior to recombination, the computation of the Boltzmann equations for the CMB becomes infeasible using standard methods: the equations describing the velocity fluctuations of the fluids become nonlinear, resulting in the coupling of individual Fourier modes.
In an attempt to address this issue when computing CMB limits on DM--baryon interactions, previous studies~\cite{Dvorkin:2013cea,Xu:2018efh,Slatyer:2018aqg} used the root-mean-square (RMS) of the relative bulk velocity as a correction to the thermal velocity dispersion, suppressing the rate of momentum transfer, and thus obtaining conservative upper limits on DM--baryon interactions.
That approach has two important caveats: the RMS velocity was computed in $\Lambda$CDM, inconsistent with a cosmology that features DM--baryon interactions; and the same RMS velocity was used in the Boltzmann equations for all Fourier modes, neglecting differences in how modes contribute at a given scale.

In this work, we develop an improved treatment of the relative bulk velocity and reassess CMB limits on DM--proton scattering.
Specifically, we supplement the standard Boltzmann linear calculations with an iterative procedure that self-consistently includes the effects of the relative bulk velocity in a cosmology in which dark matter interacts with baryons.
We parameterize the momentum-transfer cross section as $\sigma_\textrm{MT} \! = \! \sigma_0 v^n$, where $v$ is the relative velocity between the scattering particles, and focus on two interaction models for which the relative bulk velocity is expected to have a substantial impact: $n\!=\!-2$ (arising in the case of, \textit{e.g.}, electric or magnetic dipole interactions through light mediators) and $n\!=\!-4$ (from, \textit{e.g.}, Coulomb-like interactions or Yukawa interactions through light mediators).
We analyze the latest public CMB data from the \Planck{} 2015 data release~\cite{2016A&A...594A...1P,2016A&A...594A..11P} and find $\sigma_0 \! < \! 2.3 \! \times \! 10^{-33}~\cm^2$ for $n\!=\!-2$ and $\sigma_0 \! < \! 1.7 \! \times \! 10^{-41}~\cm^2$ for $n\!=\!-4$ at the 95\% confidence level (C.L.) for DM masses below $10~\MeV$.
We forecast the sensitivity of the next-generation ground-based CMB experiment and find that CMB-Stage 4~\cite{2016arXiv161002743A} could deliver roughly a factor of $\sim\! 3$ improvement (not including a CMB lensing analysis), for a DM mass of $1~\MeV$.
Additionally, we report limits on $\sigma_0$ for scenarios in which only a fraction of DM interacts with protons.
For very small fractions, large values of $\sigma_0$ are allowed, and there exists a regime in which the DM and baryons are tightly coupled, such that DM behaves as baryons and experiences acoustic oscillations.
We find that the constraining power of \textit{Planck} is drastically diminished when less than $0.4\%$ of DM is interacting.

The Experiment to Detect the Global Epoch of Reionization Signature (EDGES) recently reported an anomalously large sky-averaged absorption signal~\cite{Bowman:2018yin}, which was attributed to dark matter interactions with baryons~\cite{Barkana:2018lgd}.
Our results do not rule out a phenomenological $n\!=\!-4$ interaction invoked to explain the EDGES signal~\cite{Barkana:2018lgd}; however, we do exclude a percent of DM interacting with ions only, at a level consistent with the EDGES signal~\cite{Barkana:2018qrx}.
In a separate study, we investigate the regime of subpercent fractions of millicharge-like DM and discuss the implications of our newly-derived CMB limits for the DM interpretation of EDGES~\cite{paper2:inprep}.

This paper is structured as follows.
In Section~\ref{sec:boltzmann}, we derive the Boltzmann equations that include DM--baryon scattering and present a new treatment of the relative bulk velocity.
In Section~\ref{sec:cmb}, we describe and quantify the effects of scattering on the CMB power spectra.
In Section~\ref{sec:constraints}, we describe our analysis of \Planck{} 2015 data and present new limits on the interactions with $n\!=\!-2$ and $n\!=\!-4$.
We discuss and conclude in Section~\ref{sec:conclusions}.

\section{Modified Cosmology}
\label{sec:boltzmann}

In this section, we incorporate the DM--baryon collision term into the Boltzmann equations and present an improved treatment to account for a non-negligible relative bulk velocity between baryons and DM.
Further details of our calculations are provided in Appendix~\ref{sec:app-nonlinear}.
We consider DM interactions with protons and parameterize the momentum-transfer cross section as $\sigma_\textrm{MT} \! = \! \sigma_0 v^n$, where $v$ is the relative velocity between the scattering particles.
Scattering with helium involves non-trivial form factors that depend on the specific structure of the interaction~\cite{Catena:2015uha,Boddy:2018kfv}, and it is mainly relevant for DM masses above $1~\GeV$~\cite{Gluscevic:2017ywp,Boddy:2018kfv}.
We neglect it here for simplicity; incorporating it would improve our constraints presented in Section~\ref{sec:constraints}.

\subsection{Evolution of perturbations and temperatures}

The scattering between DM and protons introduces a drag force and heat exchange between the DM and baryon fluids.
Hence, the Boltzmann equations governing the evolution of their velocity perturbations and of their temperatures must be adjusted accordingly.
We assume that the DM and baryon fluids are nonrelativistic, with energy densities $\rho_\chi$ and $\rho_b$, temperatures $T_\chi$ and $T_b$, and sound speeds $c_\chi$ and $c_b$, respectively.
The motion of the two fluids is given by their peculiar velocities $\vec{V}_\chi$ and $\vec{V}_b$, with a relative bulk velocity $\vec{V}_{\chi b} \equiv \vec{V}_\chi - \vec{V}_b$.

The linear Boltzmann equations incorporate terms only up to first order in the metric fluctuations and fluid perturbations.
However, in the presence of DM--baryon interactions, the equations become nonlinear at times when the relative bulk velocity exceeds the relative thermal velocity dispersion.
Therefore, we begin by describing the evolution of the temperatures and peculiar velocities in real space without assuming a small relative bulk velocity.
We show the full derivation in Appendix~\ref{sec:app-nonlinear}, where we present generic expressions for $n\!>\!-5$ and for scattering with multiple species of baryons.
This calculation was previously performed in Ref.~\cite{Munoz:2015bca} for the specific case of $n\!=\!-4$, and our results agree.

From Eq.~\eqref{eq:Vfull-general}, the peculiar velocities evolve as
\begin{subequations}
\begin{align}
  \frac{\partial \vec{V}_\chi}{\partial\tau}
  - c_\chi^2 \vec{\nabla} \delta_\chi + \frac{\dot{a}}{a} \vec{V}_\chi
  &= R_\chi (\vec{V}_b - \vec{V}_\chi)\;
  {}_1F_1\left(-\frac{n+1}{2},\frac{5}{2},-\frac{V_{\chi b}^2}{2\vth^2}\right)
  \\
  \frac{\partial \vec{V}_b}{\partial\tau}
  - c_b^2 \vec{\nabla} \delta_b + \frac{\dot{a}}{a} \vec{V}_b
  &= R_\gamma(\vec{V}_\gamma - \vec{V}_b)
  + \frac{\rho_\chi}{\rho_b} R_\chi (\vec{V}_\chi - \vec{V}_b)\;
  {}_1F_1\left(-\frac{n+1}{2},\frac{5}{2},-\frac{V_{\chi b}^2}{2\vth^2}\right) \ ,
\end{align}
\label{eq:Vfull}%
\end{subequations}
and from Eq.~\eqref{eq:temp-general}, the temperatures evolve as
\begin{subequations}
\begin{align}
  \dot{T}_\chi &+ 2\frac{\dot{a}}{a}T_\chi
  = 2 R_\chi^\prime \left\{(T_b - T_\chi) \phantom{\frac{1}{2}} \right.
  \nonumber\\
  & \left. \times
  \left[ {}_1F_1\left(-\frac{n+3}{2},\frac{3}{2},-\frac{V_{\chi b}^2}{2\vth^2}\right)
    -\frac{V_{\chi b}^2}{3\vth^2} {}_1F_1\left(-\frac{n+1}{2},\frac{5}{2},-\frac{V_{\chi b}^2}{2\vth^2}\right) \right]
  + \frac{m_p}{3} V_{\chi b}^2 {}_1F_1\left(-\frac{n+1}{2},\frac{5}{2},-\frac{V_{\chi b}^2}{2\vth^2}\right) \right\} \\
  \dot{T}_b &+ 2\frac{\dot{a}}{a}T_b =
  2 \frac{\mu_b}{m_e} R_\gamma (T_\gamma - T_b)
  + 2 \frac{\mu_b}{m_\chi} \frac{\rho_\chi}{\rho_b}R_\chi^\prime
  \left\{(T_\chi - T_b) \phantom{\frac{1}{2}} \right.
  \nonumber\\
  & \left. \times
  \left[ {}_1F_1\left(-\frac{n+3}{2},\frac{3}{2},-\frac{V_{\chi b}^2}{2\vth^2}\right)
    -\frac{V_{\chi b}^2}{3\vth^2} {}_1F_1\left(-\frac{n+1}{2},\frac{5}{2},-\frac{V_{\chi b}^2}{2\vth^2}\right) \right]
  + \frac{m_\chi}{3} V_{\chi b}^2 {}_1F_1\left(-\frac{n+1}{2},\frac{5}{2},-\frac{V_{\chi b}^2}{2\vth^2}\right) \right\} \ ,
\end{align}
\label{eq:temp}%
\end{subequations}
where ${}_1F_1$ is the confluent hypergeometric function of the first kind, $\vth^2 \! = \! T_\chi / m_\chi \! + \! T_b / m_p$ is the relative thermal velocity dispersion squared, $m_\chi$ is the DM mass, $m_p$ is the proton mass, $m_e$ is the electron mass, $\mu_b$ is the mean molecular weight of the baryons, and $\delta_\chi$ and $\delta_b$ are density perturbations in DM and baryons, respectively.
These equations are written in synchronous gauge, where $a$ is the scale factor, and the dot notation indicates a derivative with respect to conformal time $\tau$.
The terms proportional to $R_\gamma$ and $R_\chi$ in Eq.~\eqref{eq:Vfull} represent drag terms, which describe the transfer of momentum between the interacting fluids.
The momentum-transfer rate coefficient $R_\gamma$ arises from photon--baryon interactions through Compton scattering, while $R_\chi$ arises from the new DM--proton interactions and is given by
\begin{equation}
  R_\chi = a\rho_b \frac{Y_H \sigma_0}{m_\chi+m_p} \mathcal{N}_n \vth^{(n+1)} \ ,
  \label{eq:rate}
\end{equation}
where $\mathcal{N}_n \equiv 2^{(5+n)/2} \Gamma(3+n/2) / (3\sqrt{\pi})$ and $Y_H$ is the mass fraction of hydrogen.
The heat-transfer rate coefficient in Eq.~\eqref{eq:temp} is $R_\chi^\prime = R_\chi m_\chi/(m_\chi + m_p)$.

The competition between $R_\chi$ and the expansion rate $aH$ determines the efficiency of momentum transfer at a given redshift: when $R_\chi/aH\gg 1$, the fluids are tightly coupled and move together.
Given the current CMB limits for interactions with $n\!\geq\!0$, this regime occurs at very early times ($z\!\gg\!10^4)$, and results in dark acoustic oscillations that imprint oscillatory features in the linear matter power spectrum at small scales~\cite{Gluscevic:2017ywp,Boddy:2018kfv}.
In that case, the drag between the DM and baryon fluids couples their motion, resulting in a small relative bulk velocity---compared to the thermal particle velocities at redshifts relevant for CMB measurements---and can thus be ignored.
However, for $n\!\leq\!-2$, the two fluids have a feeble interaction rate at early times, and the relative bulk velocity is non-negligible.
As a result, Eqs.~\eqref{eq:Vfull} and \eqref{eq:temp} are nonlinear.
In the following, we present a new prescription for capturing the effects of the relative bulk velocity on the momentum-transfer rate between DM and baryons in both regimes.

\subsection{Treatment of relative bulk velocity}
\label{sec:CMB-bulkvel}

Standard CMB computations rely on linearity of the Boltzmann equations, for which it is possible to Fourier transform real-space equations and independently track the evolution of each Fourier mode with wave number $k$.
In the limit $V_{\chi b}^2 \! \ll\! \vth^2$, the ${}_1F_1$ functions in Eqs.~\eqref{eq:Vfull} and \eqref{eq:temp} asymptote to 1, and the evolution of the peculiar velocities is indeed linear (and the temperature evolution equations are independent of the relative bulk velocity).
It is then possible to take the divergence and the Fourier transform of Eq.~\eqref{eq:Vfull} to obtain the evolution equations for the velocity divergences of the DM and baryons, $\theta_\chi(k,z)$ and $\theta_b(k,z)$, respectively.
However, when this approximation breaks down, the Boltzmann equations are nonlinear, resulting in coupling of Fourier modes.

In order to bypass this difficulty, we first define
\begin{align}
  \Vflow^2(k,z) &\equiv \int_0^k \frac{dk'}{k'} \Delta_\zeta^2(k')
  \left[\frac{\theta_b(k',z)-\theta_\chi(k',z)}{k'}\right]^2
  \label{eq:Vflow} \\
  \Vrms^2(k,z) &\equiv \int_k^\infty \frac{dk'}{k'} \Delta_\zeta^2(k')
  \left[\frac{\theta_b(k',z)-\theta_\chi(k',z)}{k'}\right]^2 \ ,
  \label{eq:Vrms}
\end{align}
where $\Delta_\zeta^2(k)$ is the initial curvature perturbation variance per $\ln k$.
We then propose the following prescription to reduce Eq.~\eqref{eq:Vfull} to a linear expression, while modifying the momentum-transfer rate coefficient to reincorporate the effects of mode mixing.
For a given mode $k^\star$, the density perturbations from larger scales cause a relative bulk flow between the DM and baryon fluids that contributes to their existing relative bulk motion.%
\footnote{Reference~\cite{Tseliakhovich:2010bj} similarly had to account for the bulk flow between the DM and baryon fluids within the context of $\Lambda$CDM.
  In that study, there was a clear separation of scales such that the post-recombination Universe could be represented as individual patches across the sky, each with a particular value of the relative bulk velocity.
  Averaging over the various patches yielded a local isotropically averaged power spectrum.
  Since we do not have a similar separation of scales, it is not appropriate to follow the same technique.}
To account for the bulk flow, we absorb the ${}_1F_1$ function into the momentum-transfer rate coefficient, replacing $V_{\chi b}$ with $\Vflow(k^\star,z)$.
Meanwhile, the density perturbations from smaller scales collectively act as a source of velocity dispersion, in addition to the thermal dispersion.
We thus augment all instances of $\vth^2$ with the square of the one-dimensional RMS velocity $\Vrms^2(k^\star,z)/3$.
With this prescription, the Boltzmann equations in Fourier space become
\begin{subequations}
\begin{align}
  \dot{\delta}_\chi &= -\theta_\chi - \frac{\dot{h}}{2}\ ,
  &
  \dot{\theta}_\chi &= -\frac{\dot{a}}{a}\theta_\chi + c_\chi^2 k^2 \delta_\chi
  + \widetilde{R}_\chi(k) (\theta_b - \theta_\chi) \ ,
  \\
  \dot{\delta}_b    &= -\theta_b    - \frac{\dot{h}}{2}\ ,
  &
  \dot{\theta}_b    &= -\frac{\dot{a}}{a}\theta_b   + c_b^2 k^2 \delta_b
  + R_\gamma (\theta_\gamma - \theta_b)
  + \frac{\rho_\chi}{\rho_b} \widetilde{R}_\chi(k) (\theta_\chi - \theta_b) \ ,
\end{align}
\label{eq:fluct}%
\end{subequations}
where $h$ is the trace of the scalar metric perturbation, and the modified momentum-transfer rate coefficient is
\begin{equation}
  \widetilde{R}_\chi = R_\chi
  \left[ 1 + \frac{\Vrms^2(k,z)/3}{\vth^2} \right]^{(n+1)/2} \,
  {}_1F_1\left(-\frac{n+1}{2},\frac{5}{2},
  -\frac{\Vflow^2(k,z)}{2\left[\vth^2 + \Vrms^2(k,z)/3\right]} \right) \ .
  \label{eq:mod-rate}
\end{equation}
In the limit $\Vrms^2, \Vflow^2 \! \ll\! \vth^2$, we recover the results from Refs.~\cite{Dvorkin:2013cea,Gluscevic:2017ywp,Boddy:2018kfv,Xu:2018efh}.

\begin{figure}[t]
  \centering
  \includegraphics[width=0.48\linewidth]{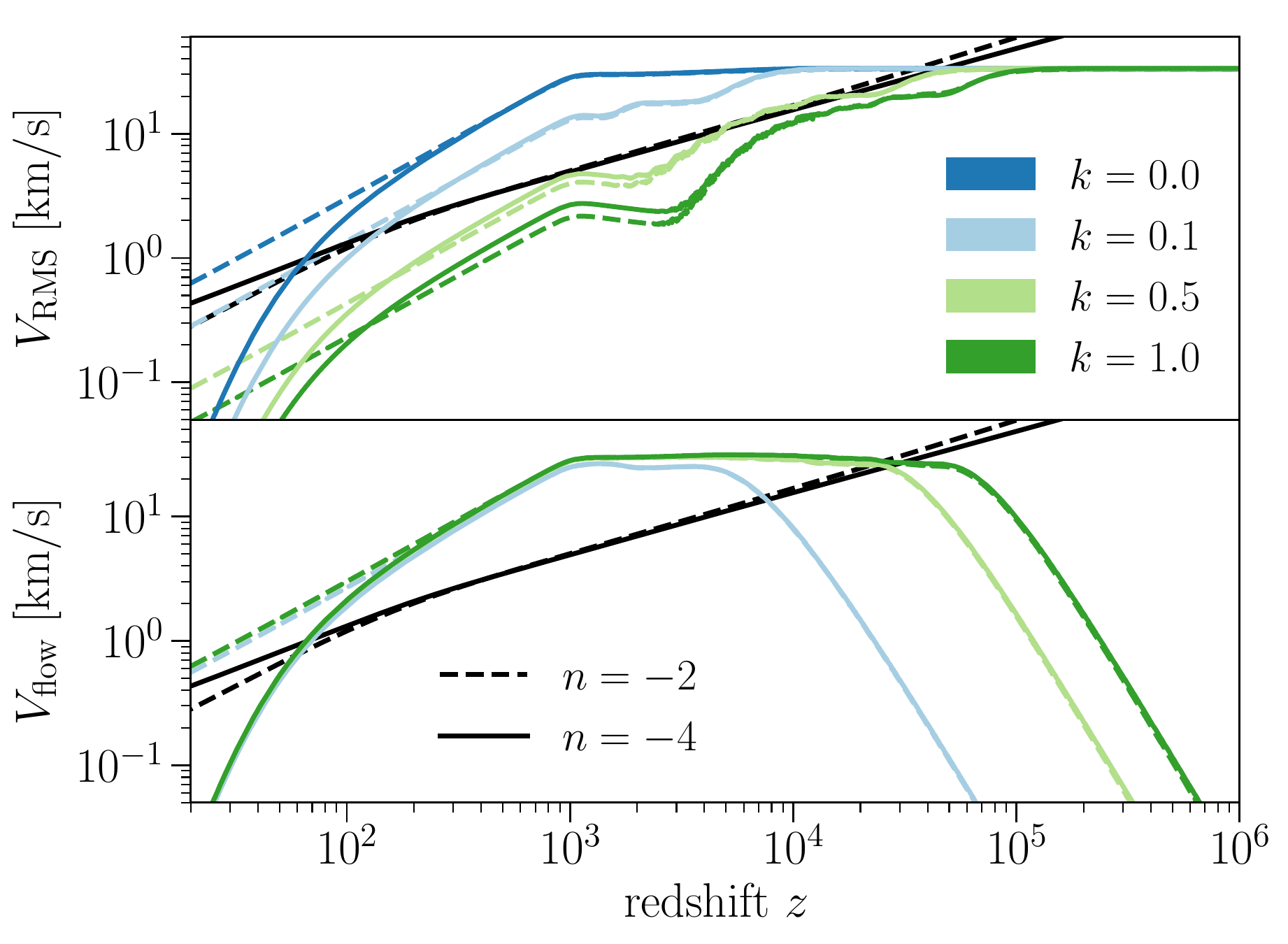}
  \hspace{0.2in}
  \includegraphics[width=0.48\linewidth]{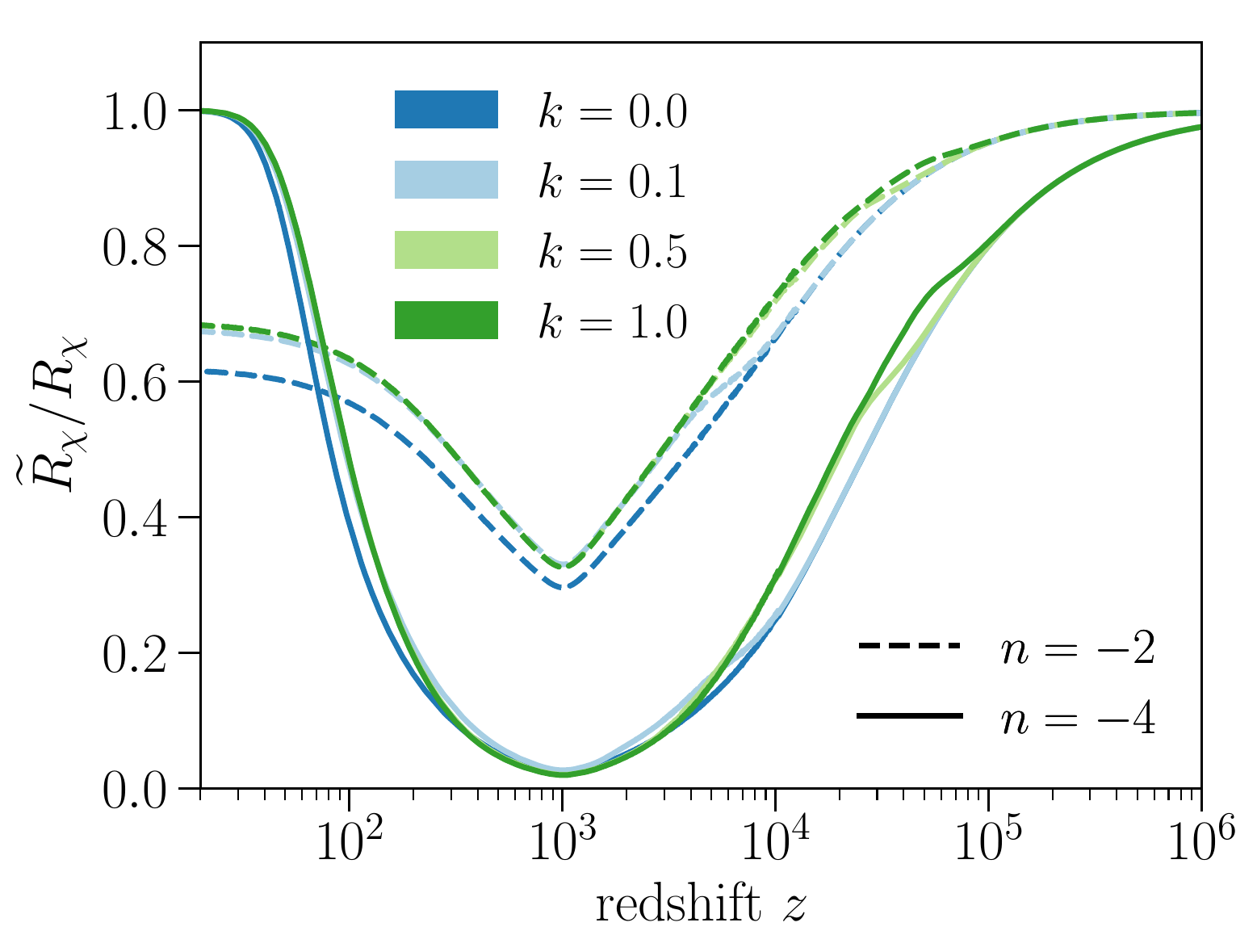}
  \caption{\textbf{[Left]} Evolution of $\Vrms$ (top panel) and $\Vflow$ (bottom panel) as a function of redshift for various wave numbers $k$, indicated in the legend.
    The thermal velocity $\vth$ (black) is shown for reference.
    \textbf{[Right]} Modification to the coefficient of the momentum-transfer rate from Eq.~\eqref{eq:mod-rate} for various $k$.
    In both panels, we show the $n\!=\!-2$ (dashed) and $n\!=\!-4$ (solid) interactions, for a DM mass of $1~\MeV$.
    We set the coefficients of the momentum-transfer cross section to their respective 95\% C.L. upper limit, derived using our ``main'' prescription, reported in Section~\ref{sec:constraints}.}
  \label{fig:mod-rate}
\end{figure}

In the left panel of Figure~\ref{fig:mod-rate}, we show the evolution of $\Vrms$ (top panel) and $\Vflow$ (bottom panel), with $k\!=\!0$ (dark blue), $k\!=\!0.1$ (light blue), $k\!=\!0.5$ (light green), and $k\!=\!1$ (dark green), for $n\!=\!-2$ (dashed lines) and $n\!=\!-4$ (solid lines).
We also show the evolution of $\vth$ (black).
Since $\Vflow(k=0)$ vanishes by definition, there is no associated curve plotted in the bottom panel.
In the right panel of Figure~\ref{fig:mod-rate}, we show the evolution for the ratio $\widetilde{R}_\chi / R_\chi$ for the same values of $k$ in the left panel.
At early times, $\widetilde{R}_\chi$ approaches $R_\chi$, since $\vth$ dominates over $\Vrms$ and $\Vflow$.
For redshifts $z\!\lesssim\! 10^5$, $\Vrms$ and $\Vflow$ become increasingly important and suppress the rate of momentum transfer.
At recombination near $z\!\sim\! 10^3$, the baryons decouple from the photons, causing $\Vrms$ and $\Vflow$ to suddenly begin decreasing adiabatically, thereby lessening the suppression of the rate.

We note that the evolution of $\widetilde{R}_\chi$ is quite similar between various $k$, indicating that incorporating the $k$ dependence via $\Vrms$ and $\Vflow$ might not play a significant role; indeed, we can understand this observation from the limiting behavior of $\widetilde{R}_\chi$.
The full variance, integrated over all $k$, of the relative bulk velocity is given by $\Vvar \equiv \Vflow^2 + \Vrms^2$.
For small values of $k$, $\Vflow^2\!\to\!0$ while $\Vrms^2\!\to\!\Vvar$; thus, the ${}_1F_1$ function in Eq.~\eqref{eq:mod-rate} approaches $1$.
For large values of $k$, $\Vrms^2\!\to\!0$ while $\Vflow^2\!\to\!\Vvar$; furthermore, if $\Vflow$ is much larger than $\vth$, the ${}_1F_1$ function in Eq.~\eqref{eq:mod-rate} asymptotes to $\sim\! \{\Vflow^2 / [2(\vth^2 + \Vrms^2 /3)]\}^{(n+1)/2}$.
In either case, the modified momentum-transfer rate coefficient has the form
\begin{equation}
  \widetilde{R}_\chi^{\chi b} = R_\chi
  \left[ 1 + \frac{\Vvar/3}{\vth^2} \right]^{(n+1)/2} \ ,
  \label{eq:mod-rate-kindep}
\end{equation}
ignoring an $n$-dependent prefactor in the large-$k$ limit.
Thus, this form of the modified rate may sufficiently capture the combined large-scale and small-scale effects of mode mixing.
In either case, for the temperature evolution equations in Eq.~\eqref{eq:temp}, we substitute $V_{\chi b}^2$ for its average value $\Vvar$.
In the limit $\Vvar \! \ll \! \vth^2$, we again recover previous results~\cite{Dvorkin:2013cea,Gluscevic:2017ywp,Boddy:2018kfv,Xu:2018efh}.

This work builds upon the mean-field approach introduced in previous studies investigating DM--baryon scattering~\cite{Dvorkin:2013cea}.
In that study, the modified momentum-transfer rate was
\begin{equation}
  \widetilde{R}_\chi^{cb} = R_\chi [1 + \Vvarcb/(3\vth^2)]^{(n+1)/2},
  \label{eq:mod-rate-lcdm}
\end{equation}
where%
\footnote{Reference~\cite{Dvorkin:2013cea} used the quantity $\Vvarcb$, but labeled it as $\Vrms^2$.
  We refer to $\Vrms(k,z)$ as a $k$-dependent quantity, calculated in the interacting theory.
  The full variances $\Vvarcb$ and $\Vvar$ are $k$-independent quantities; the former is calculated in $\Lambda$CDM and the latter in the interacting theory.}
\begin{equation}
  \Vvarcb \equiv \int_0^\infty \frac{dk'}{k'} \Delta_\zeta^2(k')
  \left[\frac{\theta_b(k',z)-\theta_c(k',z)}{k'}\right]^2
  \label{eq:Vrms-lcdm}
\end{equation}
is the variance of the relative bulk velocity in $\Lambda$CDM [which is approximately $\Vvarcb\! =\! 10^{-8}$ at $z\! >\! 10^3$ and redshifts as $(1\!+\!z)^2$ at later times~\cite{Xu:2018efh,Slatyer:2018aqg}] and $\theta_c$ is the velocity divergence for cold collisionless DM~\cite{Tseliakhovich:2010bj}.
We improve upon the previous work in two important ways.
First, we compute the variance $\Vvar$ in a consistent manner, using the values of $\theta_\chi$ and $\theta_b$ obtained in a cosmology that includes DM--baryon scattering; this improvement corresponds to using the modified momentum-transfer rate in Eq.~\eqref{eq:mod-rate-kindep}.
Second, we treat the effects of mode coupling from smaller scales ($k > k^\star$) separate from those arising from larger scales ($k < k^\star$).
These two steps constitute our main prescription captured in Eq.~\eqref{eq:mod-rate}.
When the rate of momentum exchange is sufficiently small at times relevant for \Planck{}, using $\theta_c$ within $\Lambda$CDM is a decent estimation.
This condition is satisfied for the upper limits on $\sigma_0$ derived assuming all of DM interacts with baryons, and in that case we find little difference from our improved treatments of the relative bulk velocity.
However, if the rate is moderate or large, momentum exchange drives the values of $\theta_\chi$ and $\theta_b$ closer together, such that $\Vvarcb$ computed in $\Lambda$CDM overestimates the relative bulk velocity and thus overly suppresses the interaction rate.
This situation arises if only a subcomponent of DM is allowed to couple to baryons, and it is thus essential to employ the techniques presented in this work in order to derive limits on the DM--baryon interaction for that case.

Throughout the remainder of this work, we refer to various treatments of the relative bulk velocity that enable us to explore how various aspects of our new prescription affect our constraints on the DM--baryon scattering cross section.
The ``main'' prescription is our primary treatment given by Eq.~\eqref{eq:mod-rate}, and we consider it to be the most accurate for any regime of DM--baryon coupling.
The ``$k$-independent'' prescription is the treatment given by Eq.~\eqref{eq:mod-rate-kindep}.
These two prescriptions both have the feature that the variance of the relative bulk velocity is computed self-consistently within an interacting cosmology, using the iterative procedure described in Appendix~\ref{sec:app-implementation}.
The ``cdm'' prescription is that found in previous literature~\cite{Dvorkin:2013cea,Xu:2018efh,Slatyer:2018aqg} and uses Eq.~\eqref{eq:mod-rate-lcdm}; in this case, we employ the same temperature evolution equations presented in those works and not the full expressions of our Eq.~\eqref{eq:temp}.
Finally, the ``aggressive'' prescription uses Eq.~\eqref{eq:rate} and naively ignores the relative bulk velocity entirely in both the temperature and velocity evolution equations.
The constraints on $\sigma_0$ resulting from this prescription are thus the most aggressive; their comparison with the other constraints reported in this work quantifies the importance of incuding an accurate treatment of the relative bulk velocity.

\section{The effect on cosmological observables}
\label{sec:cmb}

In this Section, we discuss the impact of DM--proton scattering on cosmological observables.
In Section~\ref{sec:thermal_history}, we show the thermal histories of the DM and baryon fluids, as well as the evolution of the free-electron fraction.
In Section~\ref{sec:power_spectra}, we describe the effects on the primary CMB anisotropies, the matter power spectrum, and the CMB lensing power spectrum.
In Section~\ref{sec:strong}, we investigate a specific regime in which DM is tightly coupled to, and oscillates together with, the baryons at some point in cosmic history; this regime is allowed by \Planck{} data for $n\!=\!-4$ if only a small fraction of DM interacts with baryons.

To compute the power spectra, we have incorporated the Boltzmann equations from Section~\ref{sec:boltzmann} into the Boltzmann solver \CLASS{}\footnote{\url{https://github.com/lesgourg/class}}~\cite{Blas:2011rf}.
We chose adiabatic initial conditions, and set the DM temperature and velocity divergences to match those of the baryons at the start of the integration ($z\!=\!10^{14}$).
For $n\!=\!-4$, the rate of heat transfer is too low to maintain thermal equilibrium with the baryons, and the temperature and velocity divergences of the DM rapidly drop from their original values.
Thus, our initial conditions are effectively equivalent to starting with vanishing temperature and velocity divergences.
In fact, for all interaction strengths relevant in this work, we have verified that the choice of initial conditions is irrelevant, as long as they are set well above $z\!\sim\!10^5$ (roughly the redshift below which modes \Planck{} is sensitive to start entering the cosmological horizon).
We present further details on our modifications to \CLASS{} in Appendix~\ref{sec:app-implementation}.

Throughout this section, we use $\Lambda$CDM parameters at their best-fit \Planck{} 2015 values~\cite{Ade:2015xua}.
Unless otherwise noted, we fix the coefficient of the momentum-transfer cross section, $\sigma_0$, to its appropriate 95\% C.L. upper limit, derived in Section~\ref{sec:constraints} using our ``main'' prescription for the relative bulk velocity.
When plotting residuals, we show the relative difference between an observable computed for the cosmology with DM--proton scattering and for the reference $\Lambda$CDM cosmology.

\subsection{Thermal history}
\label{sec:thermal_history}

\begin{figure}[t]
  \centering
  \includegraphics[width=0.48\linewidth]{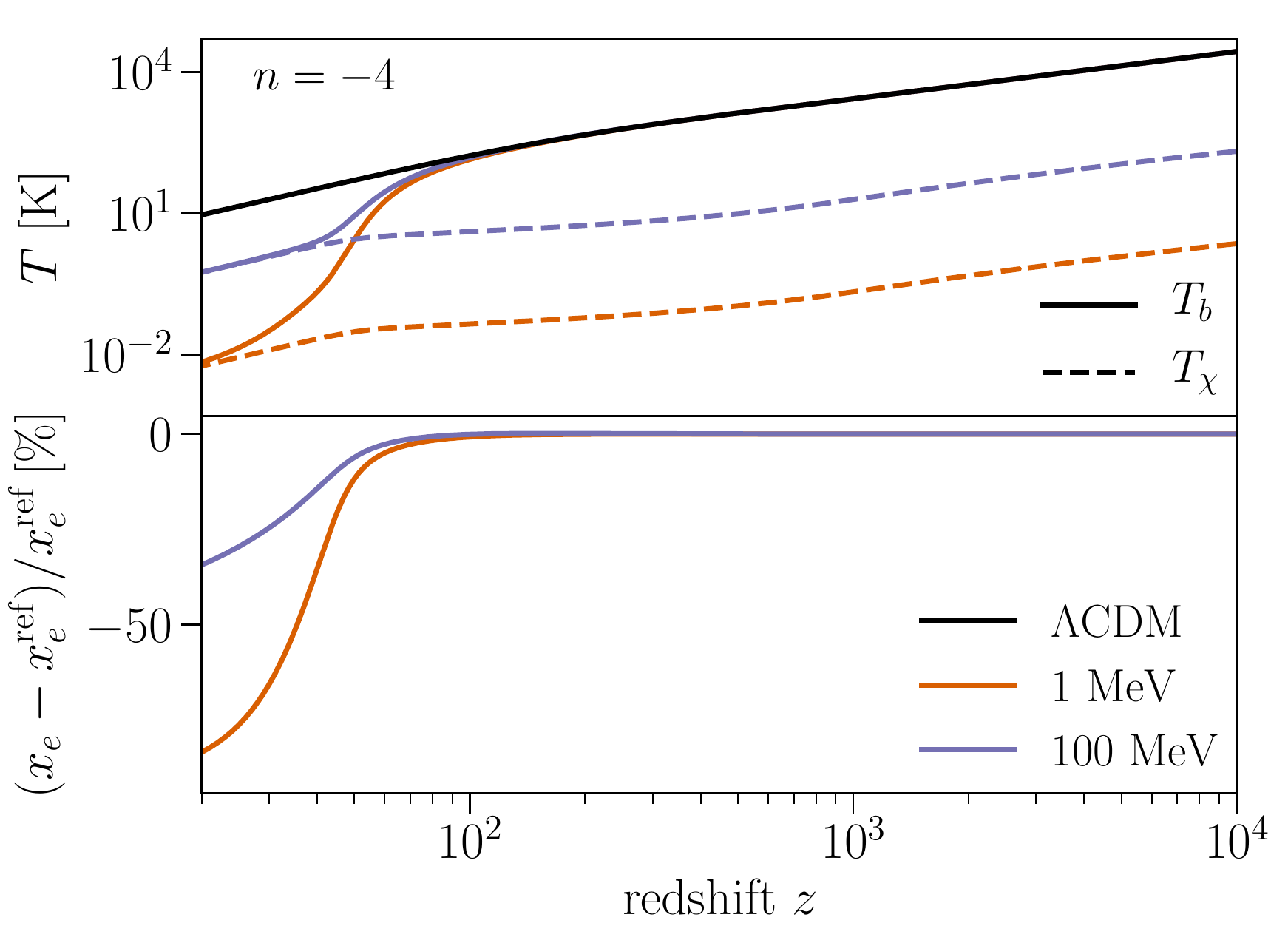}
  \hspace{0.2in}
  \includegraphics[width=0.48\linewidth]{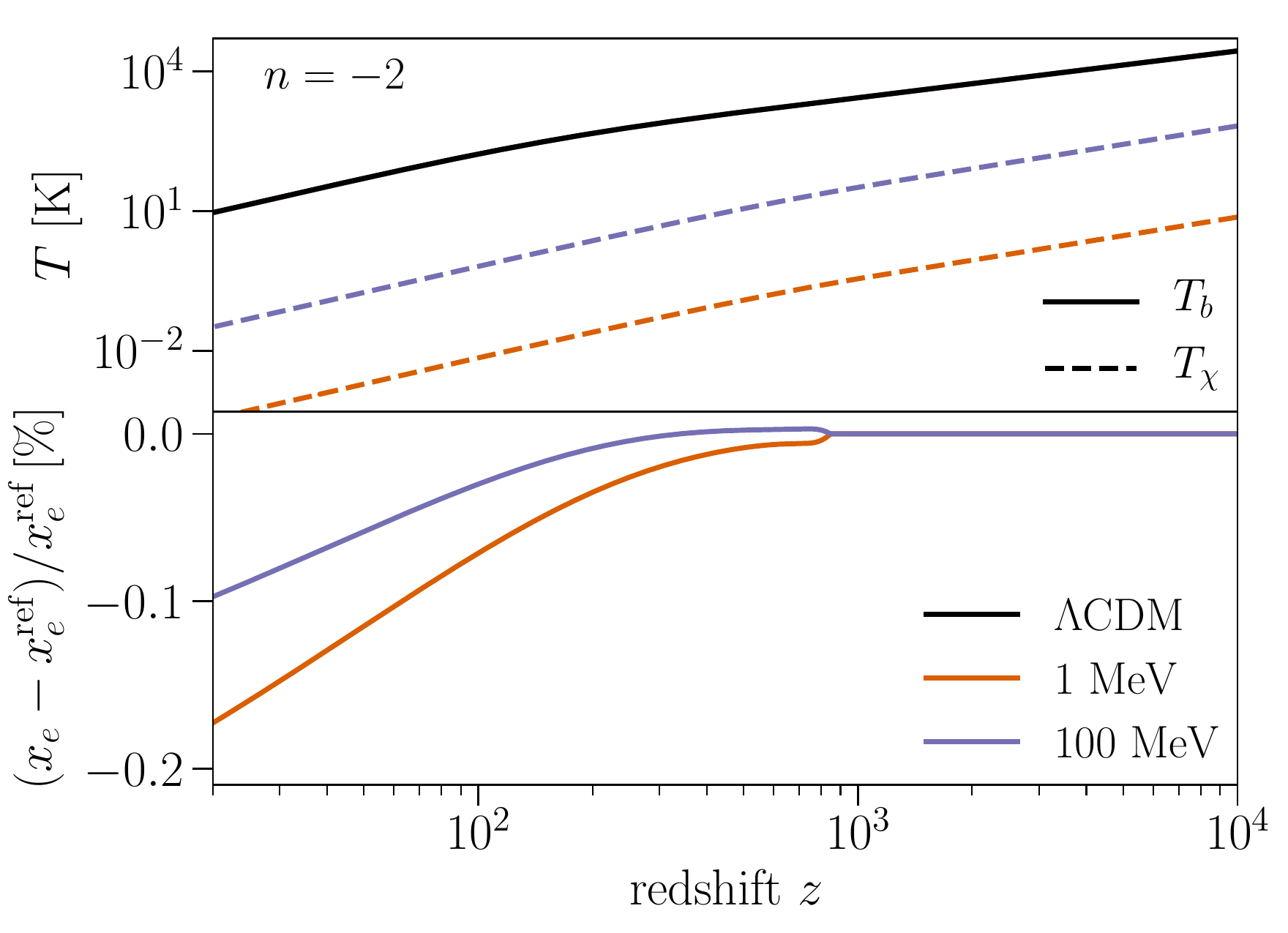}
  \caption{Temperature evolution (top panels) and residuals of $x_e$ with respect to $\Lambda$CDM (bottom panels), for the $n\!=\!-4$ (left panels) and $n\!=\!-2$ (right panels) interaction, for DM masses of $1~\MeV$ (orange) and $100~\MeV$ (purple).
    We set the coefficients of the momentum-transfer cross section to their respective 95\% C.L. upper limit, derived using our ``main'' prescription, reported in Section~\ref{sec:constraints}.
    We also show the baryon temperature in $\Lambda$CDM (black) for reference.}
  \label{fig:xe_and_Tm_n2_n4}
\end{figure}

An accurate determination of the thermal history is essential to the calculation of CMB power spectra.
The CMB is very sensitive to the number of free electrons in the plasma through the visibility function and the optical depth in the line-of-sight solution of the Boltzmann equations~\cite{Seljak:1996is}.
Scattering between DM and protons alters the temperature evolution of the baryons, which in turn influences the free-electron fraction, $x_e$.
If interactions with DM cool the baryon gas around recombination, the rate of recombination increases.
If cooling occurs at later times, it reduces the number of free electrons in a manner opposite to that of an early reionization from energy injection~\cite{Poulin:2016anj}.
In the top panels of Figure~\ref{fig:xe_and_Tm_n2_n4}, we show the evolution of the baryon (solid) and DM (dashed) temperatures as a function of redshift, comparing them to the evolution of the baryon temperature in $\Lambda$CDM (black solid), for DM masses of $1~\MeV$ (orange) and $100~\MeV$ (purple), for $n\!=\!-4$ (left panel) and $n\!=\!-2$ (right panel).
In the bottom panels, we show the residuals for the evolution of the free-electron fraction with respect to $\Lambda$CDM.
We set the values of $\sigma_0$ to their respective 95\%~C.L. upper limits, obtained using our ``main'' prescription in Section~\ref{sec:constraints}.
We find that the DM--proton interaction has no strong impact on the recombination era.
The impact on the free-electron fraction is substantial only at late times.
Since the CMB is only marginally sensitive to changes in the late-time free-electron fraction (through the low-$\ell$ $EE$ power spectrum), baryon cooling is a subdominant effect compared to the drag acceleration from scattering, and we have verified that it can be safely ignored for the purposes of this work.

\subsection{Power spectra}
\label{sec:power_spectra}

\begin{figure}[t]
  \centering
  \includegraphics[width=0.48\linewidth]{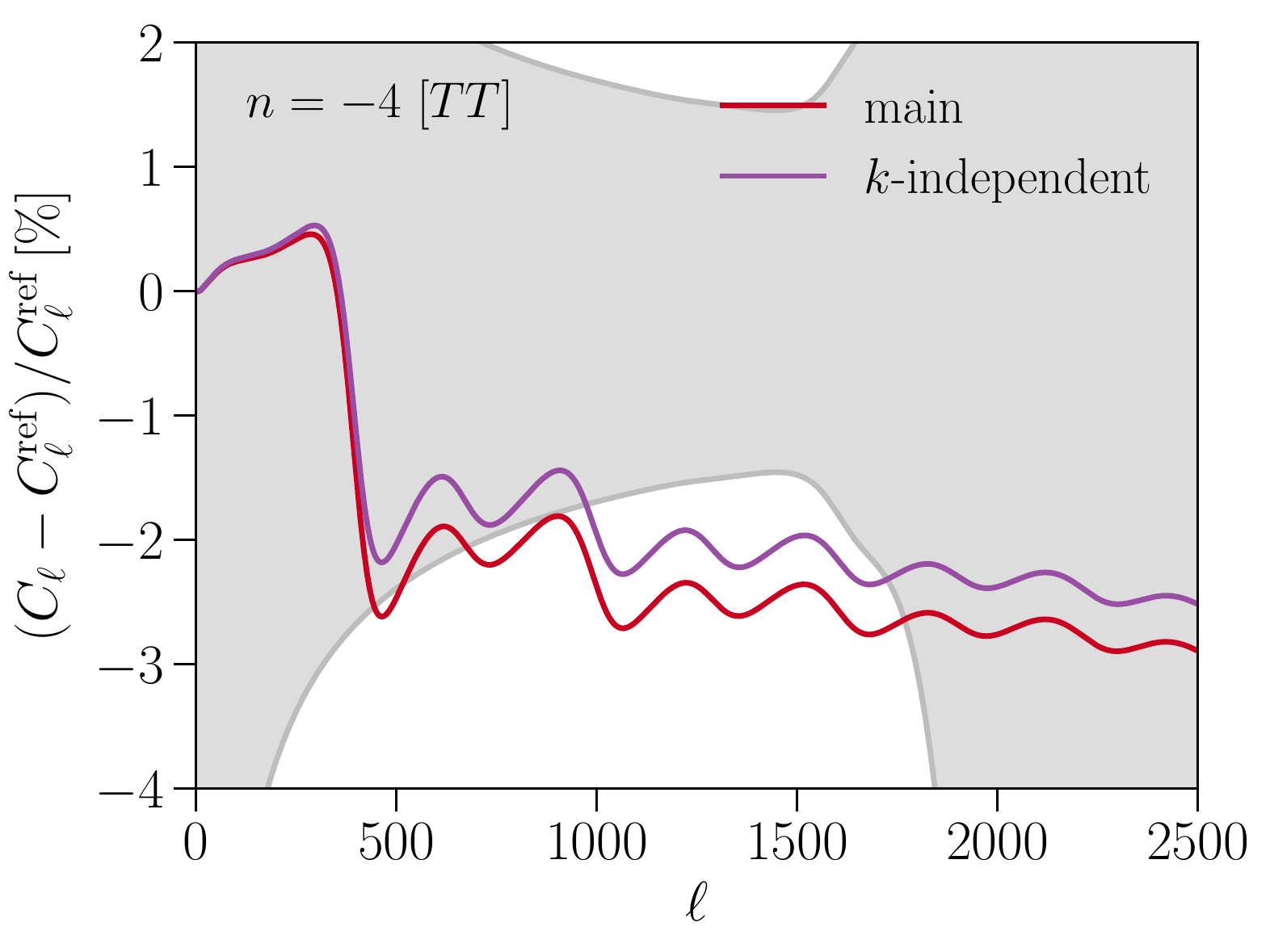}
  \hspace{0.2in}
  \includegraphics[width=0.48\linewidth]{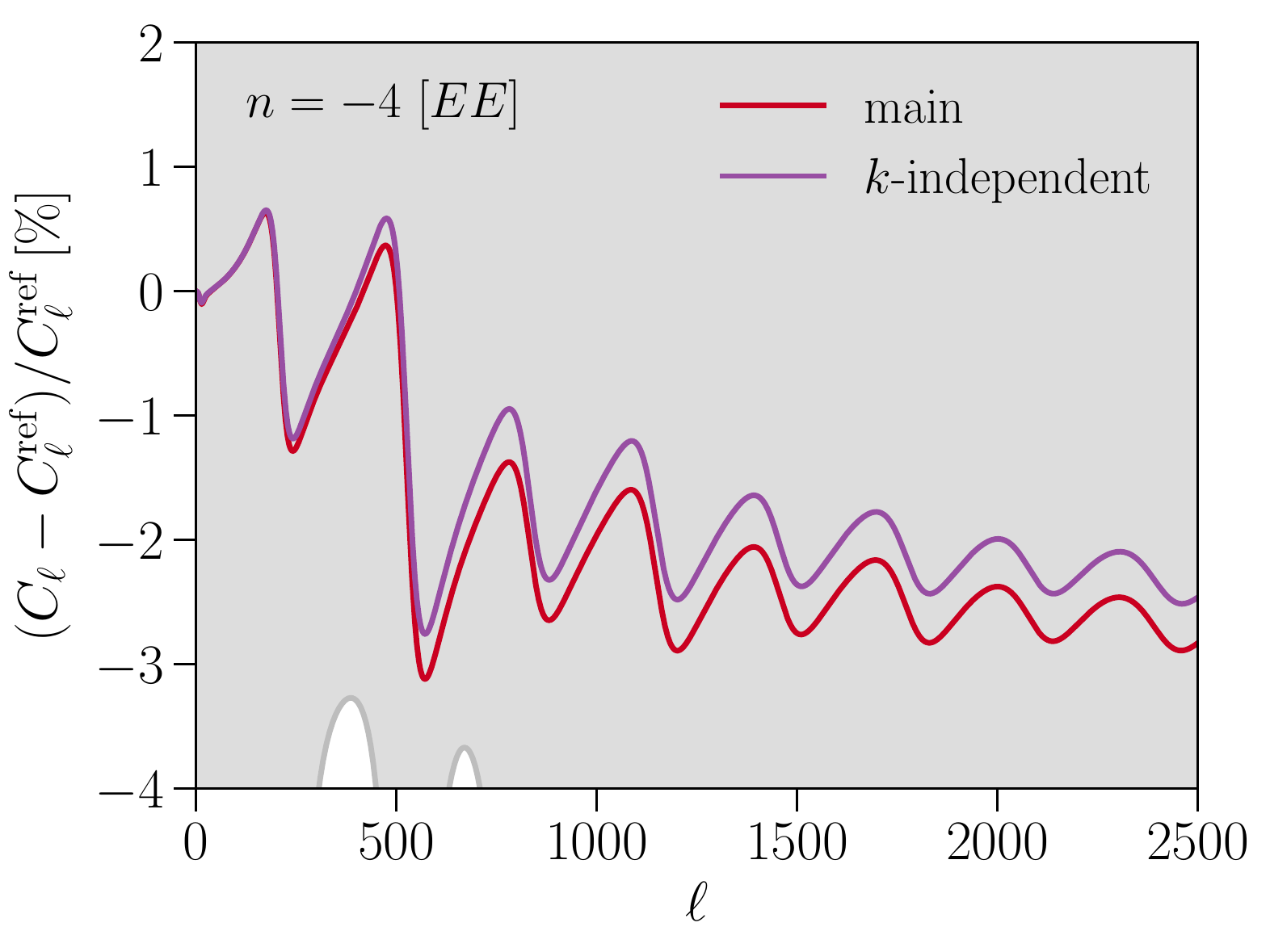} \\
  \includegraphics[width=0.48\linewidth]{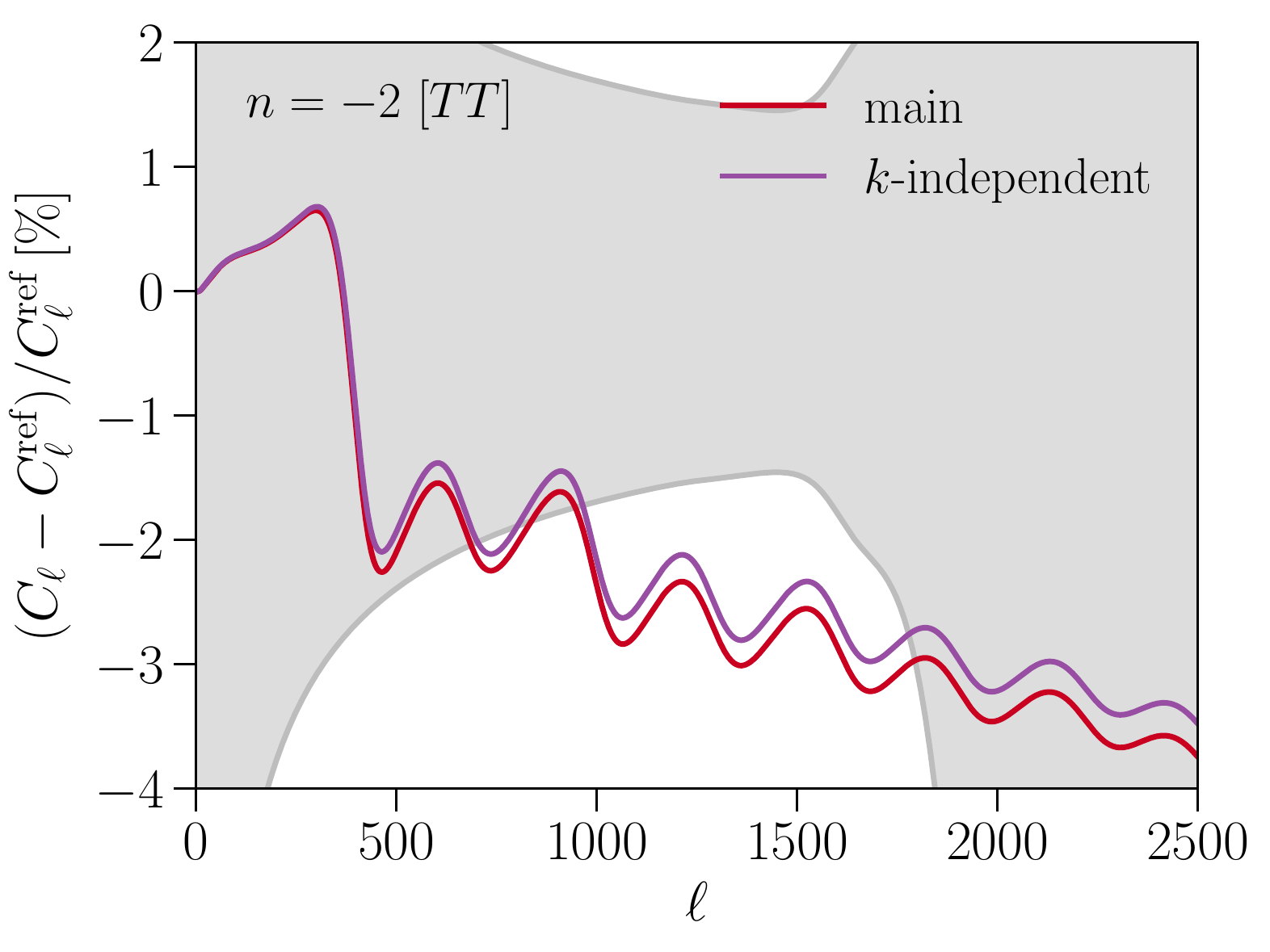}
  \hspace{0.2in}
  \includegraphics[width=0.48\linewidth]{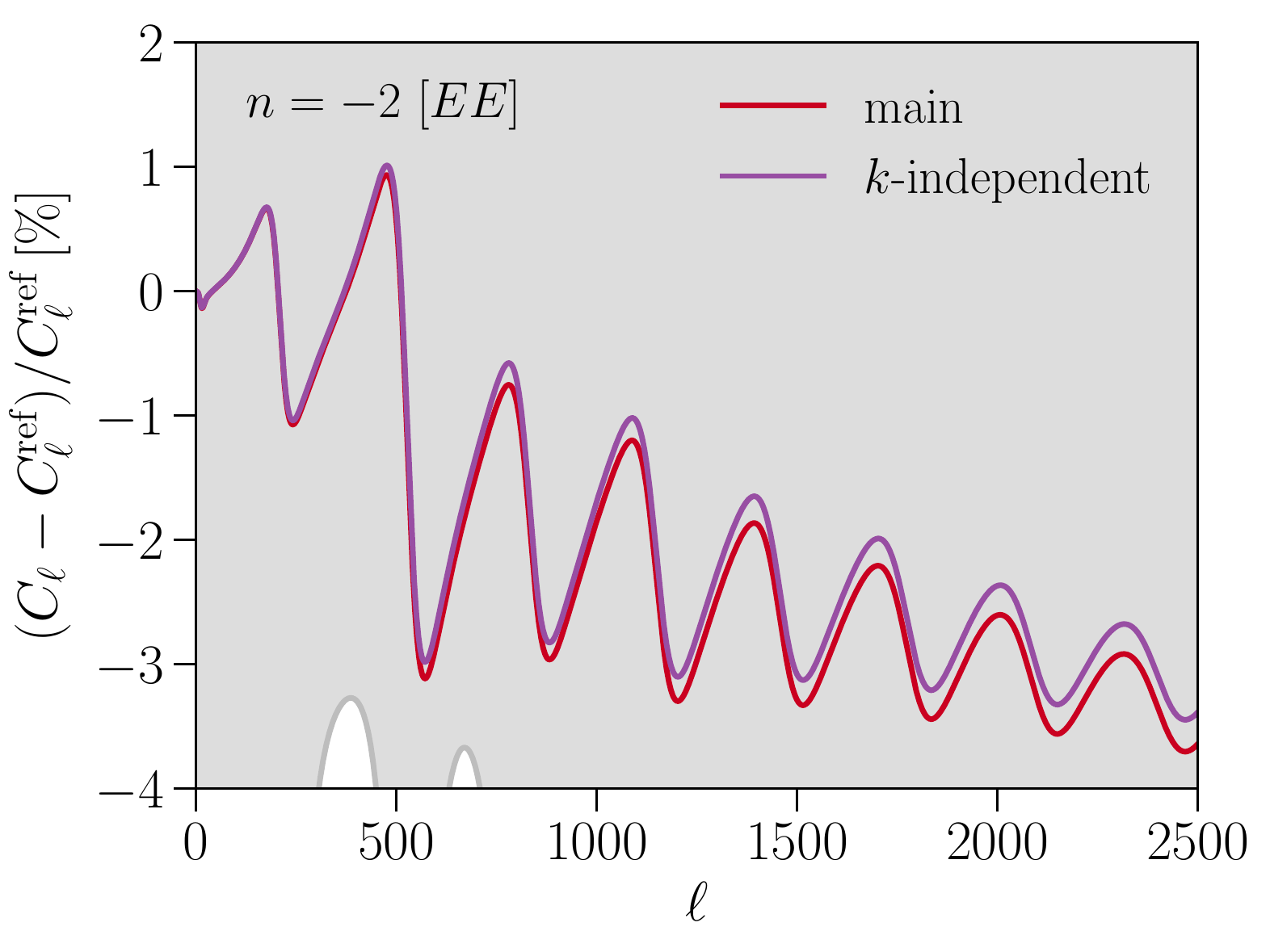}
  \caption{\textbf{[Top]:} Percent residuals (with respect to $\Lambda$CDM) of the lensed $TT$ (left panel) and $EE$ (right panel) power spectra for the case of $n\!=\!-4$ and a DM mass of $1~\MeV$.
  We show residuals for the ``main'' prescription (red) and the ``$k$-independent'' prescription (purple).
  In both cases, we set the coefficient of the momentum-transfer cross section to its 95\% C.L. upper limit, derived using the ``main'' prescription.
  Gray bands roughly represent the $2\sigma$ \Planck{} error bars, with a bin size of $\Delta \ell\!=\!50$.
  \textbf{[Bottom]:} Same as the top panels, but for the case of $n\!=\!-2$.
  The difference between the two prescriptions is less prominent in this case because of the weaker scaling of the momentum-transfer cross section with relative velocity.}
  \label{fig:ClTT_EE_dmeff}
\end{figure}

In Figure~\ref{fig:ClTT_EE_dmeff}, we illustrate the effect of DM--proton interactions on CMB temperature and polarization power spectra by showing their residuals with respect to $\Lambda$CDM.
We show the power spectra computed using two treatments of the relative bulk velocity, for comparison: the ``main'' prescription (red) and the ``$k$-independent'' prescription (purple).

In both cases, we fix $\sigma_0$ to the 95\% C.L. upper limit, derived using the ``main'' prescription.
The effects of DM--proton scattering are as follows:
\begin{itemize}
\item The dominant effect on the CMB power spectra is a scale-dependent modulation of the acoustic-oscillation amplitude, which occurs for the following reasons.
  First, small modes enter the cosmological horizon earlier and are therefore subject to damping due to friction between the two fluids for a longer time than larger modes.
  Second, the interactions reduce the overall growth of perturbations, as well as the associated metric potentials that directly affect the CMB photons~\cite{Buen-Abad:2017gxg}.
  Prior to recombination, the interactions modify the Sachs-Wolfe (SW) contribution to the metric perturbations (both in terms of the overall amplitude and zero-point of oscillations in the quantity $\delta_\gamma/4 + \psi$, where $\psi$ is the gravitational potential in the Newtonian gauge).
  Around recombination, the time evolution of the metric perturbations is affected, in turn contributing to a change in the early integrated Sachs-Wolfe (EISW) effect.
  In the $TT$ spectrum, modifications of the SW and EISW terms lead to the relative enhancement of the first acoustic peak, while other peaks are suppressed overall.
\item At early times, photons are tightly coupled to electrons such that $\theta_b \! =\! \theta_\gamma$.
  The DM--baryon interaction effectively increases the inertia of baryons, suppressing the speed of sound in the plasma and reducing the frequency of the acoustic oscillations; as a consequence, the Doppler peaks shift to smaller physical and angular scales (larger $\ell$).
  We find that this effect is subdominant in the $TT$ spectrum.
  However, the $EE$ power spectrum is mostly sourced by the quadrupolar temperature patterns close to the last scattering surface~\cite{Seljak:1996is,Kamionkowski:1996ks,Zaldarriaga:1996xe} and is thus predominantly affected by modifications to the Doppler term.
\item The sound speeds of the DM and baryon fluids depend on the fluid temperatures and are therefore affected by the heat transfer.
  Since the sound-speed terms enter Eq.~\eqref{eq:fluct} with a prefactor of $k^2$, the dynamics of small angular scales (corresponding roughly to large $k$) are affected.
  We find that these terms also have a negligible contribution to the fluid evolution equations.
\item Finally, as we detail in Section~\ref{sec:thermal_history}, post-recombination cooling of baryons decreases the number of free electrons, in turn lowering the optical depth to the surface of last scattering, as compared to the $\Lambda$CDM case.%
  \footnote{As previously mentioned, early-time cooling of baryons accelerates recombination, in turn shifting the peaks toward higher $\ell$.
    However, this effect is not present, in practice, given the strength of the CMB constraints.}
  This effect is opposite to that of an early reionization and leads to an increase of power in modes that enter the horizon before reionization (\textit{i.e.}, $\ell \! \gtrsim \! 20$ for both temperature and polarization) and to lowering of the reionization bump in the $EE$ power spectrum.
  This effect is also subdominant, as far as the CMB observables are concerned.
  However, for higher cross sections, it produces a small modulation of power at the lowest values of $\ell$ in the $EE$ power spectrum.
\end{itemize}

\begin{figure}[t]
  \centering
  \includegraphics[width=0.48\linewidth]{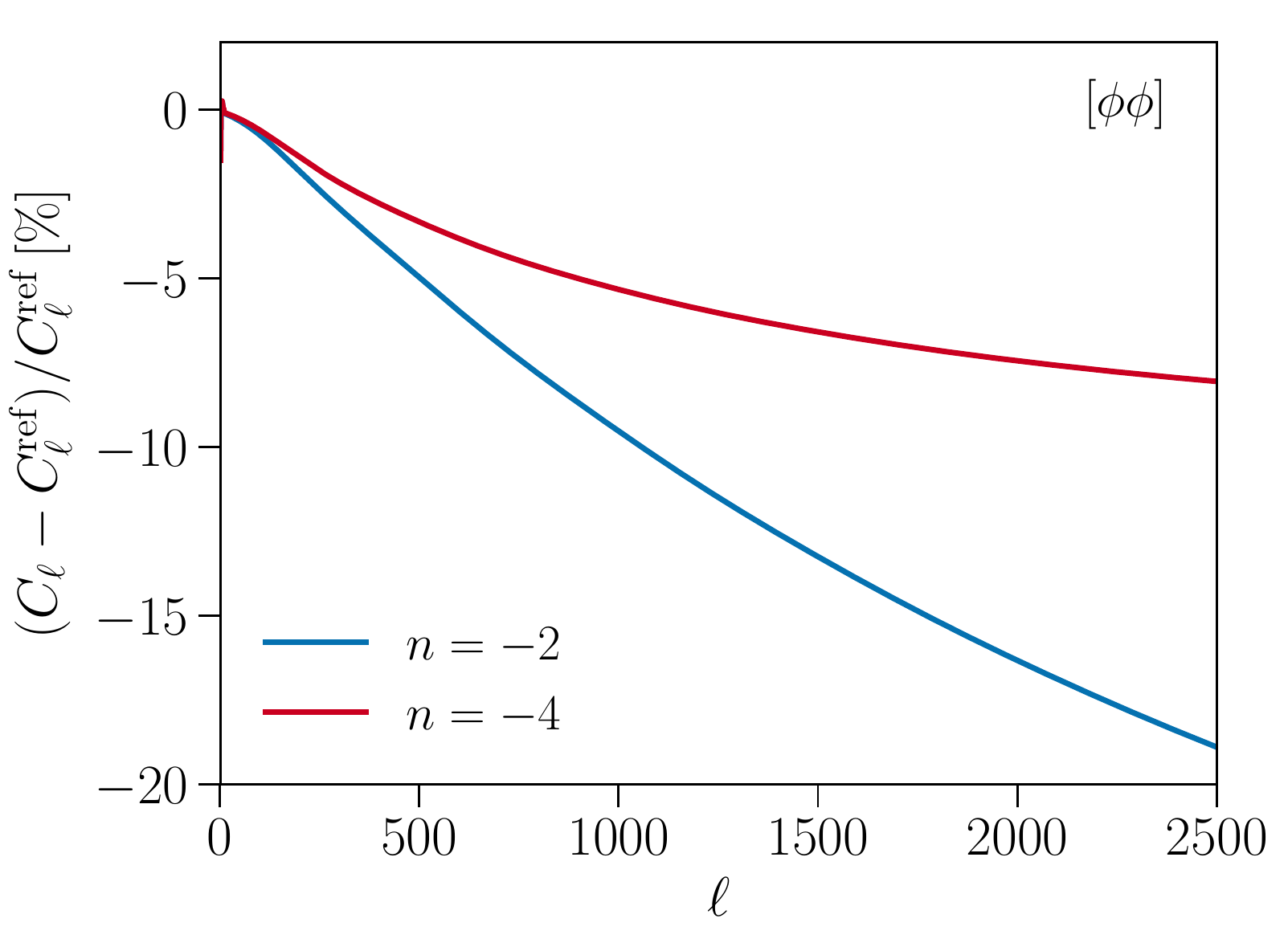}
  \hspace{0.2in}
  \includegraphics[width=0.48\linewidth]{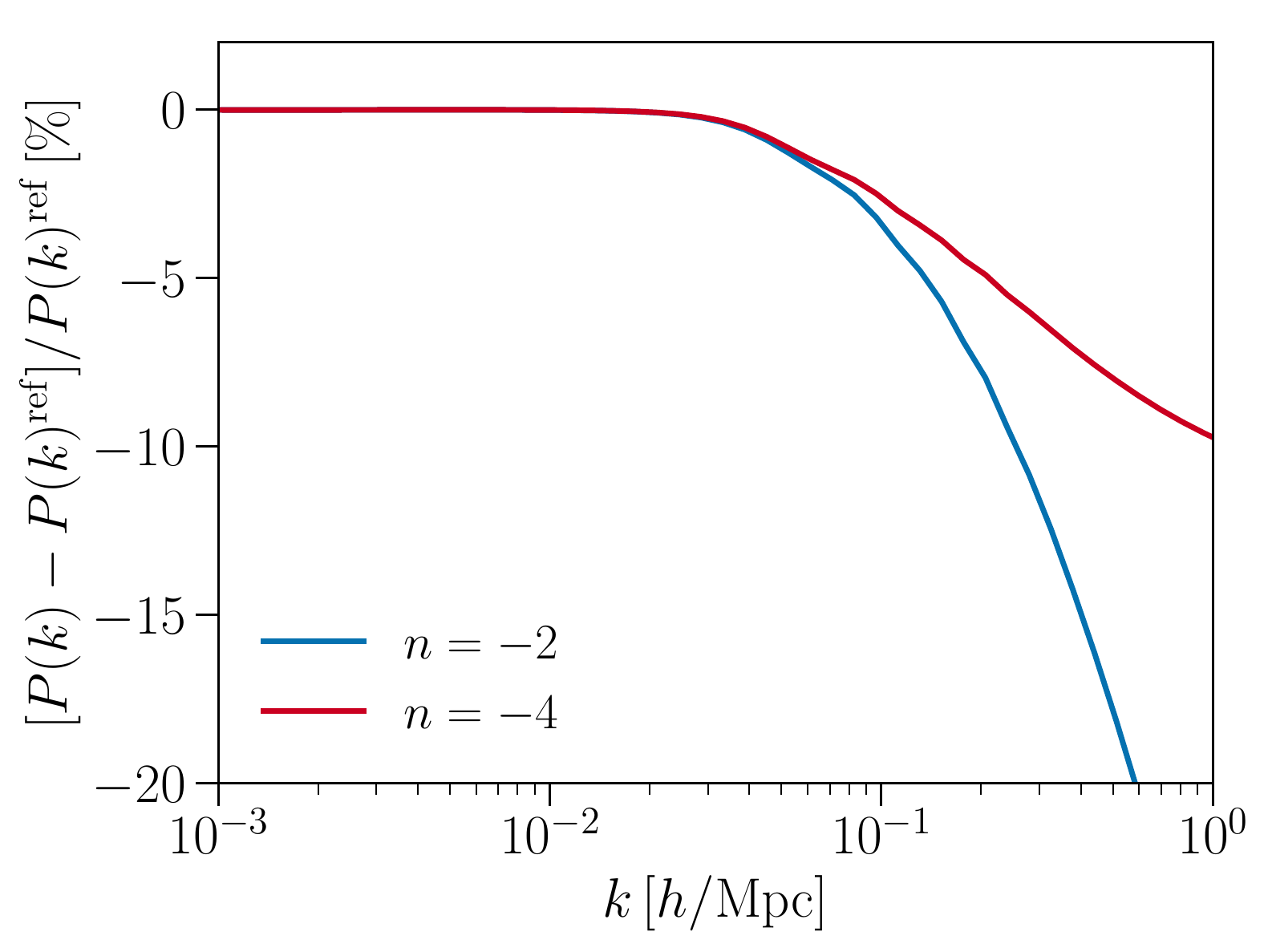}
  \caption{Percent residuals (with respect to $\Lambda$CDM) of the lensing-potential power spectrum $C_\ell^{\phi\phi}$ (left panel) and the linear matter power spectrum $P(k)$ (right panel) for the case of $n\!=\!-2$ (blue) and $n\!=\!-4$ (red), both for a DM mass of $1~\MeV$.
    We use the ``main'' prescription and set the coefficient of the momentum-transfer cross section to its 95\% C.L. upper limit, reported in Section~\ref{sec:constraints}.}
  \label{fig:Clphiphi_dmeff}
\end{figure}

The primary difference between the power spectra computed using the``main'' and the ``$k$-independent'' prescriptions is at high multipoles, where the ``main'' prescription leads to a more prominent damping tail.
For $n\!=\!-2$, the scaling of the momentum-transfer cross section with relative velocity is rather weak; the feedback of the interactions on the computation of $\Vrms$ and $\Vflow$ is small, and the power spectra of the two prescriptions look very similar.
The power spectra of the prescriptions for $n\!=\!-4$ exhibit a more noticeable difference.
For the remainder of this paper, we consider only our main ``main'' prescription.

Finally, in Figure~\ref{fig:Clphiphi_dmeff}, we show the effects of DM--proton scattering on the power spectrum of the CMB lensing potential (left) and the linear matter power spectrum (right); we plot residuals of the power spectra with respect to $\Lambda$CDM, using the ``main'' prescription, for $n\!=\!-2$ (blue) and $n\!=\!-4$ (red).
We set $\sigma_0$ to its 95\% C.L. upper limit, derived using the same prescription in Section~\ref{sec:constraints}.
The interactions suppress the growth of DM (and baryon) perturbations, resulting in a progressively larger reduction of power at smaller scales.
The suppression of lensing power manifests as a reduction of peak smearing in the $TT$ and $EE$ power spectra.

We note that it is possible to use large-scale structure data to constrain DM--baryon interactions with the matter power spectrum.
However, for the $n\!=\!-2$ and $n\!=\!-4$ models, constraints from the Lyman-$\alpha$ flux power spectrum yield a minor improvement upon CMB-only constraints at the $\mathcal{O}(1)$ level~\cite{Dvorkin:2013cea,Xu:2018efh}; and the Lyman-$\alpha$ data are subject to modeling caveats that the CMB is not.

\subsection{Strongly coupled DM fraction}
\label{sec:strong}

The discussion thus far has focused on DM comprised entirely of a single species that is only weakly coupled to baryons at all times.
It is worth noting that the CMB constraints imply weak coupling at all times only for the specific values of $n$ we are concerned with in this work, while for $n \geq 0$, CMB data constrain the interaction cross section such that the coupling is strong for $z\! \gtrsim\! 10^4-10^5$~\cite{Boddy:2018kfv,Gluscevic:2017ywp}.
In the strong-coupling case, DM is tightly coupled to baryons and behaves like an extra baryonic component, with the important caveat that it does not participate in recombination.
It does, however, experience dark acoustic oscillations, evident in the behavior of the matter power spectrum at scales $k\! \gtrsim\! 1~\Mpc^{-1}$~\cite{Boddy:2018kfv}.
It is also possible to have strong coupling for $n\!=\!-2$ and $n\!=\!-4$, without violating CMB bounds, if the interacting species represents only a fraction $f_\chi\! \equiv\! \rho_\chi/\rho_\textrm{DM}$ of the total DM density, while the remaining fraction is cold collisionless DM.
A strongly-coupled subcomponent of DM has been previously studied in the context of millicharged DM~\cite{Dubovsky:2003yn,Dolgov:2013una}.%
\footnote{We emphasize that, unlike Refs.~\cite{Dubovsky:2003yn,Dolgov:2013una}, we do not assume tight coupling between DM and baryons when analyzing data in Section~\ref{sec:constraints}; the strong coupling regime occurs as a consequence of the large value of the cross section allowed for small values of $f_\chi$.}

\begin{figure}[t]
  \includegraphics[width=0.48\linewidth]{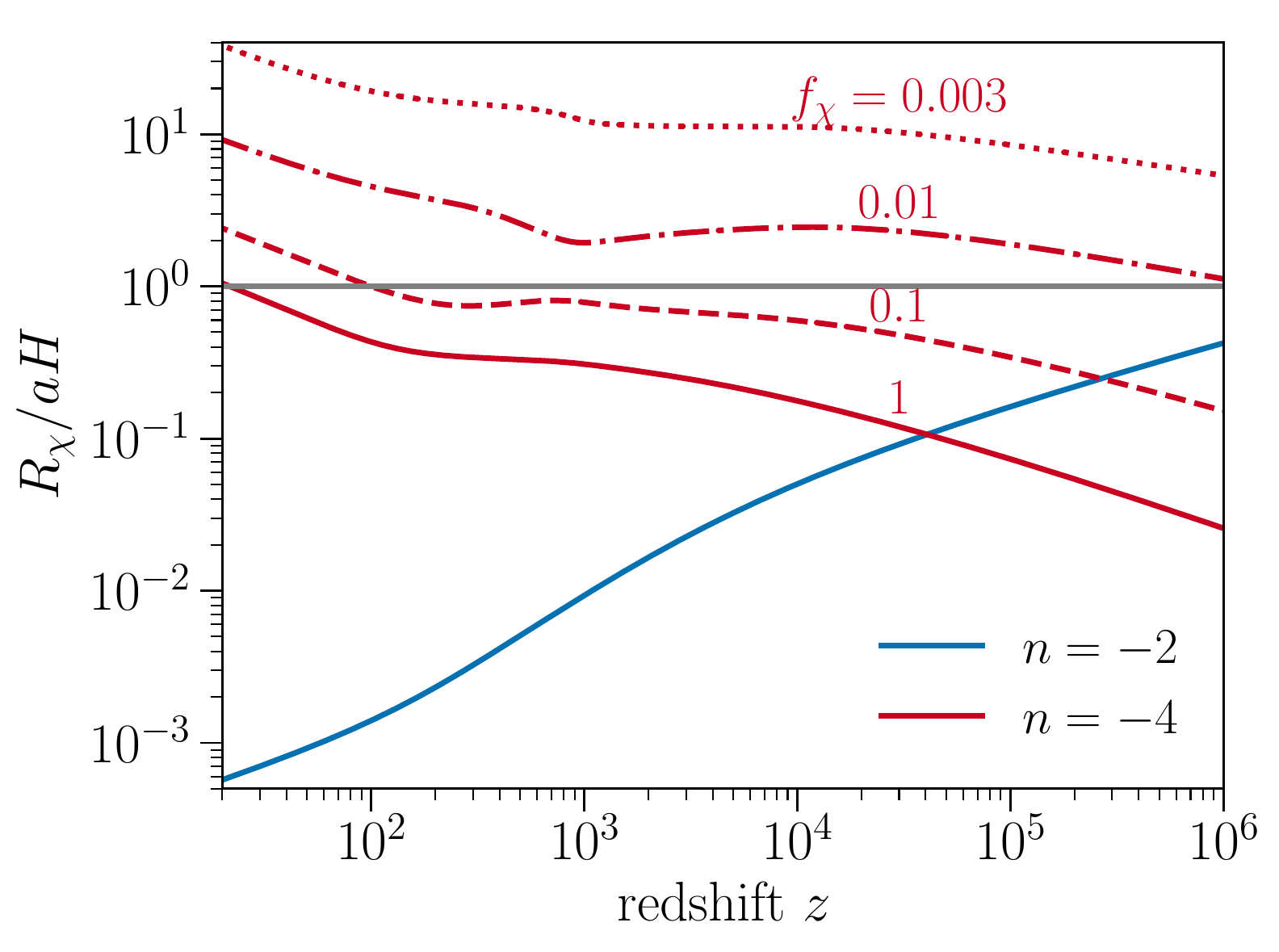}
  \hspace{0.2in}
  \includegraphics[width=0.48\linewidth]{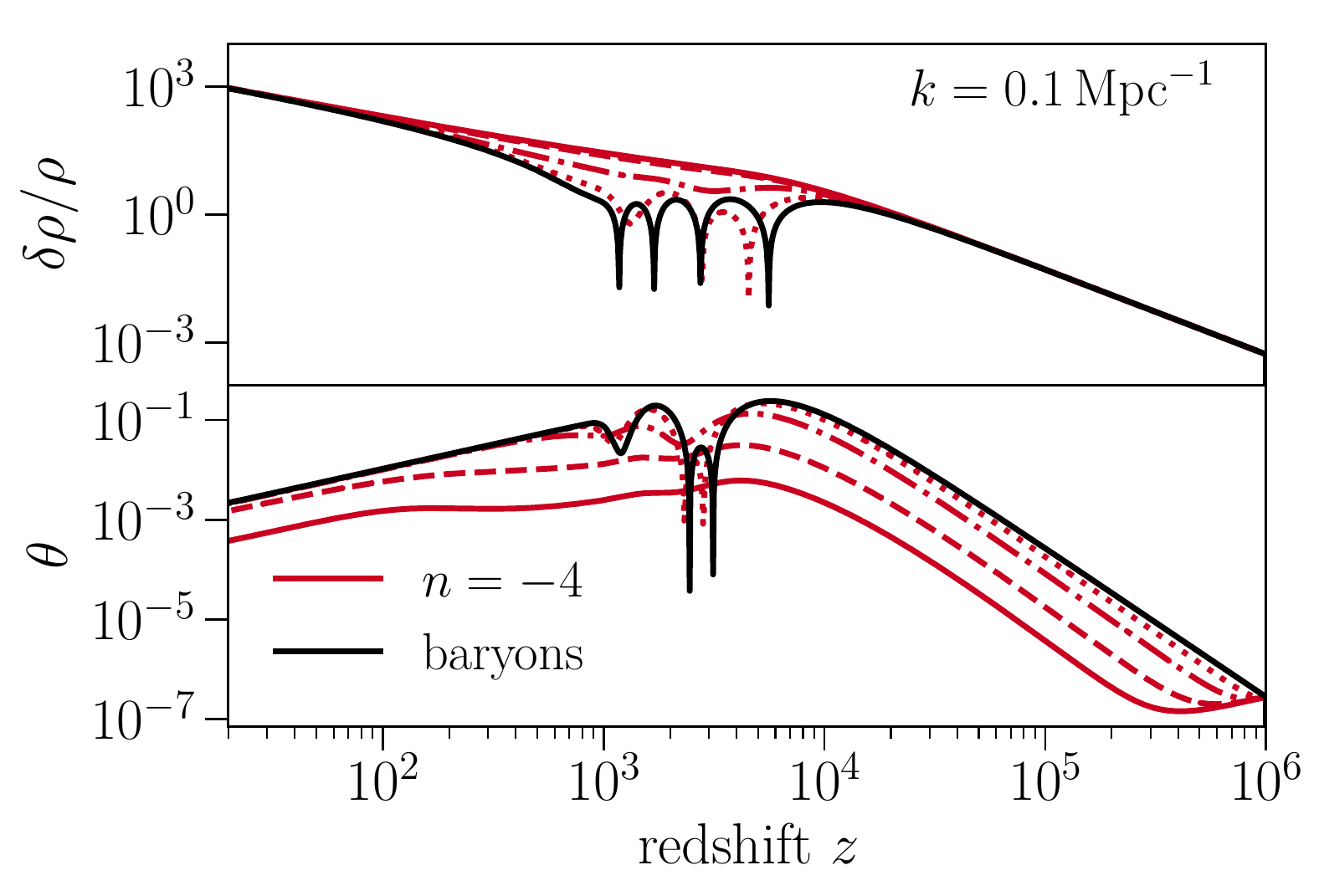}
  \caption{\textbf{[Left]:} The evolution of the ratio of the momentum-transfer rate to the expansion rate of the Universe for $n\!=\!-2$ (blue) and $n\!=\!-4$ (red), for interacting DM fractions $f_\chi \!=\! 1$ (solid), $0.1$ (dashed), $0.01$ (dot-dashed), and $0.003$ (dotted).
    We show a reference line (solid gray) where the momentum-transfer rate matches the expansion rate.
    \textbf{[Right]:} The density perturbations $\delta\rho/\rho$ (top panel) and velocity divergences (bottom panel) for the mode $k\!=\!0.1~\Mpc^{-1}$.
    We show the case of $n\!=\!-4$ (red) for the same interacting DM fractions considered in the left panel.
    For reference, we show the $\Lambda$CDM case for baryons (black); note that the line for the density perturbations of cold DM lies underneath those for the $n\!=\!-4$ interaction with fractions $f_\chi\!=\! 1$ and $f_\chi\!=\!0.1$.
    In both panels, we set the coefficient of the momentum-transfer cross section to its respective 95\% C.L. upper limit, derived using the ``main'' prescription in Section~\ref{sec:constraints}, and set the DM mass to $1~\MeV$.}
  \label{fig:fraction}
\end{figure}

In Figure~\ref{fig:fraction}, we demonstrate the behavior of a strongly-coupled DM subcomponent, for $m_\chi \! = \! 1~\MeV$.
In the left panel, we compare the evolution of the momentum-transfer rate to the expansion rate of the Universe for $n\!=\!-2$ (blue) and $n\!=\!-4$ (red), with $f_\chi \!=\! 1$ (solid), $0.1$ (dashed), $0.01$ (dot-dashed), and $0.003$ (dotted).
In the right panel, we plot the density perturbations and velocity divergences for $n\!=\!-4$ (red) and for baryons in $\Lambda$CDM.
In both panels, we set $\sigma_0$ in each case to its appropriate 95\% C.L. upper limit, derived using the ``main'' prescription.
For large values of $f_\chi$, \Planck{} constrains the momentum-transfer rate to be below the expansion rate, and the density perturbations of the interacting subcomponent of DM tracks those of cold DM in $\Lambda$CDM.
As $f_\chi$ decreases, the data allow for large momentum-transfer rates, and the interacting DM subcomponent begins to track the motion of the baryons more closely.
For $f_\chi \!=\! 0.003$, the modes that \Planck{} is sensitive to become tightly coupled upon entering the horizon, and the interacting DM subcomponent experiences acoustic oscillations.

\section{Planck Constraints}
\label{sec:constraints}

In this Section, we constrain DM--proton interactions using \Planck{} 2015 data.
We describe the data set and analysis method in Section~\ref{sec:data} and present numerical results in Section~\ref{sec:numresults}.

\subsection{Data and method}
\label{sec:data}

We analyze the \Planck{} 2015 temperature, polarization, and lensing power spectra, using the \Planck{} Likelihood Code v2.0 (\texttt{Clik/Plik})~\cite{2016A&A...594A..11P,2016A&A...594A...1P}; in particular, we use the nuisance-marginalized joint $TT$, $TE$, $EE$ likelihood, \texttt{Clik/Plik} {lite}, and the lensing likelihood with SMICA-map--based lensing reconstruction.%
\footnote{Potential issues with systematic effects in \Planck{} high-multipole polarization could, in principle, affect parameter estimation~\cite{2016A&A...594A..11P,2016A&A...594A...1P}, but Refs.~\cite{Gluscevic:2017ywp,Boddy:2018kfv} have demonstrated that exclusion of high-$\ell$ polarization degrades constraints on DM interactions by only a few percent.}
We sample the cosmological parameter space using the \texttt{MontePython}~\cite{Audren:2012wb} software package with the \texttt{PyMultinest}~\cite{2014A&A...564A.125B,Feroz:2007kg,Feroz:2008xx,Feroz:2013hea} likelihood sampler.
We verify that our sampling runs converge by evaluating the variance between several runs and by comparing a subset of results to those we obtain using a Markov chain Monte Carlo (\texttt{MCMC}) sampler.
The \texttt{MCMC} sampler implemented in \texttt{MontePython} uses the Metropolis-Hastings algorithm, and chain convergence is evaluated using the Gelman-Rubin convergence criterion $R-1<0.01$~\cite{Gelman:1992zz}.

There are nine free parameters in our interacting cosmology: the DM particle mass $m_\chi$, the fraction $f_\chi$ of the interacting subcomponent of DM, the coefficient of the momentum-transfer cross section $\sigma_0$, and the six standard $\Lambda$CDM parameters (baryon density $\Omega_bh^2$, total DM density $\Omega_\textrm{DM} h^2$, the Hubble parameter $h$, the reionization optical depth $\tau_\mathrm{reio}$, the amplitude of the scalar perturbations $A_s$, and the scalar spectral index $n_s$).
In most of our analysis runs, we fix the fraction $f_\chi$ and the mass $m_\chi$, and sample in the remaining seven free parameters using broad flat priors.
We also perform analysis runs in which we allow $f_\chi$ (or $m_\chi$) to be a free parameter, in which case we use broad log-flat priors on $f_\chi$ (or $m_\chi$) and $\sigma_0$ to sample the parameter space effectively.
We analyze the data for the $n\!=\!-4$ and $n\!=\!-2$ interaction models.

\subsection{Numerical results}
\label{sec:numresults}

\begin{figure}[t]
  \centering
  \includegraphics[scale=0.4]{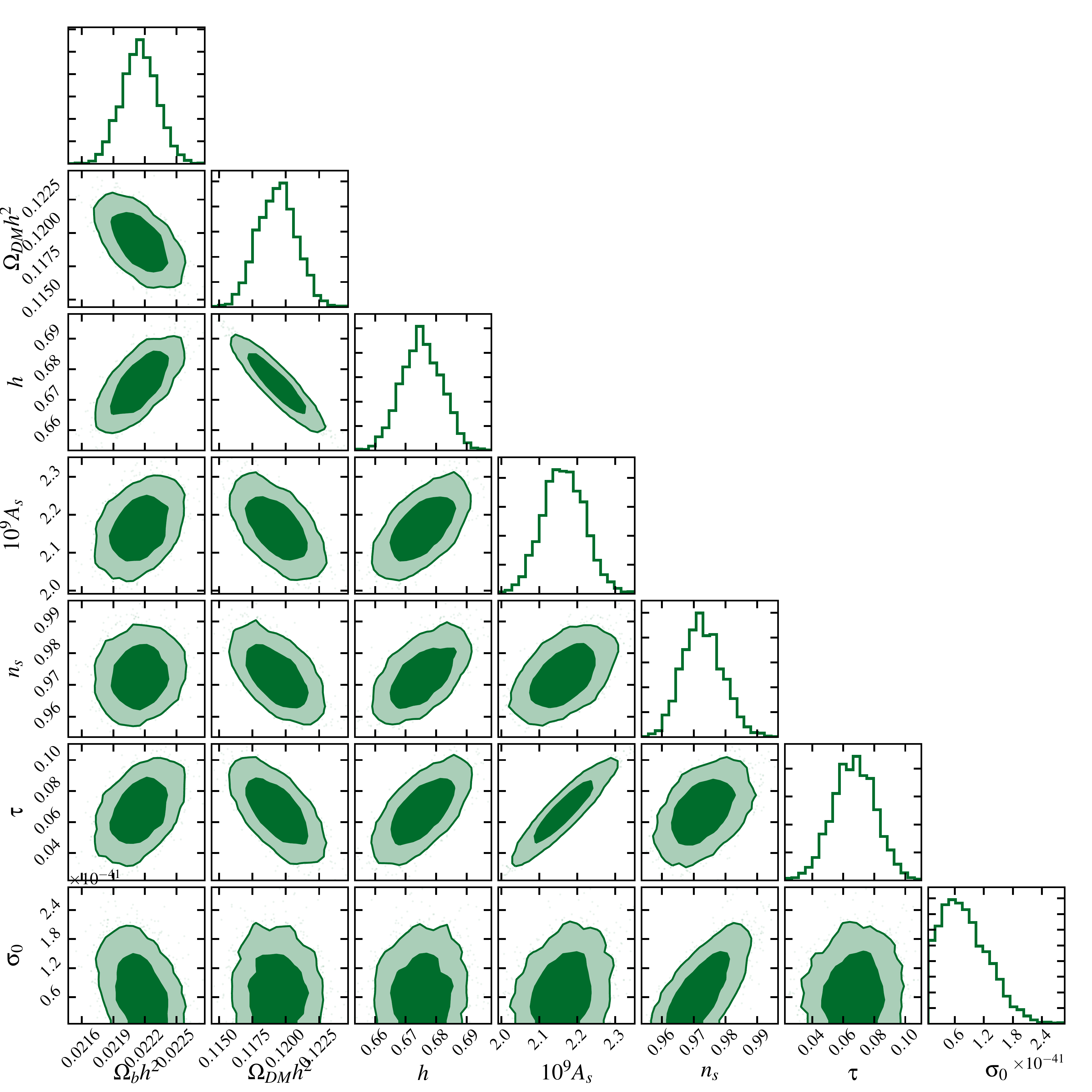}
  \caption{The posterior probability distribution for the $\Lambda$CDM parameters and the coefficient of the DM--proton momentum-transfer cross section for $n\!=\!-4$ interaction and a DM mass of $1~\MeV$.
    We show the 68\% and 95\% C.L. contours, obtained from a joint analysis of \Planck{} 2015 temperature, polarization, and lensing anisotropies.
    The one-dimensional, marginalized posteriors are shown at the top of each column.}
  \label{fig:posterior-4}
\end{figure}

We first assume that all of the DM matter is interacting ($f_\chi\!=\! 1$) and perform the likelihood analysis for $n\!=\!-4$ and $n\!=\!-2$, sampling the likelihoods in $\sigma_0$ and the six $\Lambda$CDM parameters, for seven fixed benchmark DM masses between 10 keV and 1 GeV (see Table~\ref{tab:exclusion}).
Representative examples of the reconstructed marginalized posterior probability distributions%
\footnote{Posterior probability distributions in this study were visualized using \texttt{corner.py}~\cite{corner}.}
are shown in Figure~\ref{fig:posterior-4} for the case of $n\!=\!-4$ and $m_\chi\!=\!1~\MeV$.
The general shape of the posteriors does not significantly vary as a function of DM mass and is qualitatively similar for the $n\!=\!-2$ case.
There is a prominent (positive) degeneracy between $\sigma_0$ and the scalar spectral index $n_s$: DM interactions suppress power on small scales in the CMB $TT$ power spectrum, and an increase in $n_s$ can counteract this suppression.
The mild correlations with $A_s$ and $\tau_\textrm{reio}$ are also due to the suppression of power at high values of $\ell$, but arise from a combination of the $TT$ and lensing likelihood.
The value of $A_s$ controls the overall amplitude of all power spectra, but it is modulated by a factor $\exp (-2\tau_\textrm{reio})$ above $\ell \! \simeq \! 20$ in the $TT$ power spectrum.
Increasing $A_s$ compensates for the power suppression in the lensing power spectrum, but it also requires a larger value of $\tau_\textrm{reio}$ in order to keep the combination $A_s\exp (-2\tau_\textrm{reio})$ fixed, so as not to affect the high-$\ell$ normalization of the $TT$ power spectrum.
The mild anticorrelation with $\Omega_\chi h^2$ is due to the fact that a smaller value of $\Omega_\chi h^2$ leads to a change in the expansion history that compensates for the shift of the peak positions produced by DM--baryon scattering.
However, reducing $\Omega_\chi h^2$ also delays matter--radiation equality, which boosts the amplitude of the EISW; thus, the degeneracy is very weak.
Similarly, the shift in the peak positions can be compensated by altering the value of the Hubble rate, resulting in a mild, positive correlation between $\sigma_0$ and $h$.

\begin{figure}[t]
  \centering
  \includegraphics[width=0.48\linewidth]{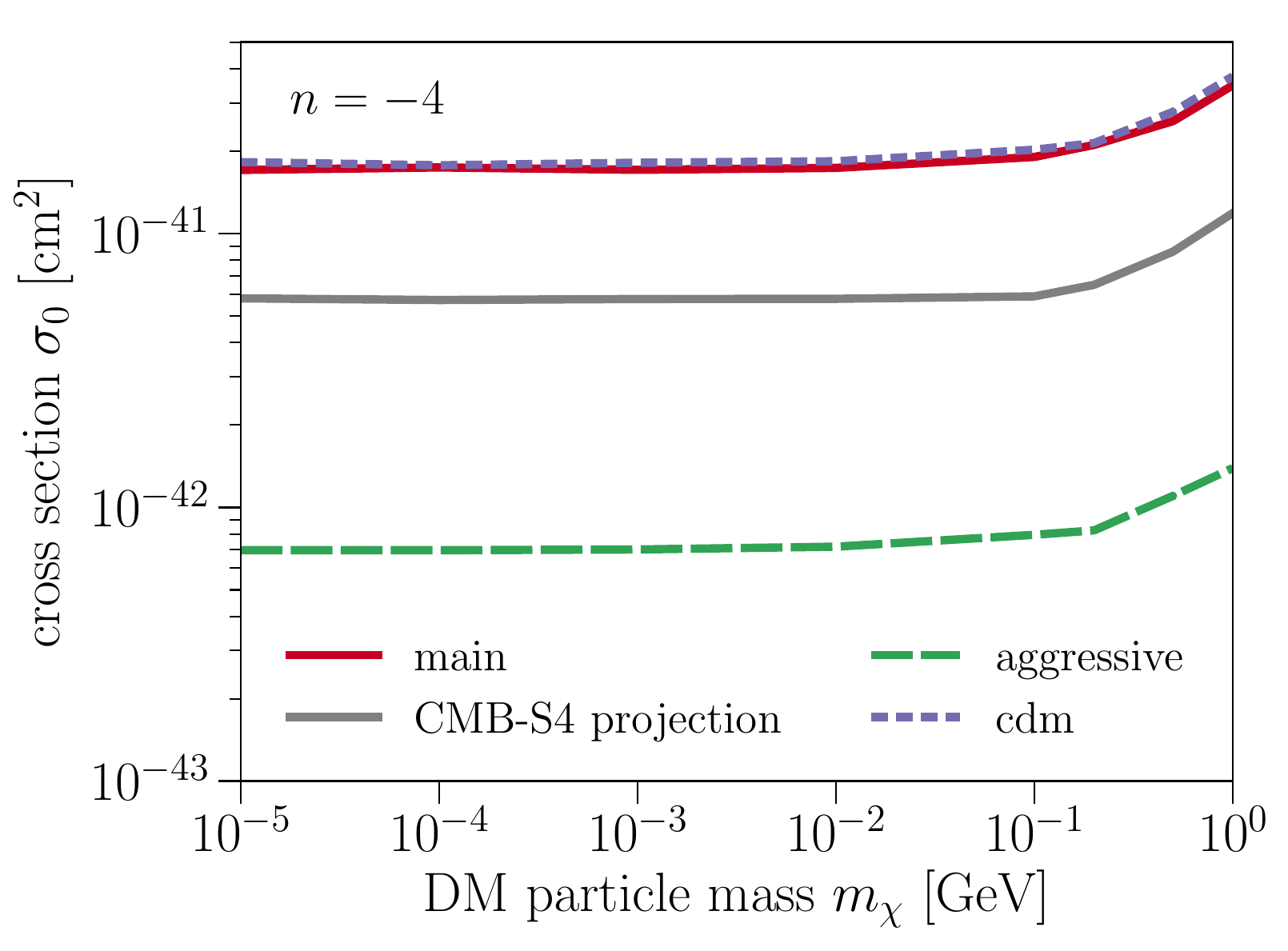}
  \hspace{0.2in}
  \includegraphics[width=0.48\linewidth]{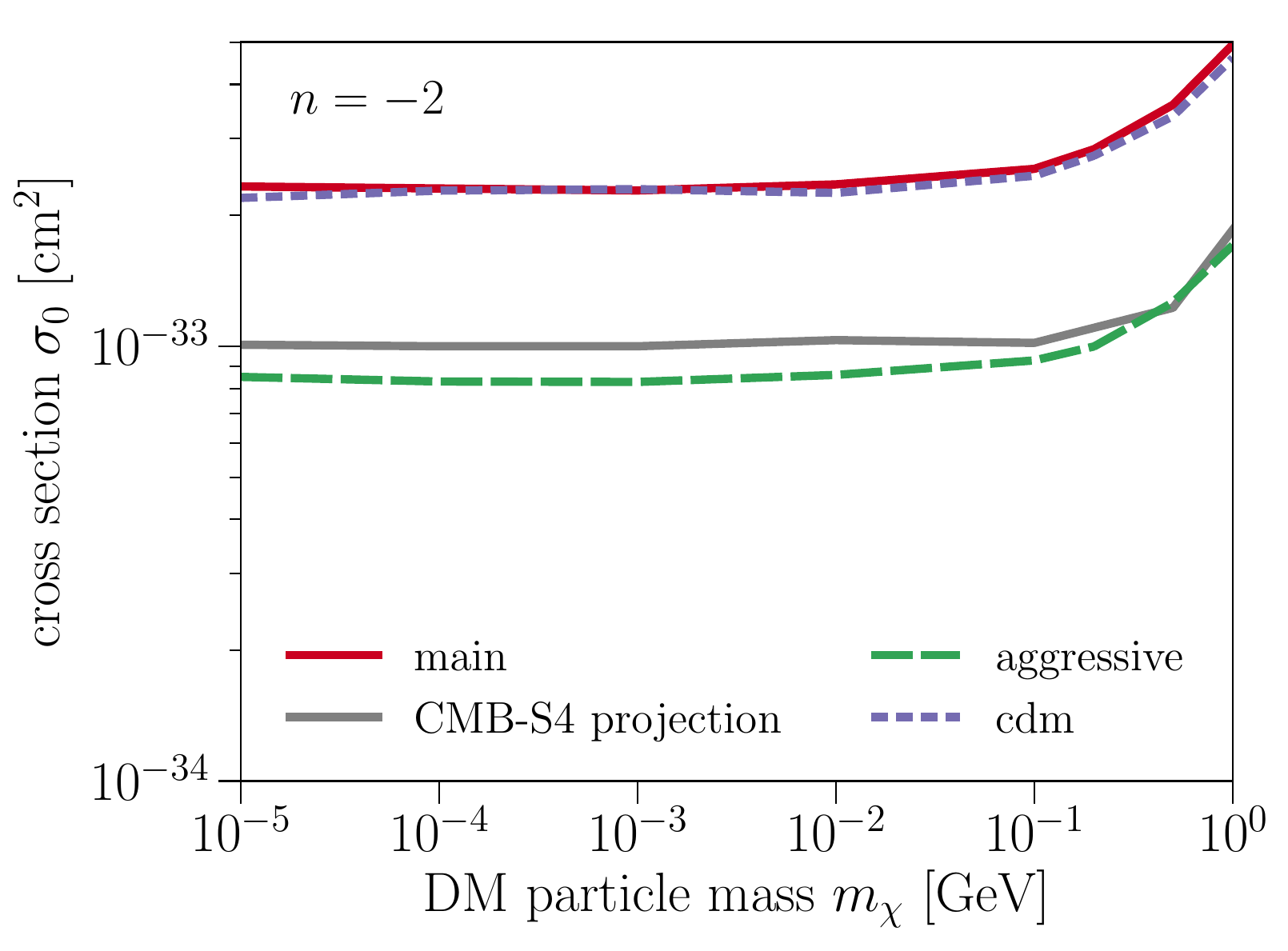}
  \caption{The 95\% C.L. upper limits for the coefficient of the DM--proton momentum-transfer cross section as a function of DM particle mass, obtained from likelihood analysis of \Planck{} 2015 temperature, polarization, and lensing anisotropies, for the $n\!=\!-4$ (left panel) and $n\!=\!-2$ (right panel) interactions.
    Results are shown for our ``main'' treatment of the relative bulk velocity between the DM and baryon fluids, described in Section~\ref{sec:boltzmann}.
    Additionally, we show the inferred limit from an ``aggressive'' assumption, which ignores the impact of the relative bulk velocity.
    For comparison, we plot the limit we obtain using the prescription proposed in previous literature (denoted as ``cdm'')~\cite{Dvorkin:2013cea}.
    We also show the projected sensitivity for a future ground-based CMB-Stage 4 experiment (obtained using the ``main'' prescription).}
  \label{fig:exclusion}
\end{figure}

We find no evidence for DM--proton scattering in the \Planck{} 2015 data---all marginalized probabilities for $\sigma_0$ are consistent with zero, and we use them to infer an upper limit on $\sigma_0$ as a function of $m_\chi$.
We present our 95\% C.L. exclusion curves in Figure~\ref{fig:exclusion} and in the corresponding Table~\ref{tab:exclusion}.
In Section~\ref{sec:cmb}, we have demonstrated the importance of accounting for the relative bulk velocity when computing the effects of scattering on CMB observables.
Using the results of our sampling runs, we show the limit (solid red) we obtain with our ``main'' prescription.
Our limits virtually have no mass dependence for $m_\chi\! \lesssim \! 10~\MeV$ (see also Ref.~\cite{Slatyer:2018aqg}): for $m_\chi\!\ll\! m_p$, $m_\chi$ appears in the Boltzmann equations via the thermal term $\vth$ in the momentum-transfer rate.
As demonstrated in Figure~\ref{fig:xe_and_Tm_n2_n4}, the DM temperature is negligible for CMB calculations, such that the thermal velocity of the baryons dominates $\vth$.
Thus, the DM mass dependence drops out of the momentum-transfer rate entirely.
On the other hand, increasing the DM mass to become comparable to or exceed the proton mass, the momentum-transfer rate scales as $R_\chi\!\sim\! \sigma_0 (m_\chi + m_p)^{-1}$, while $\vth$ continues to be dominated by the thermal velocity of the baryons and thus does not contribute to the mass scaling.
Hence, our limits on $\sigma_0$ should scale as $m_\chi$ for $m_\chi\!\gg\! m_p$; a transition towards this behavior is visible at the high-mass end of Figure~\ref{fig:exclusion}.

For comparison, we also reproduce the limit obtained using the prescription from previous literature~\cite{Dvorkin:2013cea,Xu:2018efh,Slatyer:2018aqg} (dotted line, denoted as ``cdm'').%
\footnote{We have verified that the residual spectra (calculated with respect to $\Lambda$CDM) we obtain by implementing the methods of Refs.~\cite{Xu:2018efh,Slatyer:2018aqg} align with those from Ref.~\cite{Slatyer:2018aqg}.
  However, we note that we obtain constraints that are a factor of $\sim\! 1.8$ weaker for $n\!=\!-4$ than those in Refs.~\cite{Xu:2018efh,Slatyer:2018aqg} (using the same likelihoods and sampling methods as Ref.~\cite{Slatyer:2018aqg}) and a factor of $\sim\! 1.4$ stronger for $n\!=\!-2$ than Ref.~\cite{Xu:2018efh}.}
Since the CMB constrains the cross section to be quite small, the amount of interaction does not significantly alter the evolution of $\Vvar$ from its $\Lambda$CDM counterpart $\Vvarcb$.
As we discuss at the end of Section~\ref{sec:CMB-bulkvel}, this leads to our ``main'' treatment of the relative bulk velocity to yield similar limits to the ``cdm'' prescription of previous work.
Below, we consider the case of strongly-coupled DM, where the ``cdm'' prescription of previous work is not valid.

For illustration only, in the same figure, we show ``aggressive'' constraints that are inferred when a vanishing relative bulk velocity is assumed in Eqs.~\eqref{eq:Vfull} and \eqref{eq:temp}, and thus the momentum-transfer rate is completely unsuppressed by the bulk motions.
This assumption does not hold for the case of $f_\chi \!=\! 1$: at the level of the upper limit on $\sigma_0$, there is not enough friction between the DM and baryon fluids to entirely dissipate the relative bulk velocity.
The ``aggressive'' constraint demonstrates the importance of properly incorporating the relative bulk velocity, especially for $n\!=\!-4$, where the difference in the limit is more than an order of magnitude.
Additionally, while we expect our ``main'' prescription to well-represent the exact solution, the ``aggressive'' constraint gives an absolute floor on the possible improvement that an exact treatment of the relative bulk velocity could potentially achieve.

\begin{table}[t]
  \begin{tabular}{ |c|c|c|c|c|c|c|c| }
    \hline
    & \textbf{10 keV}& \textbf{1 MeV}& \textbf{10 MeV}& \textbf{100 MeV}& \textbf{200 MeV}& \textbf{500 MeV}& \textbf{1 GeV}\\
    \hline
    $n=-4$ & 1.7e-41 & 1.7e-41 & 1.7e-41 & 1.9e-41 & 2.1e-41 & 2.6e-41 & 3.5e-41\\
    \hline
    $n=-2$ & 2.3e-33 & 2.3e-33 & 2.4e-33 & 2.6e-33 & 2.8e-33 & 3.6e-33 & 4.9e-33\\
    \hline
  \end{tabular}
  \caption{A list of the 95\% C.L. exclusion limits on coefficient of the DM--proton momentum-transfer cross section, $\sigma_0$, given in units of $\cm^2$ and obtained from \Planck{} 2015 temperature, polarization, and lensing anisotropy measurements, for the $n\!=\!-4$ and $n\!=\!-2$ interactions.
    DM masses are listed along the top row.
    The limits correspond to those in Figure~\ref{fig:exclusion} and are computed using our ``main'' prescription to account for the relative bulk velocity of the DM and baryon fluids.}
  \label{tab:exclusion}
\end{table}

We further perform a forecast of the sensitivity to $n\!=\!-4$ and $n\!=\!-2$ scattering for a future ground-based CMB-Stage 4 experiment~\cite{2016arXiv161002743A}.
We consider (in combination with \Planck{} data) an experiment with noise levels of 1~$\mu$K-arcmin and a beam size of 1~arcmin, with a survey covering 40\% of the sky, assuming $\ell_\textrm{min}\!=\!30$ and $\ell_\textrm{max}\!=\!3000$.
We do not consider CMB lensing in this analysis, which may substantially improve sensitivity~\cite{Li:2018zdm}; thus, our result is a conservative projection.
For $n\!=\!-4$, we find an improvement over the current constraints from \Planck{} 2015 by a factor of $\sim\! 2.9$ for a DM mass of $1~\MeV$, giving $\sigma_0 \! < \! 5.8 \times 10^{-42}$ at 95\% C.~L.
For $n\!=\!-2$, we find $\sigma_0 \! < \! 1.0\times10^{-33}$ for the same mass, which is a factor of $\sim\! 2.3$ improvement over \textit{Planck}.
In both cases, we use the ``main'' prescription for the relative bulk velocity.
We show the corresponding projected exclusion curves in Figure~\ref{fig:exclusion}.

\begin{figure}
  \centering
  \includegraphics[width=0.48\linewidth]{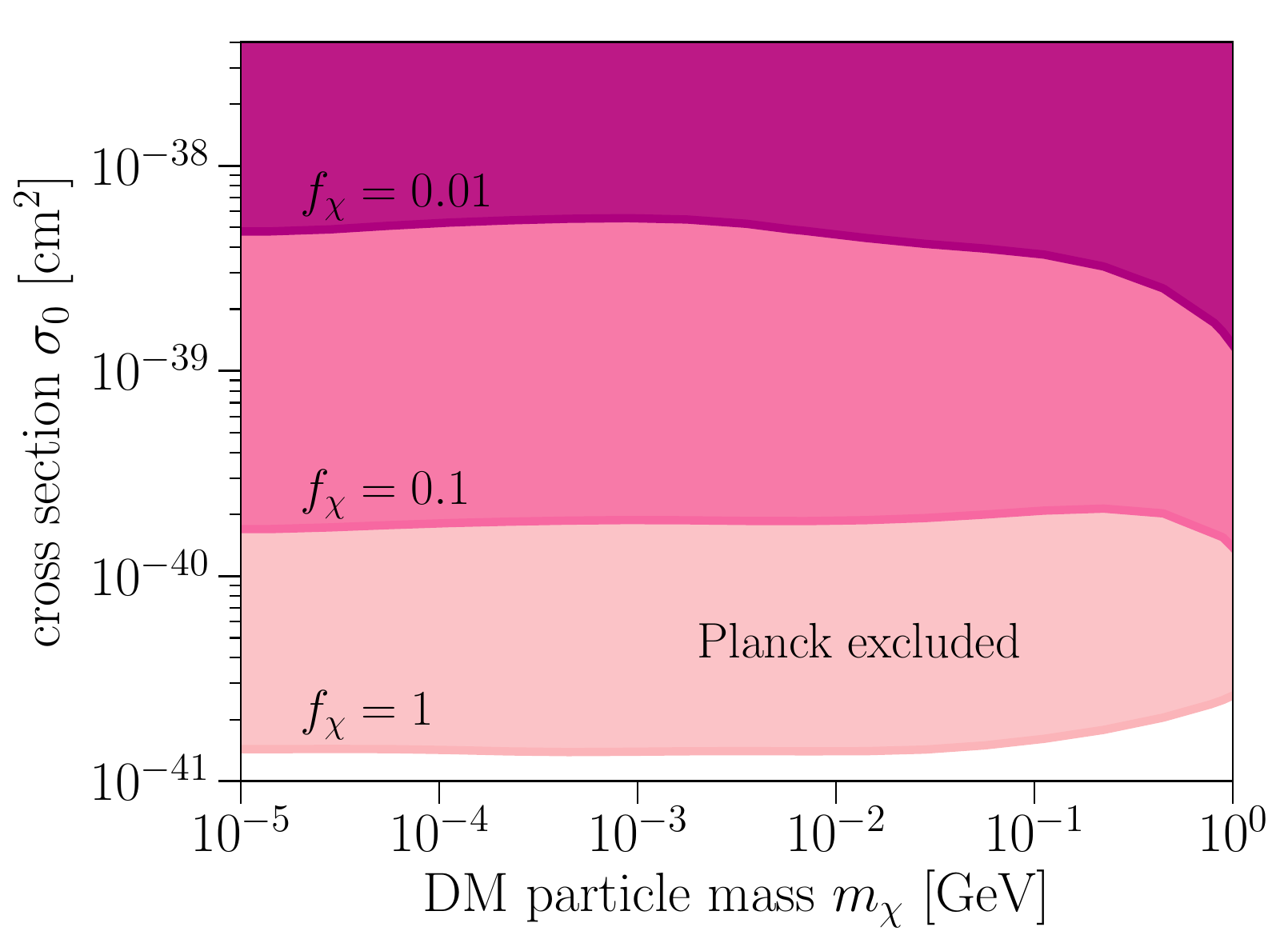}
  \hspace{0.2in}
  \includegraphics[width=0.48\linewidth]{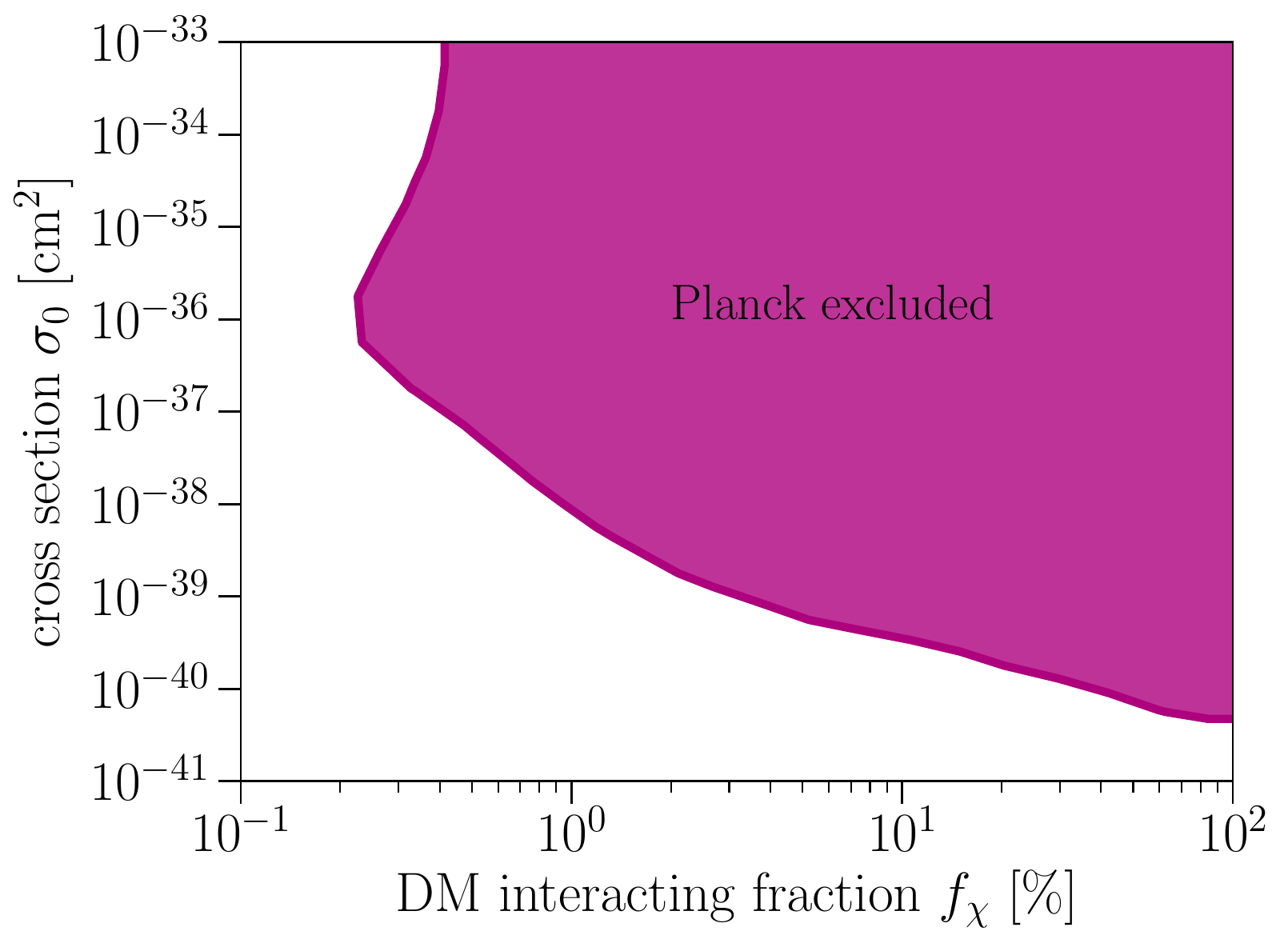}
  \caption{\textbf{[Left]:} The 95\% C.L. excluded region for the coefficient of the DM--proton momentum-transfer cross section as a function of DM mass, obtained by analyzing \Planck{} 2015 data, when the interacting fraction $f_\chi$ of the total DM energy density is fixed to (from the lightest to the darkest pink): $1$, $0.1$, and $0.01$.
    \textbf{[Right]:} The 95\% C.L. excluded region for the coefficient of the DM--proton momentum-transfer cross section as a function of the interacting fraction $f_\chi$ of the total DM energy density, for a fixed DM mass $m_\chi\! =\! 1~\MeV$; it is obtained by analyzing \Planck{} 2015 temperature, polarization, and lensing anisotropy measurements, for the $n\!=\!-4$ interaction, allowing $f_\chi$ to vary as a free sampling parameter.
    The exclusion is bimodal for fractions $0.2\%\!\lesssim\!f_\chi\!\lesssim\!0.4$\% and constraints completely relax for $f_\chi\!\lesssim\!0.2$\% (where DM becomes cosmologically indistinguishable from a small additional amount of baryons).}
  \label{fig:exclusion-fraction}
\end{figure}

To investigate how the limits presented above may change when only a fraction of DM interacts with baryons, we reanalyze \Planck{} 2015 data for the case of $n\!=\!-4$.
In the left panel of Figure~\ref{fig:exclusion-fraction}, we show the upper limits on $\sigma_0$ as a function of the DM mass $m_\chi$, fixing the interacting fraction to $f_\chi\!=\! 1$, $0.1$, and $0.01$.
The mass dependence of the constraint for $f_\chi\!=\! 1$  differs from that for $0.1$ and $0.01$ at high masses.
The temperature of the interacting DM subcomponent is negligible compared to the baryon temperature for DM masses $m_\chi\!\lesssim\! 10~\MeV$; the momentum-transfer rate is essentially independent of the DM mass, and thus so is the limit on $\sigma_0$.
At higher DM masses, however, the heat-exchange rate becomes larger for a fixed $\sigma_0$: still neglecting the DM mass dependence of $\vth$ in the expression for $R_\chi$, the heat-exchange rate coefficient scales as $R_\chi^\prime \!\sim\! m_\chi/(m_\chi + m_p)^2$.
Hence, at the higher end of the mass region in the right panel of Figure~\ref{fig:exclusion-fraction}, $T_\chi$ is no longer negligible compared to $T_b$, and the momentum-transfer rate scales as $R_\chi\!\sim\! \sigma_0 (m_\chi + m_p)^{-1} (T_b/m_p \! + \! T_\chi/m_\chi)^{-3/2}$.
It is thus reasonable to expect the limit on $\sigma_0$ to strengthen over a range of intermediate DM masses (\textit{i.e.}, near the proton mass).

To capture the $f_\chi$ dependence of the limits in further detail, we again reanalyze \Planck{} 2015 data for the case of $n\!=\!-4$, this time fixing the DM mass to $m_\chi\! =\! 1~\MeV$ and sampling the fraction $f_\chi$ as a free parameter.
In the right panel of Figure~\ref{fig:exclusion-fraction}, we show the resulting marginalized 2d posterior probability distribution for $\sigma_0$ versus $f_\chi$; the shaded region represents the outside of the 95\% C.L. contour and is thus excluded.
For $f_\chi \! \gtrsim\! 2\%$, the limit roughly scales with $f_\chi$, independent of the DM mass.
This scaling no longer holds for smaller values of $f_\chi$: as a smaller fraction of DM particles scatters with baryons, CMB bounds permit a larger value of $\sigma_0$.
At sufficiently large $\sigma_0$, the small interacting DM subcomponent allowed by data is tightly coupled to the baryons and undergoes (and amplifies) acoustic oscillations, and the effect of DM--baryon scattering on the power spectra saturates.
Specifically, when the energy density of the interacting DM subcomponent approaches the uncertainty on the baryon energy density, the interacting DM subcomponent becomes cosmologically indistinguishable from baryons, and the limits on $\sigma_0$ entirely relax.%
\footnote{The caveat is that DM does not participate in recombination and is thus not entirely degenerate with baryons.}
We find that the relaxation occurs for fractions $f_\chi\!\lesssim\!0.4\%$, which is consistent with the value $f_\chi\!\lesssim\!0.6\%$, derived using the current constraint on the helium fraction~\cite{dePutter:2018xte}.

\begin{figure}
  \includegraphics[width=0.48\linewidth]{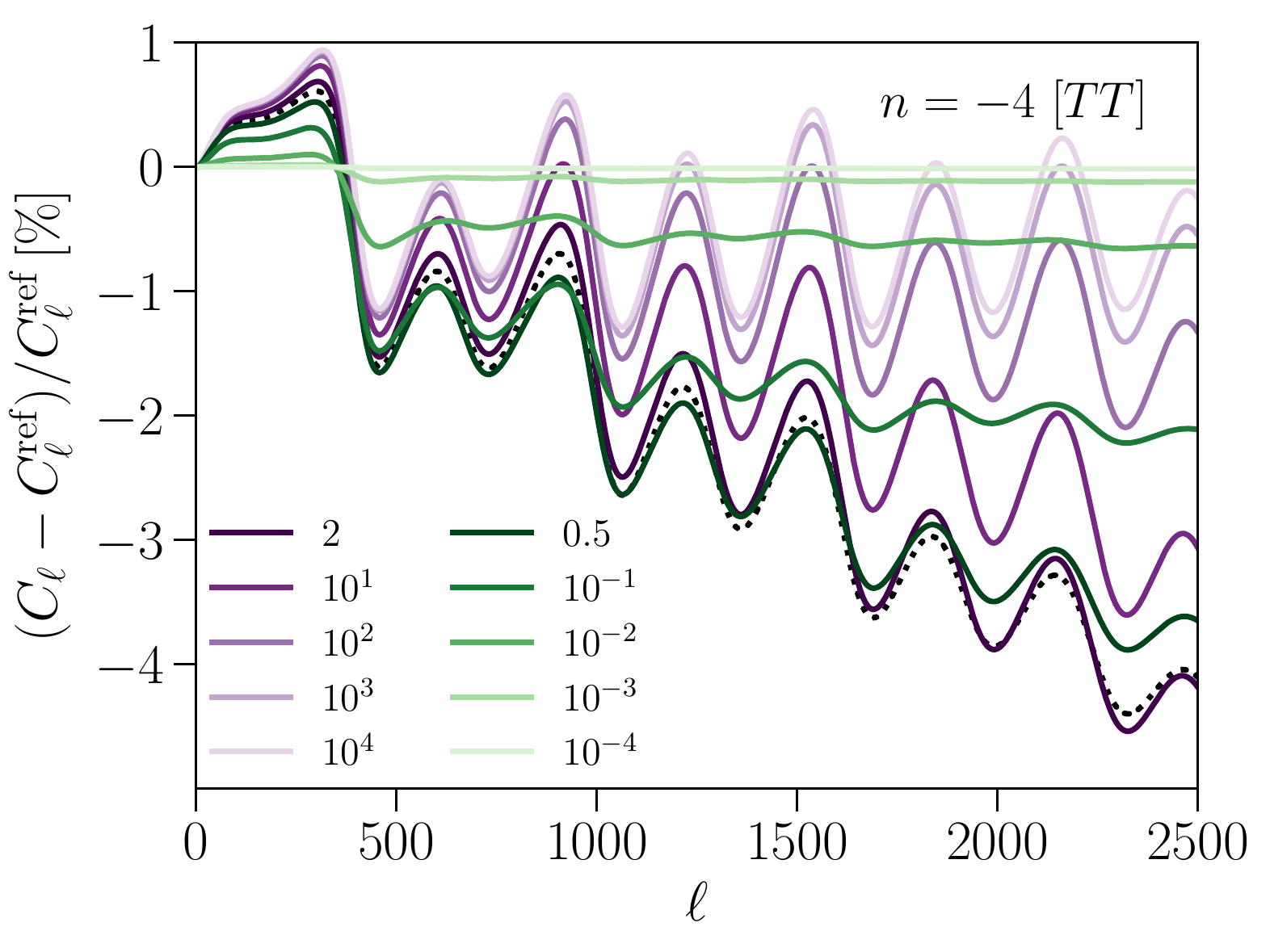}
  \caption{Percent residuals (with respect to $\Lambda$CDM) of the lensed $TT$ power spectra for the case of $n\!=\!-4$.
    The fraction of the total DM energy density in the interacting subcomponent is fixed to $0.3$\%, for a DM mass of $1~\MeV$, and we use the ``main'' prescription.
    We show the residuals with the coefficient of the momentum-transfer cross section $\sigma_0$ set to its 95\% C.L. the upper limit (dotted black) and the residuals with $\sigma_0$ set to be a numerical factor---given in the legend---above (purple) and below (green) the upper limit.
    We see the non-monotonic behavior of the residuals as a function of $\sigma_0$, which leads to the bimodality in the inferred posterior of Figure~\ref{fig:exclusion-fraction} around this value of $f_\chi$, since the regime of the turnover is detectable with \Planck{}.}
  \label{fig:bimodal}
\end{figure}

We illustrate this effect in Figure~\ref{fig:bimodal}, where we show the non-monotonic behavior of the residual in the $TT$ spectra (with respect to $\Lambda$CDM) as a function of increasing $\sigma_0$, for $m_\chi\!=\! 1~\MeV$ and $f_\chi\!=\!0.3\%$ in the $n\!=\!-4$ case.
The residuals are shown at its 95\% C.L. upper limit, derived using the ``main'' prescription (black dotted) and for $\sigma_0$ above (purple) and below (green) the upper limit.
Decreasing $\sigma_0$ well below its upper limit effectively turns off interactions between DM and baryons, and the residuals disappear.
On the other hand, a sufficient increase in $\sigma_0$ brings the spectrum closer to the reference $\Lambda$CDM spectrum, and the residuals saturate as the DM and baryon fluids become as tightly coupled as possible.%
\footnote{Complications occur if the strength of DM interaction with baryons approaches that of the Compton interaction between baryons and photons before CMB decoupling.
  This regime significantly impacts acoustic oscillations and leads to strong numerical instabilities; thus, we do not consider it further.}
If the turnover in the residuals is detectable, the inferred posterior is bimodal; we see this behavior for fractions in the range $0.2\% \!\lesssim\! f_\chi \!\lesssim\!0.4\%$ in Figure~\ref{fig:exclusion-fraction}.

\section{Discussion and Conclusions}
\label{sec:conclusions}

We have conducted a comprehensive study of the impact of scattering between DM and protons on the CMB power spectra.
In particular, we adopted a phenomenological approach of parameterizing the momentum-transfer cross section as $\sigma_\textrm{MT} \! = \! \sigma_0 v^n$ (where $v$ is the relative velocity between the scattering particles) and focused on negative powers of velocity dependence that arise in well-motivated simplified models of DM interactions: $n\!=\!-2$ and $n\!=\!-4$.
Such interactions are cosmologically important at times close to recombination, unlike the class of models with $n\! \geq\! 0$, for which scattering has the largest impact in the early Universe.

We assessed the impact of the relative bulk velocity between the DM and baryon fluids that may arise in the pre-recombination Universe, when the relative bulk velocity surpasses the relative thermal velocity dispersion.
A large relative bulk velocity results in nonlinear Boltzmann equations and the mixing of Fourier modes.
We presented a new treatment to sidestep these difficulties, while capturing the physics behind mode coupling: we introduce a mode-dependent RMS velocity dispersion as a proxy for the bulk relative velocity, and we incorporated it into the computation of the linear Boltzmann equations in a self-consistent manner appropriate for a cosmology that includes DM--baryon scattering.

We analyzed \Planck{} 2015 temperature, polarization, and lensing data to search for evidence of DM--proton scattering.
We found that the data are consistent with no interactions and use our results to produce upper limits on the coefficient of the momentum-transfer cross section as function of DM mass, shown in Figure~\ref{fig:exclusion} and provided in Table~\ref{tab:exclusion}.
Additionally, we considered the case in which only a small fraction of DM interacts with protons; for a DM mass of $1~\MeV$, we constrained the DM--proton interaction as a function of the fraction in Figure~\ref{fig:exclusion-fraction}.

We discussed two regimes of DM--baryon scattering: weak-coupling regime, in which the momentum-transfer rate is inefficient due to Hubble expansion and damps acoustic oscillations on small scales; and a strong-coupling regime, in which DM and baryons are tightly coupled, resulting in DM undergoing acoustic oscillations with the baryons.
If all the DM is allowed to interact with baryons, \Planck{} data constrain the interaction to be in the weak-coupling regime.
However, if only a fraction of DM interacts with baryons, constraints on the cross section progressively weaken as the fraction decreases.
For fractions below $\sim\! 0.4\%$, we find that \Planck{} constraints significantly degrade: the DM and baryons are allowed to be so tightly coupled that DM essentially becomes cosmologically indistinguishable from a small additional amount of baryons.

Interestingly, such strongly-coupled dark matter could alleviate the mild tension between CMB and Big Bang Nucleosynthesis (BBN) measurement of the energy density of baryons.
Recent BBN measurements of the deuterium abundance yield the values $\Omega_b h^2\!=\!0.02166\!\pm\!0.00015\!\pm\!0.00011$~\cite{Cooke:2017cwo} and $\Omega_b h^2\!=\!0.02174\!\pm\!0.00025$ \cite{Zavarygin:2018dbk}, whereas the latest \Planck{} 2018 data yield $\Omega_b h^2\!=\!0.02237\!\pm\!0.00015$~\cite{Aghanim:2018eyx}.
These BBN values for the baryon density are lower than the CMB value by the equivalent of $0.5$\%--$0.6$\% of the DM energy density, and a strongly-coupled DM subcomponent with $f_\chi\!\sim\!0.4$\% could largely account for this discrepancy.
This feature is not unique to $n\!=\!-4$; any interaction that tightly couples this subcomponent of DM to baryons around the time of recombination could be interpreted as an additional contribution to baryons.

The $n\!=\!-4$ interaction has received a fair amount of recent attention in light of DM interpretation of the EDGES signal~\cite{Barkana:2018lgd}.
The initial claim of Ref.~\cite{Barkana:2018lgd} was that a phenomenological $v^{-4}$ interaction could explain the EDGES signal---and our CMB analysis does not rule out such a possibility.
However, our results do exclude a percent of millicharge-like DM scattering only with ions at the level needed to explain EDGES~\cite{Barkana:2018qrx}.

Finally, we presented a conservative forecast for the next-generation ground-based CMB-Stage 4 experiment, and showed a factor of $\sim\! 3$ improvement over \Planck{} limits on $\sigma_0$ for the $n\!=\!-4$ interaction.
Next-generation ground-based CMB surveys optimized for high-multipole science (where the signals of DM scattering may be particularly prominent) thus have a bright future in terms of DM searches with cosmological data.

\acknowledgments

K.~B. acknowledges the Aspen Center for Physics for their hospitality and support under NSF grant \#PHY-1066293, where part of this work was completed.
K.~B. and V.~G. acknowledge KITP and \textit{The Small-Scale Structure of Cold(?) Dark Matter} workshop for their hospitality and support under NSF grant \#PHY-1748958, during the completion of this work.
V.~G. acknowledges the support of the Eric Schmidt fellowship at the Institute for Advanced Study.
This work was supported at Johns Hopkins in part by NSF Grant No. 1519353, NASA NNX17AK38G, and the Simons Foundation.
For R.~B., this publication was made possible through the support of a grant from the John Templeton Foundation; the opinions expressed in this publication are those of the author and do not necessarily reflect the views of the John Templeton Foundation.
R.~B. is also supported by the ISF-NSFC joint research program (grant No. 2580/17).

The authors thank Yacine Ali-Ha\"{i}moud and Julian Mu\~{n}oz for useful discussions, and particularly Tracy Slatyer and Chih-Liang Wu for their help in cross-comparing residual CMB spectra between this work and Ref.~\cite{Slatyer:2018aqg}.

\appendix
\section{Derivation of nonlinear terms}
\label{sec:app-nonlinear}

In this appendix, we derive the nonlinear terms that appear in the evolution of the DM and baryon peculiar velocities, $\vec{V}_\chi$ and $\vec{V}_b$, in Eq.~\eqref{eq:Vfull} and in the evolution of their temperatures, $T_\chi$ and $T_b$, in Eq.~\eqref{eq:temp}.
We drop all terms involving collisions between the baryons and photons for clarity, with the understanding that they must be incorporated to obtain the complete expressions for the baryons.
The baryon fluid consists of various species (\textit{i.e.}, electrons, protons, and helium nuclei), which are characterized by the same peculiar velocity and temperature.
We denote a particular baryonic species with $B$ and denote properties of baryon fluid as a whole with $b$.
The species $B$ has an energy density $\rho_B = Y_B \rho_b$, where $Y_B$ is its mass fraction and $\sum_B Y_B = 1$.
We assume the phase space distribution functions of the DM and baryons are Gaussian, given by
\begin{equation}
  f_\chi(\vec{v}_\chi) = \frac{1}{(2\pi)^{3/2}\bar{v}_\chi^3}
  \exp\left[-\frac{(\vec{v}_\chi - \vec{V}_\chi)^2}{2\bar{v}_\chi^2}\right]
  \quad\textrm{and}\quad
  f_B(\vec{v}_B) = \frac{1}{(2\pi)^{3/2}\bar{v}_B^3}
  \exp\left[-\frac{(\vec{v}_B - \vec{V}_b)^2}{2\bar{v}_B^2}\right] \ ,
  \label{eq:dists}
\end{equation}
where $\bar{v}_\chi^2 = T_\chi / m_\chi$ and $\bar{v}_B^2 = T_b / m_B$ are the thermal velocity dispersions, and $m_\chi$ and $m_B$ are the particle masses.
Deriving the rates of momentum and heat transfer involves integrating the distribution functions over the velocities of the DM and baryon particles, $\vec{v}_\chi$ and $\vec{v}_B$, in the center-of-mass frame.
Anticipating these calculations, it is convenient to make a change of variables to
\begin{equation}
  \vec{v}_m \equiv \frac{\bar{v}_B^2\vec{v}_\chi + \bar{v}_\chi^2\vec{v}_B}{\bar{v}_B^2 + \bar{v}_\chi^2}
  \quad\textrm{and}\quad
  \vec{v}_r \equiv \vec{v}_\chi - \vec{v}_b
\end{equation}
such that the distribution functions remain factorizable~\cite{Munoz:2015bca}:
\begin{equation}
  \int d^3 v_\chi\; f_\chi(\vec{v}_\chi) \int d^3 v_B\; f_B(\vec{v}_B)
  = \int d^3 v_r\; f_r (\vec{v}_r) \int d^3 v_m\; f_m (\vec{v}_m) \ .
  \label{eq:change-variables}
\end{equation}
These new distribution functions have the Gaussian forms
\begin{equation}
  f_m(\vec{v}_m) = \frac{1}{(2\pi)^{3/2}\bar{v}_m^3}
  \exp\left[-\frac{(\vec{v}_m - \vec{V}_m)^2}{2\bar{v}_m^2}\right]
  \quad\textrm{and}\quad
  f_r(\vec{v}_r) = \frac{1}{(2\pi)^{3/2}\bar{v}_r^3}
  \exp\left[-\frac{(\vec{v}_r - \vec{V}_r)^2}{2\bar{v}_r^2}\right] \ ,
  \label{eq:dists-alt}
\end{equation}
where the means and dispersions are
\begin{equation}
  \vec{V}_m = \frac{\bar{v}_B^2\vec{V}_\chi + \bar{v}_\chi^2\vec{V}_b}{\bar{v}_B^2 + \bar{v}_\chi^2}
  \quad\textrm{and}\quad
  \vec{V}_r = \vec{V}_\chi - \vec{V}_b \ , \quad
  \bar{v}_m^2 = \frac{\bar{v}_\chi^2 \bar{v}_B^2}{\bar{v}_\chi^2 + \bar{v}_B^2}
  \quad\textrm{and}\quad
  \bar{v}_r^2 = \bar{v}_\chi^2 + \bar{v}_B^2 \ .
  \label{eq:dists-alt-properties}
\end{equation}

We consider interactions between DM particles and baryons $B$ with momentum-transfer cross sections given by $\sigma_\textrm{MT} = \sigma_0 v_r^n$, and we are particularly interested in $n=-2$ and $n=-4$.
As we show below, the rate calculations yield confluent hypergeometric functions of the first kind with the form
\begin{align}
  {}_1F_1\left(-\frac{n+1}{2},\frac{5}{2},-\frac{r^2}{2}\right) &=
  \begin{cases}
    \frac{3}{2r^2}\left[
      \sqrt{\frac{\pi}{2}} \left(r - \frac{1}{r}\right)
      \erf\left(\frac{r}{\sqrt{2}}\right)
      + e^{-r^2/2} \right] & \textrm{for } n=-2 \\
    \frac{3}{r^2}\left[
      \sqrt{\frac{\pi}{2}} \erf\left(\frac{r}{\sqrt{2}}\right)
      - r e^{-r^2/2}\right] & \textrm{for } n=-4
  \end{cases}\\
  {}_1F_1\left(-\frac{n+3}{2},\frac{3}{2},-\frac{r^2}{2}\right) &=
  \begin{cases}
    \frac{1}{2} \left[\sqrt{\frac{\pi}{2}} \left(r + \frac{1}{r}\right)
      \erf\left(\frac{r}{\sqrt{2}}\right)
      + e^{-r^2/2} \right] & \textrm{for } n=-2 \\
    \frac{1}{r} \left[\sqrt{\frac{\pi}{2}}
      \erf\left(\frac{r}{\sqrt{2}}\right)\right]
    & \textrm{for } n=-4
  \end{cases}
\end{align}
and an overall numerical constant
\begin{equation}
  \mathcal{N}_n \equiv \frac{2^{(5+n)/2}}{3\sqrt{\pi}}
  \Gamma\left(3+\frac{n}{2}\right) =
  \begin{cases}
    \frac{2}{3} \sqrt{\frac{2}{\pi}} & \textrm{for } n=-2 \\
    \frac{1}{3} \sqrt{\frac{2}{\pi}} & \textrm{for } n=-4 \ ,
  \end{cases}
\end{equation}
where we define $r \equiv V_r / \bar{v}_r$.

\subsection{Momentum transfer}

In a single collision between a DM particle $\chi$ and baryon $B$, the momentum of the DM particle changes by~\cite{Dvorkin:2013cea}
\begin{equation}
  \Delta\vec{p}_\chi = \frac{m_\chi m_B}{m_\chi + m_B} |\vec{v}_\chi - \vec{v}_B|
  \left(\hat{n} - \frac{\vec{v}_\chi - \vec{v}_B}{|\vec{v}_\chi - \vec{v}_B|}\right) \ ,
\end{equation}
where $\hat{n}$ is the direction of the scattered DM particle in the center-of-mass frame.
The resulting drag force per unit mass on the DM fluid is
\begin{align}
  D_{\chi}(\vec{V}_r)
  &= \frac{n_B}{m_\chi} \int d^3 v_\chi\; f_\chi(\vec{v}_\chi)
  \int d^3 v_B\; f_B(\vec{v}_B)
  \int d\Omega\frac{d\sigma}{d\Omega} v_r \Delta \vec{p}_\chi \nonumber\\
  &= -\frac{Y_B \rho_b \sigma_0}{m_\chi + m_B}
  \int d^3 v_r\; f_r(\vec{v}_r)\, v_r^{n+1} \vec{v}_r
  \int d^3 v_m\; f_m(\vec{v}_m) \ ,
\end{align}
where we obtain the second line by completing the integration over angles to obtain the momentum-transfer cross section and by utilizing Eq.~\eqref{eq:change-variables}.
The integral over $\vec{v}_m$ simply evaluates to $1$, and the remaining integral over $\vec{v}_r$ yields the result
\begin{equation}
  D_{\chi}(\vec{V}_r)
  =-\frac{Y_B \rho_b \sigma_0 \mathcal{N}_n}{m_\chi + m_B}\bar{v}_r^{n+1} \vec{V}_r\;
  {}_1F_1\left(-\frac{n+1}{2},\frac{5}{2},-\frac{r^2}{2}\right) \ .
\end{equation}
The evolution of the DM and baryon peculiar velocities obeys
\begin{align}
  \frac{\partial \vec{V}_\chi}{\partial\tau}
  - c_\chi^2 \vec{\nabla} \delta_\chi + \frac{\dot{a}}{a} \vec{V}_\chi
  &= -a\sum_B \frac{Y_B \rho_b \sigma_0 \mathcal{N}_n}{m_\chi + m_B}\bar{v}_r^{n+1}
  \vec{V}_r\; {}_1F_1\left(-\frac{n+1}{2},\frac{5}{2},-\frac{r^2}{2}\right)
  \\
  \frac{\partial \vec{V}_b}{\partial\tau}
  - c_b^2 \vec{\nabla} \delta_b + \frac{\dot{a}}{a} \vec{V}_b
  &= +a \sum_B \frac{Y_B\rho_\chi\sigma_0 \mathcal{N}_n}{m_\chi + m_B}\bar{v}_r^{n+1}
  \vec{V}_r\; {}_1F_1\left(-\frac{n+1}{2},\frac{5}{2},-\frac{r^2}{2}\right) \ .
  \label{eq:Vfull-general}
\end{align}

If we neglect the terms proportional to the speeds of sound, $c_\chi$ and $c_b$, we may combine these equations to obtain the following expression for the evolution of the relative bulk velocity:
\begin{equation}
  \frac{\partial \vec{V}_r}{\partial\tau} + \frac{\dot{a}}{a} \vec{V}_r
  = -a \sum_B \frac{Y_B (\rho_b+\rho_\chi) \sigma_0 \mathcal{N}_n}{m_\chi + m_B}
  \bar{v}_r^{n+1} \vec{V}_r\;
  {}_1F_1\left(-\frac{n+1}{2},\frac{5}{2},-\frac{r^2}{2}\right) \ .
  \label{eq:Vrfull-general}
\end{equation}
By further assuming the baryon fluid is comprised of a single species with a mass given by the mean molecular weight $\mu_b$ and plugging in $n=-4$, we find
\begin{equation}
  \frac{\partial \vec{V}_r}{\partial\tau} + \frac{\dot{a}}{a} \vec{V}_r
  = -a \frac{(\rho_b+\rho_\chi) \sigma_0}{m_\chi + \mu_b}
  \frac{\hat{V}_r}{V_r^2} \left[\mathrm{Erf}\left(\frac{r}{\sqrt{2}}\right)
    - \sqrt{\frac{2}{\pi}} r e^{-r^2/2} \right] \ ,
  \label{eq:vrel}
\end{equation}
which agrees with Ref.~\cite{Munoz:2015bca}.

\subsection{Heat transfer}

The amount of energy transferred to the DM fluid is
\begin{equation}
  \Delta E_\chi = \Delta\vec{p}_\chi \cdot \vec{v}_\textrm{cm} \ ,
\end{equation}
where the center-of-mass velocity may be written as
\begin{equation}
  \vec{v}_\textrm{cm} = \vec{v}_m
  + \frac{m_\chi\bar{v}_\chi^2 - m_B\bar{v}_B^2}{(m_\chi+m_B)(\bar{v}_\chi^2 + \bar{v}_B^2)} \vec{v}_r \ .
\end{equation}
In order to find the amount of heat transfer, we work in the instantaneous rest frame of the fluid so as not to incorporate its kinetic energy due to its bulk motion~\cite{Munoz:2015bca}.
In Eqs.~\eqref{eq:dists} and \eqref{eq:dists-alt-properties}, for the DM fluid, we set $\vec{V}_\chi = 0$ and $\vec{V}_b = -\vec{V}_r$; for the baryon fluid, we set $\vec{V}_b = 0$ and $\vec{V}_\chi = \vec{V}_r$.
The rate of heat exchange for the DM fluid is
\begin{align}
  \frac{dQ_\chi}{dt}
  &= \sum_B n_B \int d^3 v_\chi\; f_\chi(\vec{v}_\chi)
  \int d^3 v_B\; f_B(\vec{v}_B)
  \int d\Omega\frac{d\sigma}{d\Omega}\; v_r\; \Delta E_\chi \nonumber\\
  &= -\sum_B \frac{m_\chi Y_B \rho_b \sigma_0}{m_\chi + m_B}
  \int d^3 v_r\; f_r(\vec{v}_r) \, v_r^{n+1} \vec{v}_r \cdot
  \int d^3 v_m\; f_m(\vec{v}_m) \, \vec{v}_\textrm{cm} \ ,
\end{align}
where we obtain the second line by completing the integration over angles to obtain the momentum-transfer cross section and by utilizing Eq.~\eqref{eq:change-variables}.
For the first term in $\vec{v}_\textrm{cm}$, the integration over $\vec{v}_m$ simply returns the mean $\vec{V}_m$; while for the second term, there is no $\vec{v}_m$ dependence and, after factoring out constants, the integration over $\vec{v}_m$ returns $1$.
The remaining integral over $\vec{v}_r$ yields the result
\begin{align}
  \frac{dQ_\chi}{dt} = 3 \sum_B
  \frac{m_\chi Y_B \rho_b \sigma_0 \mathcal{N}_n}{(m_\chi + m_B)^2} \bar{v}_r^{n+1}
  & \left\{(T_b - T_\chi)
  \left[ {}_1F_1\left(-\frac{n+3}{2},\frac{3}{2},-\frac{r^2}{2}\right)
    -\frac{r^2}{3} {}_1F_1\left(-\frac{n+1}{2},\frac{5}{2},-\frac{r^2}{2}\right) \right]
  \right. \nonumber\\
  &\quad \left. + \frac{m_B}{3} V_r^2 {}_1F_1\left(-\frac{n+1}{2},\frac{5}{2},-\frac{r^2}{2}\right) \right\} \ .
\end{align}
Similarly, the rate of heat exchange for the baryon fluid is
\begin{align}
  \frac{dQ_b}{dt} = 3 \sum_B
  \frac{\mu_b \rho_\chi \sigma_0 \mathcal{N}_n}{(m_\chi + m_B)^2} \bar{v}_r^{n+1}
  & \left\{(T_\chi - T_b)
  \left[ {}_1F_1\left(-\frac{n+3}{2},\frac{3}{2},-\frac{r^2}{2}\right)
    -\frac{r^2}{3} {}_1F_1\left(-\frac{n+1}{2},\frac{5}{2},-\frac{r^2}{2}\right) \right]
  \right. \nonumber\\
  &\quad \left. + \frac{m_\chi}{3} V_r^2 {}_1F_1\left(-\frac{n+1}{2},\frac{5}{2},-\frac{r^2}{2}\right) \right\} \ .
\end{align}
The temperature evolution equations are
\begin{equation}
  \dot{T}_\chi + 2\frac{\dot{a}}{a}T_\chi= \frac{2}{3}a \frac{dQ_\chi}{dt}
  \qquad\textrm{and}\qquad
  \dot{T}_b + 2\frac{\dot{a}}{a}T_b = \frac{2}{3}a \frac{dQ_b}{dt} \ .
  \label{eq:temp-general}
\end{equation}
Assuming the baryon fluid is comprised of a single species with a mass given by the mean molecular weight $\mu_b$ and plugging in $n=-4$, we find
\begin{align}
  \dot{T}_\chi + 2\frac{\dot{a}}{a} T_\chi
  &= \frac{2}{3}a\frac{m_\chi \rho_b \sigma_0}{(m_\chi + \mu_b)^2} \bar{v}_r^{-3}
  \left\{\sqrt{\frac{2}{\pi}} (T_b - T_\chi) e^{-r^2/2}
  + \mu_b V_r^2 \frac{1}{r^3} \left[\mathrm{Erf}\left(\frac{r}{\sqrt{2}}\right) - \sqrt{\frac{2}{\pi}} r e^{-r^2/2}\right] \right\}\\
  \dot{T}_b + 2\frac{\dot{a}}{a} T_b
  &= \frac{2}{3}a\frac{\mu_b \rho_\chi \sigma_0}{(m_\chi + \mu_b)^2} \bar{v}_r^{-3}
  \left\{\sqrt{\frac{2}{\pi}} (T_\chi - T_b) e^{-r^2/2}
  + m_\chi V_r^2 \frac{1}{r^3} \left[\mathrm{Erf}\left(\frac{r}{\sqrt{2}}\right) - \sqrt{\frac{2}{\pi}} r e^{-r^2/2}\right] \right\} \ ,
\end{align}
which agrees with Ref.~\cite{Munoz:2015bca}.

\section{Implementation in \CLASS{}}
\label{sec:app-implementation}

The implementation of the DM--baryon interaction requires modifying both the \texttt{thermodynamics} and the \texttt{perturbations} module of \CLASS{}.
In the \texttt{perturbations} module, we have incorporated the full Boltzmann system of equations~\eqref{eq:fluct}.
In principle, \CLASS{} is able to solve for each $k$ mode in parallel.
However, the ``main'' prescription described in Section~\ref{sec:CMB-bulkvel} introduces mode mixing; thus, we must determine $\Vrms(k^\star,z)$ and $\Vflow(k^\star,z)$ [which requires us to know the values of $\theta_b(k,z)$ and $\theta_\chi (k,z)$ at all $k$ to perform the integration in Eqs.~\eqref{eq:Vflow} and \eqref{eq:Vrms}] while concurrently solving the Boltzmann equations at a given redshift $z$ and Fourier mode $k^\star$ [which requires us to know $\Vrms(k^\star,z)$ and $\Vflow(k^\star,z)$ for the rate of momentum transfer in Eq.~\eqref{eq:mod-rate}].
To circumvent this difficulty, we repeat the calculations of the \texttt{perturbations} module, iteratively updating the values of $\Vrms(k^\star,z)$ and $\Vflow(k^\star,z)$ until we achieve convergence at the 1\% level.
In practice, we initialize $\Vrms(k^\star,z)$ and $\Vflow(k^\star,z)$ to the variance of the relative bulk velocity in $\Lambda$CDM given by Eq.~\eqref{eq:Vrms-lcdm} for all $k$; the choice of the initial condition merely affects the number of iterations required to achieve convergence.
This full procedure requires starting with sufficiently large $k_\textrm{max}$ such that the relative bulk velocities are not coherent~\cite{Tseliakhovich:2010bj} for the highest range of redshifts that \Planck{} is sensitive to.
We find that $k_\textrm{max}\! =\! 5$ is sufficient for $\Vrms(k^\star,z)$ to reach its plateau at high $z$, shown in the left panel of Figure~\ref{fig:mod-rate}.

In the \texttt{thermodynamics} module, we have included the modified thermal evolution of the baryon and DM fluids, given by Eq.~\eqref{eq:temp}.
These evolution equations depend on the relative bulk velocity between the DM and baryon fluids, which we take to be the square root of the variance $\Vvar$ to obtain the average evolution.
We thus include the \texttt{thermodynamics} module within the iterative procedure, and recalculate the recombination and thermal history for each iteration.
To compute the recombination of hydrogen and helium, \CLASS{} can call either \Recfast{}~\cite{Seager:1999bc} or \hyrec{}~\cite{AliHaimoud:2010dx} code.
Although \Recfast{} provides a slight improvement in computational speed, it uses various ``fudge functions'' to achieve sub-percent accuracy~\cite{RubinoMartin:2009ry} (established through validation against more accurate codes such as \hyrec{} and \texttt{CosmoRec}~\cite{Chluba:2010ca} within $\Lambda$CDM), and thus may be unreliable for calculations within a modified cosmology.
We have modified both recombination codes to include DM--proton scattering and find that our modified version of \Recfast{} performs with very good accuracy.
Given that \Recfast{} is slightly faster, we use it to produce all numerical results shown in this work.

In \Recfast, the recombination equation (before the introduction of ``fudge functions'') takes the form of an effective three-level atom approximation: the evolution equation of the free-electron fraction $x_e$, which directly depends on the baryon temperature $T_b$, takes the form%
\footnote{Following Ref.~\cite{Chluba:2015lpa}, we explicitly evaluate the photoionization rate as a function of $T_\gamma$, instead of $T_b$ as done in the original version of \Recfast.}
\cite{Seager:1999bc}
\begin{equation}
  \frac{dx_{e}(z)}{dz} = \frac{C}{(1+z)H(z)} \left[\alpha_H x_e^2 n_H
    - \beta_H (1-x_e)e^{-\frac{h\nu_\alpha}{k_bT_\gamma}}\right] \ ,
  \label{eq:x_e&T_M}
\end{equation}
where $T_b$ is governed by Eq.~\eqref{eq:temp}.
The coefficients $\alpha_H(T_b,T_\gamma)$ and $\beta_H(T_\gamma)$ are the effective recombination and photoionization rates, $\nu_\alpha$ is the Lyman-$\alpha$ frequency, and $C$ is the Peebles factor, representing the probability for an electron in the $n \! = \! 2$ state to relax to the ground state before being ionized.
An accurate calculation of the baryon temperature is essential, because it enters in the recombination rate; a smaller $T_b$ can accelerate recombination and decrease the number of free electrons in the remaining plateau.
However, both \hyrec{} and \Recfast{} do not follow the full evolution of $T_b$ up to the initial redshift of the calculation.
Above the redshift $z\!\sim\!850$, the codes assume that $T_b \! = \! T_\gamma - \epsilon$, and a linearized system of equation is solved instead~\cite{Scott:2009sz,AliHaimoud:2010dx}.
In \Recfast{}, the switch to solving the full evolution equation is set by the ratio $r_{\rm CH}\equiv t_C/t_H\sim10^{-3}$, where $t_C$ is the Compton interaction time and $t_H$ the Hubble time~\cite{Scott:2009sz}.
We find that an incorrect generalization of the linearized steady-state approximation to include DM--proton interactions can lead to significant numerical glitches in the evolution of $T_b$, which artificially enhance the impact of DM--proton scattering on the CMB power spectra.
We thus adapt the value of the switch such that we compute the full evolution equation up to $z\! \sim\! 10000$, before recombination starts.
The impact of the DM--proton heat exchange on the baryon temperature is typically negligible before $z \! \sim \! 850$, because for any value of the momentum-transfer cross section not already excluded by \Planck{} with a negative power-law scaling of the relative velocity, the cooling time $t_\chi$ is always negligible before the Compton time $t_C$ until recombination.
Hence, we neglect the modification to the baryon temperature at early times, when both codes solve the linearized system of equations.
We correctly include DM--proton scattering at and below $z \! \simeq \! 850$, when the free-electron fraction and the Compton rate begin to drop.
We have verified that both methods are in excellent agreement.

\bibliography{biblio}

\begin{thebibliography}{45}%
\makeatletter
\providecommand \@ifxundefined [1]{%
 \@ifx{#1\undefined}
}%
\providecommand \@ifnum [1]{%
 \ifnum #1\expandafter \@firstoftwo
 \else \expandafter \@secondoftwo
 \fi
}%
\providecommand \@ifx [1]{%
 \ifx #1\expandafter \@firstoftwo
 \else \expandafter \@secondoftwo
 \fi
}%
\providecommand \natexlab [1]{#1}%
\providecommand \enquote  [1]{``#1''}%
\providecommand \bibnamefont  [1]{#1}%
\providecommand \bibfnamefont [1]{#1}%
\providecommand \citenamefont [1]{#1}%
\providecommand \href@noop [0]{\@secondoftwo}%
\providecommand \href [0]{\begingroup \@sanitize@url \@href}%
\providecommand \@href[1]{\@@startlink{#1}\@@href}%
\providecommand \@@href[1]{\endgroup#1\@@endlink}%
\providecommand \@sanitize@url [0]{\catcode `\\12\catcode `\$12\catcode
  `\&12\catcode `\#12\catcode `\^12\catcode `\_12\catcode `\%12\relax}%
\providecommand \@@startlink[1]{}%
\providecommand \@@endlink[0]{}%
\providecommand \url  [0]{\begingroup\@sanitize@url \@url }%
\providecommand \@url [1]{\endgroup\@href {#1}{\urlprefix }}%
\providecommand \urlprefix  [0]{URL }%
\providecommand \Eprint [0]{\href }%
\providecommand \doibase [0]{http://dx.doi.org/}%
\providecommand \selectlanguage [0]{\@gobble}%
\providecommand \bibinfo  [0]{\@secondoftwo}%
\providecommand \bibfield  [0]{\@secondoftwo}%
\providecommand \translation [1]{[#1]}%
\providecommand \BibitemOpen [0]{}%
\providecommand \bibitemStop [0]{}%
\providecommand \bibitemNoStop [0]{.\EOS\space}%
\providecommand \EOS [0]{\spacefactor3000\relax}%
\providecommand \BibitemShut  [1]{\csname bibitem#1\endcsname}%
\let\auto@bib@innerbib\@empty
\bibitem [{\citenamefont {Chen}\ \emph {et~al.}(2002)\citenamefont {Chen},
  \citenamefont {Hannestad},\ and\ \citenamefont {Scherrer}}]{Chen:2002yh}%
  \BibitemOpen
  \bibfield  {author} {\bibinfo {author} {\bibfnamefont {Xue-lei}\ \bibnamefont
  {Chen}}, \bibinfo {author} {\bibfnamefont {Steen}\ \bibnamefont {Hannestad}},
  \ and\ \bibinfo {author} {\bibfnamefont {Robert~J.}\ \bibnamefont
  {Scherrer}},\ }\bibfield  {title} {\enquote {\bibinfo {title} {{Cosmic
  microwave background and large scale structure limits on the interaction
  between dark matter and baryons}},}\ }\href {\doibase
  10.1103/PhysRevD.65.123515} {\bibfield  {journal} {\bibinfo  {journal}
  {Phys.~Rev.~D}\ }\textbf {\bibinfo {volume} {65}},\ \bibinfo {pages} {123515}
  (\bibinfo {year} {2002})},\ \Eprint {http://arxiv.org/abs/astro-ph/0202496}
  {astro-ph/0202496} \BibitemShut {NoStop}%
\bibitem [{\citenamefont {Sigurdson}\ \emph {et~al.}(2004)\citenamefont
  {Sigurdson}, \citenamefont {Doran}, \citenamefont {Kurylov}, \citenamefont
  {Caldwell},\ and\ \citenamefont {Kamionkowski}}]{Sigurdson:2004zp}%
  \BibitemOpen
  \bibfield  {author} {\bibinfo {author} {\bibfnamefont {Kris}\ \bibnamefont
  {Sigurdson}}, \bibinfo {author} {\bibfnamefont {Michael}\ \bibnamefont
  {Doran}}, \bibinfo {author} {\bibfnamefont {Andriy}\ \bibnamefont {Kurylov}},
  \bibinfo {author} {\bibfnamefont {Robert~R.}\ \bibnamefont {Caldwell}}, \
  and\ \bibinfo {author} {\bibfnamefont {Marc}\ \bibnamefont {Kamionkowski}},\
  }\bibfield  {title} {\enquote {\bibinfo {title} {{Dark-matter electric and
  magnetic dipole moments}},}\ }\href {\doibase 10.1103/PhysRevD.70.083501}
  {\bibfield  {journal} {\bibinfo  {journal} {Phys.~Rev.~D}\ }\textbf {\bibinfo
  {volume} {70}},\ \bibinfo {eid} {083501} (\bibinfo {year} {2004})},\ \bibinfo
  {note} {[Erratum: Phys.~Rev.~D 73, 089903 (2006)]},\ \Eprint
  {http://arxiv.org/abs/astro-ph/0406355} {arXiv:astro-ph/0406355 [astro-ph]}
  \BibitemShut {NoStop}%
\bibitem [{\citenamefont {Dvorkin}\ \emph {et~al.}(2014)\citenamefont
  {Dvorkin}, \citenamefont {Blum},\ and\ \citenamefont
  {Kamionkowski}}]{Dvorkin:2013cea}%
  \BibitemOpen
  \bibfield  {author} {\bibinfo {author} {\bibfnamefont {Cora}\ \bibnamefont
  {Dvorkin}}, \bibinfo {author} {\bibfnamefont {Kfir}\ \bibnamefont {Blum}}, \
  and\ \bibinfo {author} {\bibfnamefont {Marc}\ \bibnamefont {Kamionkowski}},\
  }\bibfield  {title} {\enquote {\bibinfo {title} {{Constraining dark
  matter-baryon scattering with linear cosmology}},}\ }\href {\doibase
  10.1103/PhysRevD.89.023519} {\bibfield  {journal} {\bibinfo  {journal}
  {Phys.~Rev.~D}\ }\textbf {\bibinfo {volume} {89}},\ \bibinfo {eid} {023519}
  (\bibinfo {year} {2014})},\ \Eprint {http://arxiv.org/abs/1311.2937}
  {arXiv:1311.2937 [astro-ph.CO]} \BibitemShut {NoStop}%
\bibitem [{\citenamefont {Gluscevic}\ and\ \citenamefont
  {Boddy}()}]{Gluscevic:2017ywp}%
  \BibitemOpen
  \bibfield  {author} {\bibinfo {author} {\bibfnamefont {Vera}\ \bibnamefont
  {Gluscevic}}\ and\ \bibinfo {author} {\bibfnamefont {Kimberly~K.}\
  \bibnamefont {Boddy}},\ }\bibfield  {title} {\enquote {\bibinfo {title}
  {{Constraints on scattering of keV--TeV dark matter with protons in the early
  Universe}},}\ }\href@noop {} {\ }\Eprint {http://arxiv.org/abs/1712.07133}
  {arXiv:1712.07133 [astro-ph.CO]} \BibitemShut {NoStop}%
\bibitem [{\citenamefont {Boddy}\ and\ \citenamefont
  {Gluscevic}()}]{Boddy:2018kfv}%
  \BibitemOpen
  \bibfield  {author} {\bibinfo {author} {\bibfnamefont {Kimberly~K.}\
  \bibnamefont {Boddy}}\ and\ \bibinfo {author} {\bibfnamefont {Vera}\
  \bibnamefont {Gluscevic}},\ }\bibfield  {title} {\enquote {\bibinfo {title}
  {{First Cosmological Constraint on the Effective Theory of Dark Matter-Proton
  Interactions}},}\ }\href@noop {} {\ }\Eprint
  {http://arxiv.org/abs/1801.08609} {arXiv:1801.08609} \BibitemShut {NoStop}%
\bibitem [{\citenamefont {Xu}\ \emph {et~al.}(2018)\citenamefont {Xu},
  \citenamefont {Dvorkin},\ and\ \citenamefont {Chael}}]{Xu:2018efh}%
  \BibitemOpen
  \bibfield  {author} {\bibinfo {author} {\bibfnamefont {Weishuang~Linda}\
  \bibnamefont {Xu}}, \bibinfo {author} {\bibfnamefont {Cora}\ \bibnamefont
  {Dvorkin}}, \ and\ \bibinfo {author} {\bibfnamefont {Andrew}\ \bibnamefont
  {Chael}},\ }\bibfield  {title} {\enquote {\bibinfo {title} {{Probing sub-GeV
  Dark Matter-Baryon Scattering with Cosmological Observables}},}\ }\href@noop
  {} {\  (\bibinfo {year} {2018})},\ \Eprint {http://arxiv.org/abs/1802.06788}
  {arXiv:1802.06788} \BibitemShut {NoStop}%
\bibitem [{\citenamefont {Tashiro}\ \emph {et~al.}(2014)\citenamefont
  {Tashiro}, \citenamefont {Kadota},\ and\ \citenamefont
  {Silk}}]{Tashiro:2014tsa}%
  \BibitemOpen
  \bibfield  {author} {\bibinfo {author} {\bibfnamefont {Hiroyuki}\
  \bibnamefont {Tashiro}}, \bibinfo {author} {\bibfnamefont {Kenji}\
  \bibnamefont {Kadota}}, \ and\ \bibinfo {author} {\bibfnamefont {Joseph}\
  \bibnamefont {Silk}},\ }\bibfield  {title} {\enquote {\bibinfo {title}
  {{Effects of dark matter-baryon scattering on redshifted 21 cm signals}},}\
  }\href {\doibase 10.1103/PhysRevD.90.083522} {\bibfield  {journal} {\bibinfo
  {journal} {Phys.~Rev.~D}\ }\textbf {\bibinfo {volume} {90}},\ \bibinfo
  {pages} {083522} (\bibinfo {year} {2014})},\ \Eprint
  {http://arxiv.org/abs/1408.2571} {arXiv:1408.2571 [astro-ph.CO]} \BibitemShut
  {NoStop}%
\bibitem [{\citenamefont {Mu{\~n}oz}\ \emph {et~al.}(2015)\citenamefont
  {Mu{\~n}oz}, \citenamefont {Kovetz},\ and\ \citenamefont
  {Ali-Ha{\"\i}moud}}]{Munoz:2015bca}%
  \BibitemOpen
  \bibfield  {author} {\bibinfo {author} {\bibfnamefont {Julian~B.}\
  \bibnamefont {Mu{\~n}oz}}, \bibinfo {author} {\bibfnamefont {Ely~D.}\
  \bibnamefont {Kovetz}}, \ and\ \bibinfo {author} {\bibfnamefont {Yacine}\
  \bibnamefont {Ali-Ha{\"\i}moud}},\ }\bibfield  {title} {\enquote {\bibinfo
  {title} {{Heating of baryons due to scattering with dark matter during the
  dark ages}},}\ }\href {\doibase 10.1103/PhysRevD.92.083528} {\bibfield
  {journal} {\bibinfo  {journal} {Phys.~Rev.~D}\ }\textbf {\bibinfo {volume}
  {92}},\ \bibinfo {eid} {083528} (\bibinfo {year} {2015})},\ \Eprint
  {http://arxiv.org/abs/1509.00029} {arXiv:1509.00029} \BibitemShut {NoStop}%
\bibitem [{\citenamefont {Barkana}(2018)}]{Barkana:2018lgd}%
  \BibitemOpen
  \bibfield  {author} {\bibinfo {author} {\bibfnamefont {Rennan}\ \bibnamefont
  {Barkana}},\ }\bibfield  {title} {\enquote {\bibinfo {title} {{Possible
  interaction between baryons and dark-matter particles revealed by the first
  stars}},}\ }\href {\doibase 10.1038/nature25791} {\bibfield  {journal}
  {\bibinfo  {journal} {Nature}\ }\textbf {\bibinfo {volume} {555}},\ \bibinfo
  {pages} {71--74} (\bibinfo {year} {2018})},\ \Eprint
  {http://arxiv.org/abs/1803.06698} {arXiv:1803.06698 [astro-ph.CO]}
  \BibitemShut {NoStop}%
\bibitem [{\citenamefont {Tseliakhovich}\ and\ \citenamefont
  {Hirata}(2010)}]{Tseliakhovich:2010bj}%
  \BibitemOpen
  \bibfield  {author} {\bibinfo {author} {\bibfnamefont {Dmitriy}\ \bibnamefont
  {Tseliakhovich}}\ and\ \bibinfo {author} {\bibfnamefont {Christopher}\
  \bibnamefont {Hirata}},\ }\bibfield  {title} {\enquote {\bibinfo {title}
  {{Relative velocity of dark matter and baryonic fluids and the formation of
  the first structures}},}\ }\href {\doibase 10.1103/PhysRevD.82.083520}
  {\bibfield  {journal} {\bibinfo  {journal} {Phys.~Rev.~D}\ }\textbf {\bibinfo
  {volume} {82}},\ \bibinfo {eid} {083520} (\bibinfo {year} {2010})},\ \Eprint
  {http://arxiv.org/abs/1005.2416} {arXiv:1005.2416} \BibitemShut {NoStop}%
\bibitem [{\citenamefont {Slatyer}\ and\ \citenamefont
  {Wu}(2018)}]{Slatyer:2018aqg}%
  \BibitemOpen
  \bibfield  {author} {\bibinfo {author} {\bibfnamefont {Tracy~R.}\
  \bibnamefont {Slatyer}}\ and\ \bibinfo {author} {\bibfnamefont {Chih-Liang}\
  \bibnamefont {Wu}},\ }\bibfield  {title} {\enquote {\bibinfo {title}
  {{Early-Universe Constraints on Dark Matter-Baryon Scattering and their
  Implications for a Global 21cm Signal}},}\ }\href@noop {} {\  (\bibinfo
  {year} {2018})},\ \Eprint {http://arxiv.org/abs/1803.09734} {arXiv:1803.09734
  [astro-ph.CO]} \BibitemShut {NoStop}%
\bibitem [{\citenamefont {{Adam}}\ \emph {et~al.}(2016)\citenamefont {{Adam}}
  \emph {et~al.}}]{2016A&A...594A...1P}%
  \BibitemOpen
  \bibfield  {author} {\bibinfo {author} {\bibfnamefont {R.}~\bibnamefont
  {{Adam}}} \emph {et~al.} (\bibinfo {collaboration} {{Planck
  Collaboration}}),\ }\bibfield  {title} {\enquote {\bibinfo {title} {{Planck
  2015 results. I. Overview of products and scientific results}},}\ }\href
  {\doibase 10.1051/0004-6361/201527101} {\bibfield  {journal} {\bibinfo
  {journal} {Astron.~Astrophys.}\ }\textbf {\bibinfo {volume} {594}},\ \bibinfo
  {eid} {A1} (\bibinfo {year} {2016})},\ \Eprint
  {http://arxiv.org/abs/1502.01582} {arXiv:1502.01582} \BibitemShut {NoStop}%
\bibitem [{\citenamefont {{Aghanim}}\ \emph {et~al.}(2016)\citenamefont
  {{Aghanim}} \emph {et~al.}}]{2016A&A...594A..11P}%
  \BibitemOpen
  \bibfield  {author} {\bibinfo {author} {\bibfnamefont {N.}~\bibnamefont
  {{Aghanim}}} \emph {et~al.} (\bibinfo {collaboration} {{Planck
  Collaboration}}),\ }\bibfield  {title} {\enquote {\bibinfo {title} {{Planck
  2015 results. XI. CMB power spectra, likelihoods, and robustness of
  parameters}},}\ }\href {\doibase 10.1051/0004-6361/201526926} {\bibfield
  {journal} {\bibinfo  {journal} {Astron.~Astrophys.}\ }\textbf {\bibinfo
  {volume} {594}},\ \bibinfo {eid} {A11} (\bibinfo {year} {2016})},\ \Eprint
  {http://arxiv.org/abs/1507.02704} {arXiv:1507.02704} \BibitemShut {NoStop}%
\bibitem [{\citenamefont {{Abazajian}}\ \emph {et~al.}()\citenamefont
  {{Abazajian}} \emph {et~al.}}]{2016arXiv161002743A}%
  \BibitemOpen
  \bibfield  {author} {\bibinfo {author} {\bibfnamefont {K.~N.}\ \bibnamefont
  {{Abazajian}}} \emph {et~al.} (\bibinfo {collaboration} {CMB-S4}),\
  }\bibfield  {title} {\enquote {\bibinfo {title} {{CMB-S4 Science Book, First
  Edition}},}\ }\href@noop {} {\ }\Eprint {http://arxiv.org/abs/1610.02743}
  {arXiv:1610.02743} \BibitemShut {NoStop}%
\bibitem [{\citenamefont {Bowman}\ \emph {et~al.}(2018)\citenamefont {Bowman},
  \citenamefont {Rogers}, \citenamefont {Monsalve}, \citenamefont {Mozdzen},\
  and\ \citenamefont {Mahesh}}]{Bowman:2018yin}%
  \BibitemOpen
  \bibfield  {author} {\bibinfo {author} {\bibfnamefont {Judd~D.}\ \bibnamefont
  {Bowman}}, \bibinfo {author} {\bibfnamefont {Alan E.~E.}\ \bibnamefont
  {Rogers}}, \bibinfo {author} {\bibfnamefont {Raul~A.}\ \bibnamefont
  {Monsalve}}, \bibinfo {author} {\bibfnamefont {Thomas~J.}\ \bibnamefont
  {Mozdzen}}, \ and\ \bibinfo {author} {\bibfnamefont {Nivedita}\ \bibnamefont
  {Mahesh}},\ }\bibfield  {title} {\enquote {\bibinfo {title} {{An absorption
  profile centred at 78 megahertz in the sky-averaged spectrum}},}\ }\href
  {\doibase 10.1038/nature25792} {\bibfield  {journal} {\bibinfo  {journal}
  {Nature}\ }\textbf {\bibinfo {volume} {555}},\ \bibinfo {pages} {67--70}
  (\bibinfo {year} {2018})}\BibitemShut {NoStop}%
\bibitem [{\citenamefont {Barkana}\ \emph {et~al.}(2018)\citenamefont
  {Barkana}, \citenamefont {Outmezguine}, \citenamefont {Redigolo},\ and\
  \citenamefont {Volansky}}]{Barkana:2018qrx}%
  \BibitemOpen
  \bibfield  {author} {\bibinfo {author} {\bibfnamefont {Rennan}\ \bibnamefont
  {Barkana}}, \bibinfo {author} {\bibfnamefont {Nadav~Joseph}\ \bibnamefont
  {Outmezguine}}, \bibinfo {author} {\bibfnamefont {Diego}\ \bibnamefont
  {Redigolo}}, \ and\ \bibinfo {author} {\bibfnamefont {Tomer}\ \bibnamefont
  {Volansky}},\ }\bibfield  {title} {\enquote {\bibinfo {title} {{Signs of Dark
  Matter at 21-cm?}}}\ }\href@noop {} {\  (\bibinfo {year} {2018})},\ \Eprint
  {http://arxiv.org/abs/1803.03091} {arXiv:1803.03091 [hep-ph]} \BibitemShut
  {NoStop}%
\bibitem [{\citenamefont {Kovetz}\ \emph {et~al.}()\citenamefont {Kovetz},
  \citenamefont {Poulin}, \citenamefont {Gluscevic}, \citenamefont {Boddy},
  \citenamefont {Barkana},\ and\ \citenamefont {Kamionkowski}}]{paper2:inprep}%
  \BibitemOpen
  \bibfield  {author} {\bibinfo {author} {\bibfnamefont {Ely~D.}\ \bibnamefont
  {Kovetz}}, \bibinfo {author} {\bibfnamefont {Vivian}\ \bibnamefont {Poulin}},
  \bibinfo {author} {\bibfnamefont {Vera}\ \bibnamefont {Gluscevic}}, \bibinfo
  {author} {\bibfnamefont {Kimberly~K.}\ \bibnamefont {Boddy}}, \bibinfo
  {author} {\bibfnamefont {Rennan}\ \bibnamefont {Barkana}}, \ and\ \bibinfo
  {author} {\bibfnamefont {Marc}\ \bibnamefont {Kamionkowski}},\ }\href@noop {}
  {\ }\bibinfo {note} {Submitted in tandem with this work.}\BibitemShut {Stop}%
\bibitem [{\citenamefont {Catena}\ and\ \citenamefont
  {Schwabe}(2015)}]{Catena:2015uha}%
  \BibitemOpen
  \bibfield  {author} {\bibinfo {author} {\bibfnamefont {Riccardo}\
  \bibnamefont {Catena}}\ and\ \bibinfo {author} {\bibfnamefont {Bodo}\
  \bibnamefont {Schwabe}},\ }\bibfield  {title} {\enquote {\bibinfo {title}
  {{Form factors for dark matter capture by the Sun in effective theories}},}\
  }\href {\doibase 10.1088/1475-7516/2015/04/042} {\bibfield  {journal}
  {\bibinfo  {journal} {J.~Cosmol.~Astropart.~Phys.}\ }\textbf {\bibinfo
  {volume} {4}},\ \bibinfo {pages} {042} (\bibinfo {year} {2015})},\ \Eprint
  {http://arxiv.org/abs/1501.03729} {arXiv:1501.03729 [hep-ph]} \BibitemShut
  {NoStop}%
\bibitem [{\citenamefont {Blas}\ \emph {et~al.}(2011)\citenamefont {Blas},
  \citenamefont {Lesgourgues},\ and\ \citenamefont {Tram}}]{Blas:2011rf}%
  \BibitemOpen
  \bibfield  {author} {\bibinfo {author} {\bibfnamefont {Diego}\ \bibnamefont
  {Blas}}, \bibinfo {author} {\bibfnamefont {Julien}\ \bibnamefont
  {Lesgourgues}}, \ and\ \bibinfo {author} {\bibfnamefont {Thomas}\
  \bibnamefont {Tram}},\ }\bibfield  {title} {\enquote {\bibinfo {title} {{The
  Cosmic Linear Anisotropy Solving System (CLASS). Part II: Approximation
  schemes}},}\ }\href {\doibase 10.1088/1475-7516/2011/07/034} {\bibfield
  {journal} {\bibinfo  {journal} {J.~Cosmol.~Astropart.~Phys.}\ }\textbf
  {\bibinfo {volume} {7}},\ \bibinfo {eid} {034} (\bibinfo {year} {2011})},\
  \Eprint {http://arxiv.org/abs/1104.2933} {arXiv:1104.2933 [astro-ph.CO]}
  \BibitemShut {NoStop}%
\bibitem [{\citenamefont {{Ade}}\ \emph {et~al.}(2016)\citenamefont {{Ade}}
  \emph {et~al.}}]{Ade:2015xua}%
  \BibitemOpen
  \bibfield  {author} {\bibinfo {author} {\bibfnamefont {P.~A.~R.}\
  \bibnamefont {{Ade}}} \emph {et~al.} (\bibinfo {collaboration} {Planck}),\
  }\bibfield  {title} {\enquote {\bibinfo {title} {{Planck 2015 results. XIII.
  Cosmological parameters}},}\ }\href {\doibase 10.1051/0004-6361/201525830}
  {\bibfield  {journal} {\bibinfo  {journal} {Astron.~Astrophys.}\ }\textbf
  {\bibinfo {volume} {594}},\ \bibinfo {pages} {A13} (\bibinfo {year}
  {2016})},\ \Eprint {http://arxiv.org/abs/1502.01589} {arXiv:1502.01589
  [astro-ph.CO]} \BibitemShut {NoStop}%
\bibitem [{\citenamefont {Seljak}\ and\ \citenamefont
  {Zaldarriaga}(1996)}]{Seljak:1996is}%
  \BibitemOpen
  \bibfield  {author} {\bibinfo {author} {\bibfnamefont {Uros}\ \bibnamefont
  {Seljak}}\ and\ \bibinfo {author} {\bibfnamefont {Matias}\ \bibnamefont
  {Zaldarriaga}},\ }\bibfield  {title} {\enquote {\bibinfo {title} {{A Line of
  sight integration approach to cosmic microwave background anisotropies}},}\
  }\href {\doibase 10.1086/177793} {\bibfield  {journal} {\bibinfo  {journal}
  {Astrophys.~J.}\ }\textbf {\bibinfo {volume} {469}},\ \bibinfo {pages}
  {437--444} (\bibinfo {year} {1996})},\ \Eprint
  {http://arxiv.org/abs/astro-ph/9603033} {arXiv:astro-ph/9603033 [astro-ph]}
  \BibitemShut {NoStop}%
\bibitem [{\citenamefont {Poulin}\ \emph {et~al.}(2017)\citenamefont {Poulin},
  \citenamefont {Lesgourgues},\ and\ \citenamefont {Serpico}}]{Poulin:2016anj}%
  \BibitemOpen
  \bibfield  {author} {\bibinfo {author} {\bibfnamefont {Vivian}\ \bibnamefont
  {Poulin}}, \bibinfo {author} {\bibfnamefont {Julien}\ \bibnamefont
  {Lesgourgues}}, \ and\ \bibinfo {author} {\bibfnamefont {Pasquale~D.}\
  \bibnamefont {Serpico}},\ }\bibfield  {title} {\enquote {\bibinfo {title}
  {{Cosmological constraints on exotic injection of electromagnetic energy}},}\
  }\href {\doibase 10.1088/1475-7516/2017/03/043} {\bibfield  {journal}
  {\bibinfo  {journal} {J.~Cosmol.~Astropart.~Phys.}\ }\textbf {\bibinfo
  {volume} {1703}},\ \bibinfo {pages} {043} (\bibinfo {year} {2017})},\ \Eprint
  {http://arxiv.org/abs/1610.10051} {arXiv:1610.10051 [astro-ph.CO]}
  \BibitemShut {NoStop}%
\bibitem [{\citenamefont {Buen-Abad}\ \emph {et~al.}(2018)\citenamefont
  {Buen-Abad}, \citenamefont {Schmaltz}, \citenamefont {Lesgourgues},\ and\
  \citenamefont {Brinckmann}}]{Buen-Abad:2017gxg}%
  \BibitemOpen
  \bibfield  {author} {\bibinfo {author} {\bibfnamefont {Manuel~A.}\
  \bibnamefont {Buen-Abad}}, \bibinfo {author} {\bibfnamefont {Martin}\
  \bibnamefont {Schmaltz}}, \bibinfo {author} {\bibfnamefont {Julien}\
  \bibnamefont {Lesgourgues}}, \ and\ \bibinfo {author} {\bibfnamefont {Thejs}\
  \bibnamefont {Brinckmann}},\ }\bibfield  {title} {\enquote {\bibinfo {title}
  {{Interacting Dark Sector and Precision Cosmology}},}\ }\href {\doibase
  10.1088/1475-7516/2018/01/008} {\bibfield  {journal} {\bibinfo  {journal}
  {J.~Cosmol.~Astropart.~Phys.}\ }\textbf {\bibinfo {volume} {1801}},\ \bibinfo
  {pages} {008} (\bibinfo {year} {2018})},\ \Eprint
  {http://arxiv.org/abs/1708.09406} {arXiv:1708.09406 [astro-ph.CO]}
  \BibitemShut {NoStop}%
\bibitem [{\citenamefont {Kamionkowski}\ \emph {et~al.}(1997)\citenamefont
  {Kamionkowski}, \citenamefont {Kosowsky},\ and\ \citenamefont
  {Stebbins}}]{Kamionkowski:1996ks}%
  \BibitemOpen
  \bibfield  {author} {\bibinfo {author} {\bibfnamefont {Marc}\ \bibnamefont
  {Kamionkowski}}, \bibinfo {author} {\bibfnamefont {Arthur}\ \bibnamefont
  {Kosowsky}}, \ and\ \bibinfo {author} {\bibfnamefont {Albert}\ \bibnamefont
  {Stebbins}},\ }\bibfield  {title} {\enquote {\bibinfo {title} {{Statistics of
  cosmic microwave background polarization}},}\ }\href {\doibase
  10.1103/PhysRevD.55.7368} {\bibfield  {journal} {\bibinfo  {journal}
  {Phys.~Rev.~D}\ }\textbf {\bibinfo {volume} {55}},\ \bibinfo {pages}
  {7368--7388} (\bibinfo {year} {1997})},\ \Eprint
  {http://arxiv.org/abs/astro-ph/9611125} {arXiv:astro-ph/9611125 [astro-ph]}
  \BibitemShut {NoStop}%
\bibitem [{\citenamefont {Zaldarriaga}\ and\ \citenamefont
  {Seljak}(1997)}]{Zaldarriaga:1996xe}%
  \BibitemOpen
  \bibfield  {author} {\bibinfo {author} {\bibfnamefont {Matias}\ \bibnamefont
  {Zaldarriaga}}\ and\ \bibinfo {author} {\bibfnamefont {Uros}\ \bibnamefont
  {Seljak}},\ }\bibfield  {title} {\enquote {\bibinfo {title} {{An all sky
  analysis of polarization in the microwave background}},}\ }\href {\doibase
  10.1103/PhysRevD.55.1830} {\bibfield  {journal} {\bibinfo  {journal}
  {Phys.~Rev.~D}\ }\textbf {\bibinfo {volume} {55}},\ \bibinfo {pages}
  {1830--1840} (\bibinfo {year} {1997})},\ \Eprint
  {http://arxiv.org/abs/astro-ph/9609170} {arXiv:astro-ph/9609170 [astro-ph]}
  \BibitemShut {NoStop}%
\bibitem [{\citenamefont {Dubovsky}\ \emph {et~al.}(2004)\citenamefont
  {Dubovsky}, \citenamefont {Gorbunov},\ and\ \citenamefont
  {Rubtsov}}]{Dubovsky:2003yn}%
  \BibitemOpen
  \bibfield  {author} {\bibinfo {author} {\bibfnamefont {S.~L.}\ \bibnamefont
  {Dubovsky}}, \bibinfo {author} {\bibfnamefont {D.~S.}\ \bibnamefont
  {Gorbunov}}, \ and\ \bibinfo {author} {\bibfnamefont {G.~I.}\ \bibnamefont
  {Rubtsov}},\ }\bibfield  {title} {\enquote {\bibinfo {title} {{Narrowing the
  window for millicharged particles by CMB anisotropy}},}\ }\href {\doibase
  10.1134/1.1675909} {\bibfield  {journal} {\bibinfo  {journal} {JETP Lett.}\
  }\textbf {\bibinfo {volume} {79}},\ \bibinfo {pages} {1--5} (\bibinfo {year}
  {2004})},\ \bibinfo {note} {[Pisma Zh. Eksp. Teor. Fiz.79,3(2004)]},\ \Eprint
  {http://arxiv.org/abs/hep-ph/0311189} {arXiv:hep-ph/0311189 [hep-ph]}
  \BibitemShut {NoStop}%
\bibitem [{\citenamefont {Dolgov}\ \emph {et~al.}(2013)\citenamefont {Dolgov},
  \citenamefont {Dubovsky}, \citenamefont {Rubtsov},\ and\ \citenamefont
  {Tkachev}}]{Dolgov:2013una}%
  \BibitemOpen
  \bibfield  {author} {\bibinfo {author} {\bibfnamefont {A.~D.}\ \bibnamefont
  {Dolgov}}, \bibinfo {author} {\bibfnamefont {S.~L.}\ \bibnamefont
  {Dubovsky}}, \bibinfo {author} {\bibfnamefont {G.~I.}\ \bibnamefont
  {Rubtsov}}, \ and\ \bibinfo {author} {\bibfnamefont {I.~I.}\ \bibnamefont
  {Tkachev}},\ }\bibfield  {title} {\enquote {\bibinfo {title} {{Constraints on
  millicharged particles from Planck data}},}\ }\href {\doibase
  10.1103/PhysRevD.88.117701} {\bibfield  {journal} {\bibinfo  {journal} {Phys.
  Rev.}\ }\textbf {\bibinfo {volume} {D88}},\ \bibinfo {pages} {117701}
  (\bibinfo {year} {2013})},\ \Eprint {http://arxiv.org/abs/1310.2376}
  {arXiv:1310.2376 [hep-ph]} \BibitemShut {NoStop}%
\bibitem [{\citenamefont {Audren}\ \emph {et~al.}(2013)\citenamefont {Audren},
  \citenamefont {Lesgourgues}, \citenamefont {Benabed},\ and\ \citenamefont
  {Prunet}}]{Audren:2012wb}%
  \BibitemOpen
  \bibfield  {author} {\bibinfo {author} {\bibfnamefont {Benjamin}\
  \bibnamefont {Audren}}, \bibinfo {author} {\bibfnamefont {Julien}\
  \bibnamefont {Lesgourgues}}, \bibinfo {author} {\bibfnamefont {Karim}\
  \bibnamefont {Benabed}}, \ and\ \bibinfo {author} {\bibfnamefont {Simon}\
  \bibnamefont {Prunet}},\ }\bibfield  {title} {\enquote {\bibinfo {title}
  {{Conservative constraints on early cosmology with MONTE PYTHON}},}\ }\href
  {\doibase 10.1088/1475-7516/2013/02/001} {\bibfield  {journal} {\bibinfo
  {journal} {J.~Cosmol.~Astropart.~Phys.}\ }\textbf {\bibinfo {volume} {2}},\
  \bibinfo {eid} {001} (\bibinfo {year} {2013})},\ \Eprint
  {http://arxiv.org/abs/1210.7183} {arXiv:1210.7183 [astro-ph.CO]} \BibitemShut
  {NoStop}%
\bibitem [{\citenamefont {{Buchner}}\ \emph {et~al.}(2014)\citenamefont
  {{Buchner}}, \citenamefont {{Georgakakis}}, \citenamefont {{Nandra}},
  \citenamefont {{Hsu}}, \citenamefont {{Rangel}}, \citenamefont {{Brightman}},
  \citenamefont {{Merloni}}, \citenamefont {{Salvato}}, \citenamefont
  {{Donley}},\ and\ \citenamefont {{Kocevski}}}]{2014A&A...564A.125B}%
  \BibitemOpen
  \bibfield  {author} {\bibinfo {author} {\bibfnamefont {J.}~\bibnamefont
  {{Buchner}}}, \bibinfo {author} {\bibfnamefont {A.}~\bibnamefont
  {{Georgakakis}}}, \bibinfo {author} {\bibfnamefont {K.}~\bibnamefont
  {{Nandra}}}, \bibinfo {author} {\bibfnamefont {L.}~\bibnamefont {{Hsu}}},
  \bibinfo {author} {\bibfnamefont {C.}~\bibnamefont {{Rangel}}}, \bibinfo
  {author} {\bibfnamefont {M.}~\bibnamefont {{Brightman}}}, \bibinfo {author}
  {\bibfnamefont {A.}~\bibnamefont {{Merloni}}}, \bibinfo {author}
  {\bibfnamefont {M.}~\bibnamefont {{Salvato}}}, \bibinfo {author}
  {\bibfnamefont {J.}~\bibnamefont {{Donley}}}, \ and\ \bibinfo {author}
  {\bibfnamefont {D.}~\bibnamefont {{Kocevski}}},\ }\bibfield  {title}
  {\enquote {\bibinfo {title} {{X-ray spectral modelling of the AGN obscuring
  region in the CDFS: Bayesian model selection and catalogue}},}\ }\href
  {\doibase 10.1051/0004-6361/201322971} {\bibfield  {journal} {\bibinfo
  {journal} {Astron.~Astrophys.}\ }\textbf {\bibinfo {volume} {564}},\ \bibinfo
  {eid} {A125} (\bibinfo {year} {2014})},\ \Eprint
  {http://arxiv.org/abs/1402.0004} {arXiv:1402.0004 [astro-ph.HE]} \BibitemShut
  {NoStop}%
\bibitem [{\citenamefont {{Feroz}}\ and\ \citenamefont
  {{Hobson}}(2008)}]{Feroz:2007kg}%
  \BibitemOpen
  \bibfield  {author} {\bibinfo {author} {\bibfnamefont {F.}~\bibnamefont
  {{Feroz}}}\ and\ \bibinfo {author} {\bibfnamefont {M.~P.}\ \bibnamefont
  {{Hobson}}},\ }\bibfield  {title} {\enquote {\bibinfo {title} {{Multimodal
  nested sampling: an efficient and robust alternative to Markov Chain Monte
  Carlo methods for astronomical data analyses}},}\ }\href {\doibase
  10.1111/j.1365-2966.2007.12353.x} {\bibfield  {journal} {\bibinfo  {journal}
  {Mon.~Not.~R.~Astron.~Soc.}\ }\textbf {\bibinfo {volume} {384}},\ \bibinfo
  {pages} {449--463} (\bibinfo {year} {2008})},\ \Eprint
  {http://arxiv.org/abs/0704.3704} {arXiv:0704.3704} \BibitemShut {NoStop}%
\bibitem [{\citenamefont {{Feroz}}\ \emph {et~al.}(2009)\citenamefont
  {{Feroz}}, \citenamefont {{Hobson}},\ and\ \citenamefont
  {{Bridges}}}]{Feroz:2008xx}%
  \BibitemOpen
  \bibfield  {author} {\bibinfo {author} {\bibfnamefont {F.}~\bibnamefont
  {{Feroz}}}, \bibinfo {author} {\bibfnamefont {M.~P.}\ \bibnamefont
  {{Hobson}}}, \ and\ \bibinfo {author} {\bibfnamefont {M.}~\bibnamefont
  {{Bridges}}},\ }\bibfield  {title} {\enquote {\bibinfo {title} {{MULTINEST:
  an efficient and robust Bayesian inference tool for cosmology and particle
  physics}},}\ }\href {\doibase 10.1111/j.1365-2966.2009.14548.x} {\bibfield
  {journal} {\bibinfo  {journal} {Mon.~Not.~R.~Astron.~Soc.}\ }\textbf
  {\bibinfo {volume} {398}},\ \bibinfo {pages} {1601--1614} (\bibinfo {year}
  {2009})},\ \Eprint {http://arxiv.org/abs/0809.3437} {arXiv:0809.3437}
  \BibitemShut {NoStop}%
\bibitem [{\citenamefont {{Feroz}}\ \emph {et~al.}()\citenamefont {{Feroz}},
  \citenamefont {{Hobson}}, \citenamefont {{Cameron}},\ and\ \citenamefont
  {{Pettitt}}}]{Feroz:2013hea}%
  \BibitemOpen
  \bibfield  {author} {\bibinfo {author} {\bibfnamefont {F.}~\bibnamefont
  {{Feroz}}}, \bibinfo {author} {\bibfnamefont {M.~P.}\ \bibnamefont
  {{Hobson}}}, \bibinfo {author} {\bibfnamefont {E.}~\bibnamefont {{Cameron}}},
  \ and\ \bibinfo {author} {\bibfnamefont {A.~N.}\ \bibnamefont {{Pettitt}}},\
  }\bibfield  {title} {\enquote {\bibinfo {title} {{Importance Nested Sampling
  and the MultiNest Algorithm}},}\ }\href@noop {} {\ }\Eprint
  {http://arxiv.org/abs/1306.2144} {arXiv:1306.2144 [astro-ph.IM]} \BibitemShut
  {NoStop}%
\bibitem [{\citenamefont {Gelman}\ and\ \citenamefont
  {Rubin}(1992)}]{Gelman:1992zz}%
  \BibitemOpen
  \bibfield  {author} {\bibinfo {author} {\bibfnamefont {Andrew}\ \bibnamefont
  {Gelman}}\ and\ \bibinfo {author} {\bibfnamefont {Donald~B.}\ \bibnamefont
  {Rubin}},\ }\bibfield  {title} {\enquote {\bibinfo {title} {{Inference from
  Iterative Simulation Using Multiple Sequences}},}\ }\href {\doibase
  10.1214/ss/1177011136} {\bibfield  {journal} {\bibinfo  {journal} {Statist.
  Sci.}\ }\textbf {\bibinfo {volume} {7}},\ \bibinfo {pages} {457--472}
  (\bibinfo {year} {1992})}\BibitemShut {NoStop}%
\bibitem [{\citenamefont {Foreman-Mackey}(2016)}]{corner}%
  \BibitemOpen
  \bibfield  {author} {\bibinfo {author} {\bibfnamefont {Daniel}\ \bibnamefont
  {Foreman-Mackey}},\ }\bibfield  {title} {\enquote {\bibinfo {title}
  {corner.py: Scatterplot matrices in python},}\ }\href {\doibase
  10.21105/joss.00024} {\bibfield  {journal} {\bibinfo  {journal} {The Journal
  of Open Source Software}\ }\textbf {\bibinfo {volume} {24}} (\bibinfo {year}
  {2016}),\ 10.21105/joss.00024}\BibitemShut {NoStop}%
\bibitem [{\citenamefont {Li}\ \emph {et~al.}(2018)\citenamefont {Li},
  \citenamefont {Gluscevic}, \citenamefont {Boddy},\ and\ \citenamefont
  {Madhavacheril}}]{Li:2018zdm}%
  \BibitemOpen
  \bibfield  {author} {\bibinfo {author} {\bibfnamefont {Zack}\ \bibnamefont
  {Li}}, \bibinfo {author} {\bibfnamefont {Vera}\ \bibnamefont {Gluscevic}},
  \bibinfo {author} {\bibfnamefont {Kimberly~K.}\ \bibnamefont {Boddy}}, \ and\
  \bibinfo {author} {\bibfnamefont {Mathew~S.}\ \bibnamefont {Madhavacheril}},\
  }\bibfield  {title} {\enquote {\bibinfo {title} {{Disentangling Dark Physics
  with Cosmic Microwave Background Experiments}},}\ }\href@noop {} {\
  (\bibinfo {year} {2018})},\ \Eprint {http://arxiv.org/abs/1806.10165}
  {arXiv:1806.10165 [astro-ph.CO]} \BibitemShut {NoStop}%
\bibitem [{\citenamefont {de~Putter}\ \emph {et~al.}(2018)\citenamefont
  {de~Putter}, \citenamefont {Dor{\'e}}, \citenamefont {Gleyzes}, \citenamefont
  {Green},\ and\ \citenamefont {Meyers}}]{dePutter:2018xte}%
  \BibitemOpen
  \bibfield  {author} {\bibinfo {author} {\bibfnamefont {Roland}\ \bibnamefont
  {de~Putter}}, \bibinfo {author} {\bibfnamefont {Olivier}\ \bibnamefont
  {Dor{\'e}}}, \bibinfo {author} {\bibfnamefont {J{\'e}r{\^o}me}\ \bibnamefont
  {Gleyzes}}, \bibinfo {author} {\bibfnamefont {Daniel}\ \bibnamefont {Green}},
  \ and\ \bibinfo {author} {\bibfnamefont {Joel}\ \bibnamefont {Meyers}},\
  }\bibfield  {title} {\enquote {\bibinfo {title} {{Dark Matter Interactions,
  Helium, and the CMB}},}\ }\href@noop {} {\  (\bibinfo {year} {2018})},\
  \Eprint {http://arxiv.org/abs/1805.11616} {arXiv:1805.11616 [astro-ph.CO]}
  \BibitemShut {NoStop}%
\bibitem [{\citenamefont {Cooke}\ \emph {et~al.}(2018)\citenamefont {Cooke},
  \citenamefont {Pettini},\ and\ \citenamefont {Steidel}}]{Cooke:2017cwo}%
  \BibitemOpen
  \bibfield  {author} {\bibinfo {author} {\bibfnamefont {Ryan~J.}\ \bibnamefont
  {Cooke}}, \bibinfo {author} {\bibfnamefont {Max}\ \bibnamefont {Pettini}}, \
  and\ \bibinfo {author} {\bibfnamefont {Charles~C.}\ \bibnamefont {Steidel}},\
  }\bibfield  {title} {\enquote {\bibinfo {title} {{One Percent Determination
  of the Primordial Deuterium Abundance}},}\ }\href {\doibase
  10.3847/1538-4357/aaab53} {\bibfield  {journal} {\bibinfo  {journal}
  {Astrophys. J.}\ }\textbf {\bibinfo {volume} {855}},\ \bibinfo {pages} {102}
  (\bibinfo {year} {2018})},\ \Eprint {http://arxiv.org/abs/1710.11129}
  {arXiv:1710.11129 [astro-ph.CO]} \BibitemShut {NoStop}%
\bibitem [{\citenamefont {Zavarygin}\ \emph {et~al.}(2018)\citenamefont
  {Zavarygin}, \citenamefont {Webb}, \citenamefont {Riemer-Sørensen},\ and\
  \citenamefont {Dumont}}]{Zavarygin:2018dbk}%
  \BibitemOpen
  \bibfield  {author} {\bibinfo {author} {\bibfnamefont {E.~O.}\ \bibnamefont
  {Zavarygin}}, \bibinfo {author} {\bibfnamefont {J.~K.}\ \bibnamefont {Webb}},
  \bibinfo {author} {\bibfnamefont {S.}~\bibnamefont {Riemer-Sørensen}}, \
  and\ \bibinfo {author} {\bibfnamefont {V.}~\bibnamefont {Dumont}},\
  }\bibfield  {title} {\enquote {\bibinfo {title} {{Primordial deuterium
  abundance at $z_{abs}$ = 2:504 towards Q1009+2956}},}\ }\bibfield
  {booktitle} {\emph {\bibinfo {booktitle} {{Proceedings, International Youth
  Conference PhysicA.SPb/2017: Saint-Petersburg, Russia, October 24-26,
  2014}}},\ }\href {\doibase 10.1088/1742-6596/1038/1/012012} {\bibfield
  {journal} {\bibinfo  {journal} {J. Phys. Conf. Ser.}\ }\textbf {\bibinfo
  {volume} {1038}},\ \bibinfo {pages} {012012} (\bibinfo {year} {2018})},\
  \Eprint {http://arxiv.org/abs/1801.04704} {arXiv:1801.04704 [astro-ph.CO]}
  \BibitemShut {NoStop}%
\bibitem [{\citenamefont {Aghanim}\ \emph {et~al.}(2018)\citenamefont {Aghanim}
  \emph {et~al.}}]{Aghanim:2018eyx}%
  \BibitemOpen
  \bibfield  {author} {\bibinfo {author} {\bibfnamefont {N.}~\bibnamefont
  {Aghanim}} \emph {et~al.} (\bibinfo {collaboration} {Planck}),\ }\bibfield
  {title} {\enquote {\bibinfo {title} {{Planck 2018 results. VI. Cosmological
  parameters}},}\ }\href@noop {} {\  (\bibinfo {year} {2018})},\ \Eprint
  {http://arxiv.org/abs/1807.06209} {arXiv:1807.06209 [astro-ph.CO]}
  \BibitemShut {NoStop}%
\bibitem [{\citenamefont {Seager}\ \emph {et~al.}(1999)\citenamefont {Seager},
  \citenamefont {Sasselov},\ and\ \citenamefont {Scott}}]{Seager:1999bc}%
  \BibitemOpen
  \bibfield  {author} {\bibinfo {author} {\bibfnamefont {Sara}\ \bibnamefont
  {Seager}}, \bibinfo {author} {\bibfnamefont {Dimitar~D.}\ \bibnamefont
  {Sasselov}}, \ and\ \bibinfo {author} {\bibfnamefont {Douglas}\ \bibnamefont
  {Scott}},\ }\bibfield  {title} {\enquote {\bibinfo {title} {{A new
  calculation of the recombination epoch}},}\ }\href {\doibase 10.1086/312250}
  {\bibfield  {journal} {\bibinfo  {journal} {Astrophys.~J.}\ }\textbf
  {\bibinfo {volume} {523}},\ \bibinfo {pages} {L1--L5} (\bibinfo {year}
  {1999})},\ \Eprint {http://arxiv.org/abs/astro-ph/9909275}
  {arXiv:astro-ph/9909275 [astro-ph]} \BibitemShut {NoStop}%
\bibitem [{\citenamefont {Ali-Ha{\"\i}moud}\ and\ \citenamefont
  {Hirata}(2011)}]{AliHaimoud:2010dx}%
  \BibitemOpen
  \bibfield  {author} {\bibinfo {author} {\bibfnamefont {Yacine}\ \bibnamefont
  {Ali-Ha{\"\i}moud}}\ and\ \bibinfo {author} {\bibfnamefont {Christopher~M.}\
  \bibnamefont {Hirata}},\ }\bibfield  {title} {\enquote {\bibinfo {title}
  {{HyRec: A fast and highly accurate primordial hydrogen and helium
  recombination code}},}\ }\href {\doibase 10.1103/PhysRevD.83.043513}
  {\bibfield  {journal} {\bibinfo  {journal} {Phys.~Rev.~D}\ }\textbf {\bibinfo
  {volume} {83}},\ \bibinfo {pages} {043513} (\bibinfo {year} {2011})},\
  \Eprint {http://arxiv.org/abs/1011.3758} {arXiv:1011.3758 [astro-ph.CO]}
  \BibitemShut {NoStop}%
\bibitem [{\citenamefont {Rubino-Martin}\ \emph {et~al.}(2010)\citenamefont
  {Rubino-Martin}, \citenamefont {Chluba}, \citenamefont {Fendt},\ and\
  \citenamefont {Wandelt}}]{RubinoMartin:2009ry}%
  \BibitemOpen
  \bibfield  {author} {\bibinfo {author} {\bibfnamefont {J.~A.}\ \bibnamefont
  {Rubino-Martin}}, \bibinfo {author} {\bibfnamefont {J.}~\bibnamefont
  {Chluba}}, \bibinfo {author} {\bibfnamefont {W.~A.}\ \bibnamefont {Fendt}}, \
  and\ \bibinfo {author} {\bibfnamefont {B.~D.}\ \bibnamefont {Wandelt}},\
  }\bibfield  {title} {\enquote {\bibinfo {title} {{Estimating the impact of
  recombination uncertainties on the cosmological parameter constraints from
  cosmic microwave background experiments}},}\ }\href {\doibase
  10.1111/j.1365-2966.2009.16136.x} {\bibfield  {journal} {\bibinfo  {journal}
  {Mon.~Not.~R.~Astron.~Soc.}\ }\textbf {\bibinfo {volume} {403}},\ \bibinfo
  {pages} {439} (\bibinfo {year} {2010})},\ \Eprint
  {http://arxiv.org/abs/0910.4383} {arXiv:0910.4383 [astro-ph.CO]} \BibitemShut
  {NoStop}%
\bibitem [{\citenamefont {Chluba}\ and\ \citenamefont
  {Thomas}(2011)}]{Chluba:2010ca}%
  \BibitemOpen
  \bibfield  {author} {\bibinfo {author} {\bibfnamefont {J.}~\bibnamefont
  {Chluba}}\ and\ \bibinfo {author} {\bibfnamefont {R.~M.}\ \bibnamefont
  {Thomas}},\ }\bibfield  {title} {\enquote {\bibinfo {title} {{Towards a
  complete treatment of the cosmological recombination problem}},}\ }\href
  {\doibase 10.1111/j.1365-2966.2010.17940.x} {\bibfield  {journal} {\bibinfo
  {journal} {Mon.~Not.~R.~Astron.~Soc.}\ }\textbf {\bibinfo {volume} {412}},\
  \bibinfo {pages} {748} (\bibinfo {year} {2011})},\ \Eprint
  {http://arxiv.org/abs/1010.3631} {arXiv:1010.3631 [astro-ph.CO]} \BibitemShut
  {NoStop}%
\bibitem [{\citenamefont {Chluba}\ \emph {et~al.}(2015)\citenamefont {Chluba},
  \citenamefont {Paoletti}, \citenamefont {Finelli},\ and\ \citenamefont
  {Rubiño-Martín}}]{Chluba:2015lpa}%
  \BibitemOpen
  \bibfield  {author} {\bibinfo {author} {\bibfnamefont {Jens}\ \bibnamefont
  {Chluba}}, \bibinfo {author} {\bibfnamefont {D.}~\bibnamefont {Paoletti}},
  \bibinfo {author} {\bibfnamefont {F.}~\bibnamefont {Finelli}}, \ and\
  \bibinfo {author} {\bibfnamefont {Jose-Alberto}\ \bibnamefont
  {Rubiño-Martín}},\ }\bibfield  {title} {\enquote {\bibinfo {title} {{Effect
  of primordial magnetic fields on the ionization history}},}\ }\href {\doibase
  10.1093/mnras/stv1096} {\bibfield  {journal} {\bibinfo  {journal}
  {Mon.~Not.~R.~Astron.~Soc.}\ }\textbf {\bibinfo {volume} {451}},\ \bibinfo
  {pages} {2244--2250} (\bibinfo {year} {2015})},\ \Eprint
  {http://arxiv.org/abs/1503.04827} {arXiv:1503.04827 [astro-ph.CO]}
  \BibitemShut {NoStop}%
\bibitem [{\citenamefont {Scott}\ and\ \citenamefont
  {Moss}(2009)}]{Scott:2009sz}%
  \BibitemOpen
  \bibfield  {author} {\bibinfo {author} {\bibfnamefont {Douglas}\ \bibnamefont
  {Scott}}\ and\ \bibinfo {author} {\bibfnamefont {Adam}\ \bibnamefont
  {Moss}},\ }\bibfield  {title} {\enquote {\bibinfo {title} {{Matter
  temperature after cosmological recombination}},}\ }\href {\doibase
  10.1111/j.1365-2966.2009.14939.x} {\bibfield  {journal} {\bibinfo  {journal}
  {Mon.~Not.~R.~Astron.~Soc.}\ }\textbf {\bibinfo {volume} {397}},\ \bibinfo
  {pages} {445--446} (\bibinfo {year} {2009})},\ \Eprint
  {http://arxiv.org/abs/0902.3438} {arXiv:0902.3438 [astro-ph.CO]} \BibitemShut
  {NoStop}%
\end{thebibliography}%

\end{document}